\documentstyle[epsf]{crnart12}
\def\Journal#1#2#3#4{{#1} {\bf #2}, #3 (#4)}
\def\add#1#2#3{{\bf #1}, #2 (#3)}

\def\NPB{{\em Nucl. Phys.} B}
\def\PLB{{\em Phys. Lett.}  B}
\def\PRL{{\em Phys. Rev. Lett.}}
\def\PRD{{\em Phys. Rev.} D}
\def\PR{{\em Phys. Rev.}}
\def\ZPC{{\em Z. Phys.} C}

\def\PTPS{{\em Prog. Theor. Phys. Suppl.}}

\def\beq{\begin{equation}}
\def\eeq#1{\label{#1}\end{equation}}
\def\eeqn{\end{equation}}
\def\beqa{\begin{eqnarray}}
\def\eeqa#1{\label{#1}\end{eqnarray}}
\def\eeqan{\end{eqnarray}}
\def\CR{\nonumber \\ }
\def\leqn#1{(\ref{#1})}

\let\bar=\overbar
\def\Dslash{\not{\hbox{\kern-4pt $D$}}}
\def\dslash{\not{\hbox{\kern-2pt $\del$}}}
\def\half{\frac{1}{2}}
\def\thalf{\frac{3}{2}}

\def\E{{\cal E}}

\def\L{{\cal L}}
\def\M{{\cal M}}

\def\tr{{\mbox{\rm tr}}}
\def\del{\partial}

\def\VEV#1{\left\langle{ #1} \right\rangle}
\def\bra#1{\left\langle{ #1} \right|}
\def\ket#1{\left| {#1} \right\rangle}
\def\vev#1{\langle #1 \rangle}
\def\eff{{\mbox{\rm eff}}}
\def\hc{{\mbox{\rm h.c.}}}
\def\Pl{{\mbox{\scriptsize Pl}}}
\def\eff{{\mbox{\scriptsize eff}}}
\def\SM{{\mbox{\scriptsize SM}}}

\def\ee{e^+e^-}
\def\CM{{\mbox{\scriptsize CM}}}

\def\L{{\cal L}}
\def\M{{\cal M}}
\def\sstw{\sin^2\theta_w}
\def\cstw{\cos^2\theta_w}
\def\mz{m_Z}

\def\mw{m_W}

\def\msb{{\bar{\ssstyle M \kern -1pt S}}}

\def\ch#1{\widetilde\chi^+_{#1}}
\def\chm#1{\widetilde\chi^-_{#1}}
\def\ne#1{\widetilde\chi^0_{#1}}

\def\etal{{\it et al.}}

\newcommand\pubnumber{SLAC-PUB-7479}
\newcommand\pubdate{May, 1997}
\def\Title#1{\begin{center} {\Large #1 } \end{center}}
\def\Author#1{\begin{center}{ \sc #1} \end{center}}
\def\Address#1{\begin{center}{ \it #1} \end{center}}

\def\doeack{\footnote{Work supported by the Department of Energy,
                     contract DE--AC03--76SF00515.}}
\def\SLAC{Stanford Linear Accelerator Center\\
    Stanford University, Stanford, California 94309 USA}
\newcommand\pubblock{\rightline{\begin{tabular}{l} \pubnumber\\
         \pubdate  \end{tabular}}}
\newenvironment{Abstract}{\begin{quotation} \begin{center}
                       ABSTRACT
     \end{center}\bigskip  }{\end{quotation}}

\begin{document}

\begin{titlepage}
\pubblock

\vfill
\Title{Beyond the Standard Model}
\vfill
\Author{Michael E. Peskin\doeack}
\Address{\SLAC}
\vfill
\begin{Abstract}
These lectures constitute a short course in `Beyond the Standard Model'
for students of experimental particle physics.  I discuss the general
ideas which guide the construction of models of physics beyond the
Standard Model. The central principle, the one which most
directly motivates the  search for new physics, is the search for the
mechanism of the spontaneous symmetry breaking observed in the theory
of weak interactions.  To illustrate models of weak-interaction
symmetry breaking, I give a detailed discussion of the idea of
supersymmetry and that of new strong interactions at the
TeV energy scale. I discuss experiments that will probe
the details of these models at future $pp$ and $\ee$ colliders.
\end{Abstract}
\vfill
\begin{center}
to appear in the proceedings of the \\
1996 European School of High-Energy Physics\\
Carry-le-Rouet, France, September 1--14, 1996
\end{center}

\vfill
\end{titlepage}
\def\thefootnote{\fnsymbol{footnote}}
\setcounter{footnote}{0}
\hbox to \hsize{\hfill }
\newpage
\setcounter{page}{1}
\vspace*{2cm}
\begin{flushleft}
{\bf BEYOND THE STANDARD MODEL}\\
Michael E. Peskin\\
SLAC, Stanford University, Stanford, California USA
\vspace{2cm}
\end{flushleft}

{\rightskip=5pc
\leftskip=5pc
\noindent
{\bf Abstract}\\
These lectures constitute a short course in `Beyond the Standard Model'
for students of experimental particle physics.  I discuss the general
ideas which guide the construction of models of physics beyond the
Standard Model. The central principle, the one which most
directly motivates the  search for new physics, is the search for the
mechanism of the spontaneous symmetry breaking observed in the theory
of weak interactions.  To illustrate models of weak-interaction
symmetry breaking, I give a detailed discussion of the idea of
supersymmetry and that of new strong interactions at the
TeV energy scale. I discuss experiments that will probe
the details of these models at future $pp$ and $\ee$ colliders.
\vglue 2.0cm}

\section{Introduction}	

Every year, the wise people who organize the European School of
Particle Physics feel it necessary to subject young experimentalists to
a course of lectures on `Beyond the Standard Model'.  They treat this
subject as if it were a discipline of science that one could study and
master. Of course, it is no such thing.  If we knew what lies beyond
the Standard Model, we could teach it with some confidence.  But the
interest in this subject is precisely that we do not know what is
waiting for us there.

The confusion about `Beyond the Standard Model' goes beyond students
and summer school organizers to the senior scientists in our field.  A
theorist such as myself who claims to be able to explain things about
physics beyond the Standard Model is very often met with skepticism
that such explanations are even possible.  `Do we really have any
idea', one is told, `what we will find a higher energies?'  `Don't we
just want the highest possible energy and luminosity?'  `The Standard
Model works very well, so why must there be any new physics at all?'

And yet there are specific things that one can teach that should be
relevant to physics beyond the Standard Model. Though we do not know
what physics to expect at higher energies, the principles of
physics that we have learned in the explication of the
Standard Model should still apply there.  In addition, we hope that
some of the questions not answered by the Standard Model should be
answered there.  This course will concentrate its attention on these
two issues: What questions are likely to be addressed by new physics
beyond the Standard Model, and what general methods of analysis can we
use to create and analyze proposed answers to these questions?

A set of lectures on `Beyond the Standard Model' should have one
further goal as well.  It is possible that the first sign of physics
beyond the Standard Model could be discovered next year at LEP, or
perhaps it is already waiting in the unanalyzed data from the Fermilab
collider. On the other hand, it is possible that this discovery will
have to wait for the great machines of the next generation.
  Many people feel dismay at the fact
that the pace of discovery in high-energy physics is very slow, with
experiments operating on the time scale of a decade familiar in
planetary science rather than on the time scale of days or weeks. 
Because of the cost and complexity of modern elementary particle
experiments, these long time scales are inevitable, and we have to
adjust our expectations to them.  But the long time scales also require
that we set for ourselves very clear goals that we can try to realize a
decade in the future.  To do this, it is useful to have a concrete
understanding of what experiments will look like at the next generation
of colliders and what physics issues they address.  Even if we cannot
correctly predict what Nature will provide for us at higher energy, it
is essential to take some models as illustrative examples
and work out in complete detail how
to analyze them experimentally.  With luck, we can choose models will
have features relevant to the ultimate correct theory of the next scale
in physics.  But even if we are not sufficiently lucky or insightful to
predict what will appear, such a study will leave us prepared to solve
whatever puzzles Nature has set.

This, then, is what I would like to accomplish in these lectures.  I
will set out some questions which I feel are the most important ones at
the present stage of our  understanding, and the ones which I feel are
most likely to be addressed by the new phenomena of the next energy
scale. I will explain some theoretical ideas that have come from our
understanding of the Standard Model that I feel will play an important
role at the next level.  Building on these ideas, 
 I will describe illustrative models of
physics beyond the Standard Model.  And, for each case, I will describe
the program of
experiments that will clarify the nature of the new physics that the model
implies.

When we design a  program of future high-energy experiments, we are also
calling for the construction of new high-energy accelerators that 
would be needed to carry out this program.  I hope that 
students of high-energy physics will take an interest in this practical
or political aspect of our field of 
science.  Those who think about this seriously know
that we cannot ask society to support such expensive machines unless we 
can promise that these facilities will give back fundamental knowledge that
is of the utmost importance and that cannot be obtained in any other way.
I hope that they will be interested to see how central a role the 
CERN Large Hadron Collider (LHC) plays in each of the experimental programs
that I will describe.   Another proposed facility will also play a major
role in my discussion, a high-energy $\ee$ linear collider with 
center-of-mass energy about 1 TeV.   I will argue in these lectures that,
with these  facilities, the scientific
justification changes qualitatively from that
of the present colliders at CERN and Fermilab.  Whereas at current energies,
we search for new physics and try to place limits, at next step in energy we
must find new physics that addresses one of the major gaps in the
Standard Model.

This last issue leads to us to ask another, and perhaps unfamiliar, 
question about the colliders of the next generation.
 Much ink has been wasted in comparing hadron and lepton colliders 
on the basis of energy reach and asking which is preferable.  The real 
issue for these machines is a different one.
  We will see that illustrative models
of new physics based on simple ideas will out to have rich and 
complex phenomenological consequences.  Thus, it is a serious question whether
we will be able to understand the model that Nature has put forward for us
from experimental observations.  I will argue through my 
examples  that these two 
types of colliders, which focus on different and complementary aspects
of the high-energy phenomena, can bring back a complete picture of the new 
phenomena of a clarity that neither, working alone, could achieve.

The outline of these lectures is as follows.  In Section 2, I will
introduce the question of the mechanism of electroweak symmetric
breaking and also two related questions that influence the construction
and analysis of models of new physics.  In Sections 3 and 4, I will
give one illustrative set of answers to these questions through a
detailed discussion of models with supersymmetry at the
weak-interaction scale.  Section 3 will develop the formalism of
supersymmetry and derive its connection to the questions I have set
out.  Section 4 will discuss more detailed properties of supersymmetric
models which provide interesting experimental probes. In Section 5, I
will discuss models with new strong interactions at the TeV mass scale,
models which give very different answers to our broad questions about
physics beyond the Standard Model.  In Section 6, I will summarize the
lessons of our study of these two very different types of models and
draw some general conclusions.

\section{Three Basic Questions}

To begin our study of physics beyond the Standard Model, I will
 review some properties of the Standard Model and some insights that
it provides.  I will also discuss some questions that the Standard
Model does not answer, but which might reasonably be answered at the
next scale in fundamental physics.

\subsection{Why not just the Standard Model?}

To introduce the study of physics beyond the Standard Model, I must
first explain what is wrong with the Standard Model.  To see this, we
only have to compare the publicity for the Standard Model, what we say
about it to beginning
students and to our colleagues in other fields, with the
explicit expression for the Standard Model Lagrangian.

When we want to advertise the virtues of the Standard Model, we say
that it is a model whose foundation is symmetry.  We start from the
principle of local gauge invariance, which tells us that the
interactions of vector bosons are associated with a  global symmetry
group.  The form of these interactions is  uniquely specified by the
group structure. Thus, from the knowledge of the basic symmetry group,
we can write down the Lagrangian or the equations of motion. 
Specifying the group to be $U(1)$, we derive electromagnetism. To
create a complete theory of Nature, we choose the group, in accord with
observation,  to be $SU(3)\times SU(2)\times U(1)$.  This group is a
product, and we are free to include a different coupling constant for
each factor.  But in the ideal theory, these would be the only
parameters.  Specify to which representations of the gauge group the
matter particles belong, fix the three coupling constants, and we have
a complete theory of Nature.

This set of ideas is tantalizing because it is so close to being true.
The couplings of quarks and leptons to the strong, weak, and
electromagnetic interactions are indeed fixed correctly in terms of
three coupling constants.  From the LEP and SLC experiments, we have
learned that the pattern of weak-interaction couplings of the quarks
and leptons follows the symmetry prediction to the accuracy of a few
percent, and also that the strong-interaction coupling is universal
among quark flavors at a similar level of accuracy.

On the other hand, the Lagrangian of the Minimal Standard Model tells a rather
different story.  Let me write it here for reference:
\beqa
 \L &=& \bar q i \Dslash q + \bar\ell i \Dslash \ell -
            \frac{1}{4} (F^a_{\mu\nu})^2  \CR
    & & + \left|D_\mu\phi\right|^2 - V(\phi) \CR
    & & - \left( \lambda^{ij}_u \bar u_R^i \phi\cdot  Q_L^j
 + \lambda^{ij}_d \bar d_R^i\phi^* \cdot Q_L^j + 
 \lambda^{ij}_\ell \bar e_R^i \phi^* \cdot L_L^j + \hc \right) \ . 
\eeqa{eq:a}

The first line of \leqn{eq:a} is the pure gauge theory discussed in the
previous paragraph.  This line of the Lagrangian contains only three
parameters, the three Standard Model gauge couplings $g_s$, $g$, $g'$,
and it does correctly describe the couplings of all species of quarks
and leptons to the strong, weak, and electromagnetic gauge bosons.

The second line of \leqn{eq:a} is associated with the Higgs boson field
$\phi$.  The Minimal Standard Model introduces one scalar field, a
doublet of weak interaction $SU(2)$, so that its vacuum expectation
value can give a mass to the $W$ and $Z$ bosons.  The potential energy
of this field $V(\phi)$ contains at least two new parameters which play
a role in determining the $W$ boson mass.  At this moment, there is no
experimental evidence for the existence of the Higgs field $\phi$ and
very little evidence that constrains the form of its potential.

The third line of \leqn{eq:a} similarly gives an origin for the masses
of quarks and leptons.  In the Standard Model, the left- and
right-handed quark fields belong to different representations of
$SU(2)\times U(1)$; a similar conclusion holds for the leptons.  On the
other hand, a mass term for a fermion couples the left- and
right-handed components.  This is impossible as long as the gauge
symmetry is exact.  In the Standard Model, one can write a
trilinear term linking a left- and right-handed pair of species to the
Higgs field.  When the Higgs field acquires a vacuum expectation value,
this coupling turns into a mass term.  Unfortunately, a generic
fermion-fermion-boson coupling is restricted only rather weakly by
gauge symmetries.  The Standard Model gauge symmetry allows three
complex $3\times 3$ matrices of couplings, the 
paramaters $\lambda^{ij}$ of \leqn{eq:a}.  When $\phi$ acquires a
vacuum expectation values, these matrices become the mass matrices of
quarks and leptons.  Thus, whereas the gauge couplings of quarks and
leptons were strongly restricted by symmetry, the mass terms for these
particles can be of general and, indeed, complex, structure.

If we consider \leqn{eq:a} to be the  fundamental Lagrangian
of Nature, the
situation is even worse.  The Higgs coupling matrices $\lambda^{ij}$
are renormalizable couplings in this Lagrangian.  The property of
renormalizability implies that, once these couplings are specified, the
theory gives definite predictions. However, the specification of the
renormalizable couplings is part of the statement of the problem. 
Except in very special field theories, these couplings cannot be
determined from the internal consistency of the theory itself.  The
Standard Model Lagrangian then leaves us with the three matrices
$\lambda^{ij}$, and the parameters of the Higgs potential $V(\phi)$, as
conditions of the problem which cannot in principle be determined.  In
order to understand why the masses of the quarks, the  leptons, and the
$W$ and $Z$ bosons have their observed values, we must find a deeper
theory beyond the Standard Model from which the Lagrangian \leqn{eq:a},
or some replacement for it, can be derived.

Thus, it is a disappointing feature of the Minimal Standard Model that
it has a large number of parameter which are undetermined, and which
cannot be determined. This disappointment, though, has an interesting
converse.  Typically in physics, when we meet a system with a large
number of parameters, what stands behind it is a system with a simple
description which is realized with some complexity in its dynamics. 
The transport coefficients of fluids or the properties of electrons in
a semiconductor are described in terms of a large number of parameters,
but these parameters can be computed from an underlying atomic picture.
Through this analogy, we would conclude that the gauge couplings of
quarks and leptons are likely to reflect a fundamental structure, but
that the Higgs boson is unlikely to be simple, minimal, or elementary. 
The multiplicity of undermined couplings of the Minimal Standard Model
are precisely those of the Higgs boson. If we could break through and
discover the simple underlying picture behind the Higgs boson, or
behind the breaking of $SU(2)\times U(1)$ symmetry, we would then have
the correct deeper viewpoint from which to understand the undetermined
parameters of the Standard Model.

\subsection{Three models of electroweak symmetry breaking}

The argument given in the previous section leads us to the question:
What is actually the mechanism of electroweak symmetry breaking?  In
this section, I would like to present three possible models for this
phenomenon and to discuss their strengths and weaknesses.

The first of these is the model of electroweak symmetry breaking
contained in the Minimal Standard Model.  We introduce a Higgs field
\beq
             \phi = \pmatrix{\phi^+ \cr \phi^0}
\eeq{eq:b}
with $SU(2)\times U(1)$ quantum numbers $I= \frac{1}{2}$, $Y =
\frac{1}{2}$. I will use $\tau^a = \sigma^a/2$ to denote the generators
of $SU(2)$, and I normalize the hypercharge so that the electric charge
is $Q = I^3 + Y$.

Take the Lagrangian for the field $\phi$ to be the second line of
\leqn{eq:a}, with
\beq 
     V(\phi) = - \mu^2 \phi^\dagger \phi + \lambda (\phi^\dagger\phi)^2 \ .
\eeq{eq:c}
This potential is minimized when $\phi^\dagger\phi = \mu^2/2\lambda$. 
Thus, one particular vacuum state is given by
\beq
       \VEV\phi = \pmatrix{0 \cr \frac{1}{\sqrt{2}}v\cr} \ , 
\eeq{eq:d}
where $v^2 = \mu^2/\lambda$.

The most general $\phi$ field configuration can be written in the same
notation as
\beq
      \phi = e^{i\alpha(x)\cdot \tau}
\pmatrix{0 \cr \frac{1}{\sqrt{2}} (v + h(x))\cr} \ . 
\eeq{eq:e}
In this expression, $\alpha^a(x)$ parametrizes an $SU(2)$ gauge
transformation. The field $h(x)$ is a gauge-invariant fluctation away
from the vacuum state; this is the physical Higgs field. The mass of
this field is given by
\beq
            m_h^2 = 2\mu^2 = 2\lambda v^2 \ .
\eeq{eq:ee}
Notice that, in this model, $h(x)$ is the only gauge-invariant degree of
freedom in $\phi(x)$, and so the symmetry-breaking sector gives rise to
only one new particle, the Higgs scalar.

If we insert \leqn{eq:d} into the kinetic term for $\phi$, we obtain
a mass term for $W$ and $Z$; this is the usual Higgs mechanism for
producing these masses.  If $g$ and $g'$ are the $SU(2)\times U(1)$
coupling constants, one finds the familiar result
\beq
      \mw =  g \frac{v}{2} \ , 
\qquad \mz = \sqrt{g^2 + g^{\prime 2}} \frac{v}{2} \ .
\eeq{eq:f}
The measured values of the masses and couplings then lead to 
\beq
        v = 246 \ \mbox{GeV} \ .
\eeq{eq:g}

This is a very simple model of $SU(2)\times U(1)$ symmetry breaking.
Perhaps it is even too simple.  If we ask the question, why is
$SU(2)\times U(1)$ broken, this model gives the answer `because
$(-\mu^2) < 0$.'  This is a perfectly correct answer, but it teaches
us nothing.  Normally, the grand qualitative phenomena of physics
happen as the result of definite physical mechanisms.  But there is no
physically understandable mechanism operating here.

One often hears it said that if the minimal Higgs model is too simple,
one can make the model more complex by adding a second Higgs doublet. 
For our next case, then, let us consider a model with two Higgs
doublets $\phi_1$, $\phi_2$, both with  $I= \frac{1}{2}$, $Y =
\frac{1}{2}$.  The Lagrangian of the Higgs fields is
\beq
         \L = \left| D_\mu\phi_1\right|^2 +  
          \left| D_\mu\phi_1\right|^2 - 
                      V(\phi_1,\phi_2) \ , 
\eeq{eq:h}
with
\beq 
        V =  - \pmatrix{\phi_1^\dagger & \phi_2^\dagger\cr} M^2
 \pmatrix{\phi_1 \cr \phi_2 \cr} +  \cdots \ ,
\eeq{eq:i}
where $M^2$ is a $2\times 2$ matrix.  It is not difficult to engineer a
form for $V$ such that, at the minimum, the vacuum expectation values
of $\phi_1$ and $\phi_2$ are aligned:
\beq
        \VEV{\phi_1} = \pmatrix{0 \cr \frac{1}{\sqrt{2}}v_1\cr} \ , \qquad
     \VEV{\phi_2} = \pmatrix{0 \cr \frac{1}{\sqrt{2}}v_2\cr} \ .
\eeq{eq:j}
The ratio of the two vacuum expectation values is conventionally 
parametrized by an angle $\beta$,
\beq
           \tan\beta = \frac{v_2}{v_1}\ .
\eeq{eq:k}
To reproduce the correct values of the $W$ and $Z$ mass, 
\beq
            v_1^2 + v_2^2 = v^2 = (246\ \mbox{GeV})^2 \ .
\eeq{eq:l}

The field content of this model is considerably richer than that of the
minimal model.  An infinitesimal gauge transformation of the vacuum
configuration \leqn{eq:j} leads to a field configuration
\beq
        \delta\phi_1 =\frac{1}{2} \pmatrix{v_1(\alpha_1+i\alpha_2) \cr
                        v_1( i\alpha_3)\cr} \ , \qquad
     \delta\phi_2 = \frac{1}{2} \pmatrix{v_2(\alpha_1+i\alpha_2) \cr
                        v_2( i\alpha_3)\cr} \ .
\eeq{eq:m}
The fluctuations of the field configuration which are orthogonal to
this lead to new physical particles.  These include the motions
\beq
        \delta\phi_1 =\frac{1}{2} \pmatrix{\sin\beta\cdot (h_1+ih_2) \cr
                        \sin\beta\cdot ( ih_3)\cr} \ , \qquad
     \delta\phi_2 = \frac{1}{2} \pmatrix{-\cos\beta\cdot (h_1 + i h_2) \cr
                        -\cos\beta\cdot( ih_3)\cr} \ ,
\eeq{eq:n}
as well as the fluctuations 
$v_i \to v_i + H_i$  of the two vacuum expectation values.  Thus we
find five new particles.  The fields $h_1$ and $h_2$ combine to form
charged Higgs bosons $H^\pm$.  The field $h_3$ is a CP-odd neutral
boson, usually called $A^0$.  The two fields $H_i$ typically mix to form
mass eigenstates
called $h^0$ and $H^0$.

I have discussed this structure in some detail because we will later
see it appear in specific model contexts.  But it does nothing as far
as answering the physical question that I posed a moment ago.  Again,
if one asks what is the mechanism of weak interaction symmetry
breaking, the answer this model gives is that the matrix $(-M^2)$ has a
negative eigenvalue.

The third model I would like to discuss is a model of a very different
kind proposed in 1979 by Weinberg and Susskind \cite{Wein,Suss}. 
Imagine that the fundamental interactions include a new gauge
interaction which is almost an exact copy of QCD with two quark
flavors. The new interactions differ from QCD in only two respects: 
First, the quarks are massless; second, the nonperturbative scales
$\Lambda$ and $m_\rho$ are  much larger in the new subsection.  The two
flavors of quarks should be coupled to $SU(2)\times U(1)$ just as
$(u,d)$ are, and  I will call them $(U,D)$.

In QCD, the strong interactions between quarks and antiquarks leads to
the generation of large effective masses for the $u$ and $d$.  This
mass generation is associated with spontaneous symmetry breaking. The
strong interactions between very light quarks and antiquarks make it
energetically favorable for the vacuum of space to fill up with
quark-antiquark pairs.  This gives vacuum expectation values to
operators built from quark and antiquark fields.

The analogue of this phenomenon should occur in our theory of new
interactions---for just the same reason---and so we should find
\beq
        \VEV{\bar U U } = \VEV{\bar D D} = - \Delta \neq 0 \ .
\eeq{eq:o}
In terms of chiral components,
\beq
         \bar U U =  U_L^\dagger  U_R + U_R^\dagger U_L \ ,
\eeq{eq:p}
and similarly for $\bar D D$.  But, in the weak-interaction theory, the
left-handed quark fields transform under $SU(2)$ while the right-handed
fields do not. Thus, the vacuum expectation value in \leqn{eq:o}
signals $SU(2)$ symmetry breaking.  In fact, under $SU(2)\times U(1)$,
the operator $\bar Q_L U_R$ has  the  same quantum numbers  $I=
\frac{1}{2}$, $Y = \frac{1}{2}$ as the elementary Higgs boson that we
introduced in our earlier model. The vacuum expectation value of this
operator then has the same effect: It breaks $SU(2)\times U(1)$ to the
$U(1)$ symmetry of electromagnetism and gives mass to the three
weak-interaction bosons.

I will explain in Section 5.1  that the pion decay constant $F_\pi$ of
the new strong interaction theory plays the role of $v$ in \leqn{eq:f}
in determining the mass scale of $\mw$ and $\mz$.  If we were to set
$F_\pi$ to the value given in \leqn{eq:g}, we would need to scale up
QCD by the factor
\beq
            {246\ \mbox{GeV}\over 93\ \mbox{MeV}} = 2600.
\eeq{eq:q} 
Then the hadrons of these new strong interactions would be at TeV
energies.

For me, the Weinberg-Susskind model is much more appealing as a model
of electroweak symmetry breaking than the Minimal Standard Model.  The
reason for this is that, in the Weinberg-Susskind model, electroweak
symmetry breaking happens naturally, for a reason, rather than being
included as the result of an arbitrary choice of parameters. I would
like to emphasize especially that the Weinberg-Susskind model is
preferable even though it is more complex. In fact, this complexity is
an essential part of its foundation.  In this model, {\em something
happens}, and that physical action gives rise to a set of
consequences, of which electroweak symmetry breaking is one.

This notion that the consequences of physical theories flow from their
complexity is familiar from the theories in particle physics that we
understand well. In QCD, quark confinement, the spectrum of hadrons,
and the parton description of high-energy reactions all flow out of the
idea of a strongly-coupled non-Abelian gauge interaction.  In the weak
interactions, the $V$--$A$ structure of weak couplings and all of its
consequences for decays and asymmetries follow from the underlying
gauge structure.

Now we are faced with a new phenomenon, the symmetry breaking of
$SU(2)\times U(1)$, whose explanation  lies outside the realm of the
known gauge theories. Of course it is possible that this phenomenon
could be explained by the simplest, most minimal addition to the laws
of physics. But that is not how we have seen Nature work. In searching
for an explanation of electroweak symmetry breaking, we should not be
searching for a simplistic theory but rather for a simple idea from
which deep and rich consequences might flow.

\subsection{Questions for orientation}

The argument of the previous section gives focus to the study of
physics beyond the Standard Model.  We have a phenomenon necessary to
the working of weak-interaction theory, the symmetry-breaking of
$SU(2)\times U(1)$, which we must understand.  This symmetry-breaking
is characterized by a mass scale, $v$ in \leqn{eq:g}, which is close to
the energy scales now being probed at accelerators.  At the same time,
it is a new qualitative phenomenon which cannot originate from the
known gauge interactions. Therefore, it calls for new physics, and in
an energy region where we can hope to discover it.  For me, this is the
number one question of particle physics today:
\begin{description}
\item [*] {\bf What is the mechanism of electroweak symmetry breaking?}
\end{description}

Along with this question come two subsidiary ones.  Both of these
are connected to the fact that electroweak symmetry breaking is
necessary for the generation of masses for the weak-interaction bosons,
the quarks, and the leptons.  Perhaps there are also other particles
which cannot obtain mass until $SU(2)\times U(1)$ is broken.  Then
these particles also must have masses at the scale of a few hundred GeV
or below.  The heaviest of these particles must be especially strongly
coupled to the fields that are the basic cause of the
symmetry-breaking.  At the very least, the top quark belongs to this
class of very heavy particles, and other members of this class might
well be found.  Thus, we are also led to ask,
\begin{description}
\item [*]
{\bf What is the spectrum of elementary particles at the 1 TeV energy scale?}
\item [*]
{\bf Is the mass of the top quark generated by weak couplings or by 
new strong interactions?}
\end{description}

In the remainder of this section, I will comment on these three
questions. In the following sections, when we consider explicit models
of electroweak symmetry breaking, I will develop the models theoretically
to propose
answer these questions. At any stage in the argument, 
though, you should have firmly in mind that these answers will ultimately
come from experiment, and, in particular, from direct observations
of TeV-energy phenomena.  The goal of my theoretical arguments, then, will
be to suggest particular phenomena which
could  be observed experimentally to shed light on these questions.
We will see in Sections 4 and 5 that models which attempt to explain 
electroweak symmetry breaking typically suggest  a variety of new
experimental probes, which may allow us to 
 uncover a whole new layer of the fundamental 
interactions.

\subsection{General features of electroweak symmetry breaking}

Since the question of electroweak symmetry breaking will be our main
concern, it is important to state at the beginning what we do know
about this phenomenon.  Unfortunately, our knowledge is very limited. 
Basically it consists of only three items.

First, we know the general scale of electroweak symmetry breaking,
which is set by the scale of $\mw$ and $\mz$,
\beq
        v = 246 \ \mbox{GeV} \ .
\eeq{eq:r}
If there are new particles associated with the mechanism of electroweak
symmetry breaking, their masses should be at the scale $v$.  Of course,
this is only an order-of-magnitude estimate.  The precise relation
between $v$ and the masses of new particles depends on the specific
model of electroweak symmetry breaking.  In the course of these
lectures, I will discuss examples in which the most important new
particles lie  below $v$ and other examples in which they lie higher by
a large factor.

Second, we know that the electroweak boson masses follow the pattern
\leqn{eq:f}, that is,
\beq
         {\mw\over \mz} = \cos\theta_w \ , \qquad  m_\gamma = 0 \ .
\eeq{eq:s}
In terms of the original $SU(2)$ and $U(1)$ gauge bosons $A^a_\mu$,
$B_\mu$, this pattern tells us that the mass matrix had the form
\beq
      m^2 =  {v^2\over 2}\pmatrix{ g^2 & & &  \cr & g^2 & & \cr
                     & & g^2 & -gg' \cr  && -gg' & g^{\prime 2}\cr}
\eeq{eq:t}
acting on the vector $(A^1_\mu, A^2_\mu, A^3_\mu, B_\mu)$. Notice that
the $3\times 3$ block of this matrix acting on the $SU(2)$ bosons is
diagonal. This would naturally be a consequence of an unbroken $SU(2)$
symmetry under which $(A^1_\mu, A^2_\mu, A^3_\mu)$ form a triplet
\cite{Marvin,SSVZ}.  This strongly suggests that an unbroken $SU(2)$ symmetry,
called {\em custodial $SU(2)$}, should be included in any successful model
of electroweak symmetry breaking.

The Minimal Standard Model actually contains such a symmetry
accidentally. the  complex doublet $\phi$ can be viewed as a set of
four real-valued fields,
\beq
   \phi =  \frac{1}{\sqrt{2}} \pmatrix{ \phi^1 + i \phi^2\cr
 \phi^3 + i \phi^4\cr} \ .
\eeq{eq:tt}
The Higgs potential \leqn{eq:c} is invariant to $SO(4)$ rotations of
these fields.  The vacuum expectation value \leqn{eq:d} gives an
expectation value to one of the four components and so breaks $SO(4)$
spontaneously to $SO(3) = SU(2)$.  In the Weinberg-Susskind model,
there is also a custodial $SU(2)$ symmetry, the isospin symmetry of the
new strong interactions.  In this case, the custodial symmetry is not
an accident, but rather a component of the new idea.

Third, we know that the new interactions responsible for electroweak
symmetry breaking contribute very little to precision electroweak
observables. I will discuss this constraint in somewhat more detail in
Section 5.2.  For the moment, let me point out that, if we take the value
of the electromagnetic coupling $\alpha$ and the weak interaction
parameters $G_F$ and $\mz$ as input parameters, the value of the weak
mixing angle $\sstw$ that governs the forward-backward and polarization
asymmetries of the $Z^0$ can be shifted by radiative corrections
involving particles associated with the symmetry breaking.  In the
Minimal Standard Model, this shift is rather small,
\beq
  \delta(\sstw) = {\alpha\over \cstw - \sstw} {1 + 9 \sstw\over 24 \pi}
                  \log{m_h\over m_Z} \ .
\eeq{eq:w}
The coefficient of the logarithm has the value $6\times 10^{-4}$. 
The accuracy of the LEP and SLC experiments is such that the size of
the logarithm cannot be much larger than 1, and larger radiative
corrections from additional sources
are forbidden.  In models of electroweak symmetry breaking
based on new strong interactions, this can be an important constraint.

\subsection{The evolution of couplings}

Now I would like to comment similarly on the two subsidiary questions
that I put forward in Section 2.3.  I will begin with the first of
these  questions: What is the spectrum of elementary particles at the 1
TeV energy scale? In the discussion above, I have already argued for
the importance of this question.  Because mass generation in quantum
field theory is associated with symmetry breaking, and because one of
the major symmetries of Nature is broken at the scale $v$, we might
expect a sizeable multiplet of particles to have masses of the order of
magnitude of $v$, that is, in the range of hundreds of GeV.  Well above
the scale of $v$, these particles are effectively massless species
characterized by their definite quantum numbers under $SU(2)\times
U(1)$.

It is important to note that, at energies much higher than $v$, the
basic species are chiral. For example, the right- and left-handed
components of the $u$ quark have the following quantum numbers in this
high-energy world:
\beq
 u_R \ : \  I=0,\ Y=\frac{2}{3} \qquad    \pmatrix{u\cr d\cr}_L\ : \ 
 I=\frac{1}{2},\ Y=-\frac{1}{6} \ .
\eeq{eq:x}
There are no relations between these two  species; each half of the
low-energy $u$ quark has a completely different fundamental assignment.
And, each multiplet is prohibited from acquiring mass by $SU(2)\times
U(1)$ symmetry.

It is tempting to characterize the full set of elementary particles at 1
TeV---the particles, that is, that we  have a chance of observing at
accelerators in the foreseeable future---as precisely those which are
forbidden to acquire mass until $SU(2)\times U(1)$ is broken.  This
would explain why these particles are left over from the truly high-energy
dynamics of Nature, the dynamics which generates and perhaps unifies
the gauge and flavor interactions, to survive down to the much lower
energy scales accessible to our experiments.

%%%%%%%%%%%%%%%%%%%%%%%%%%%%%%%%%%%%%%%%%%%%%%%%%%%%%%%%%%%%%%%%%%%%%%
\begin{figure}[t]
\begin{center}
\leavevmode
\epsfbox{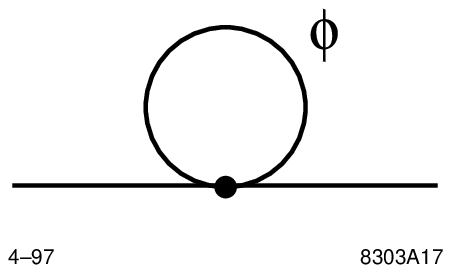}
\end{center}
 \caption{The simplest diagram which generates a Higgs boson mass term
in the Minimal Standard Model.}
\label{fig:one}
\end{figure}
%%%%%%%%%%%%%%%%%%%%%%%%%%%%%%%%%%%%%%%%%%%%%%%%%%%%%%%%%%%%%%%%%%%%%%

Before giving in to this temptation, however, I would like to point out
that the Minimal Standard Model contains a glaring counterexample to
this point of view, the Higgs boson itself.  The mass term for the
Higgs field
\beq
        \Delta\L = - \mu^2 \phi^\dagger \phi
\eeq{eq:y}
respects all of the symmetries of the Standard Model whatever the value
of $\mu$.  This model, then, gives no reason why $\mu^2$ is of order
$v$ rather than being, for example, twenty orders of magnitude larger.

Further, if we arbitrarily set $\mu^2 = 0$, the $\mu^2$ term would be
generated by radiative corrections.  The first correction to the mass
is shown in Figure \ref{fig:one}. This simple diagram is
formally infinite,
 but we might cut off its integral at a scale $\Lambda$ where
the Minimal Standard Model breaks down.  With this prescription, the
diagram contributes to the Higgs boson mass $m^2 = -\mu^2$ in the amount
\beqa
    -i m^2 &=&  - i \lambda\int {d^4k\over (2\pi)^4} {i\over k^2} \CR
           &=&  - i {\lambda\over 16\pi^2} \Lambda^2 \ .
\eeqa{eq:z}
Thus, the contribution of radiative corrections to the Higgs boson mass
is {\em nonzero}, {\em divergent}, and {\em positive}.  The last of
these properties is actually the worst.  Since electroweak symmetry
breaking requires that $m^2$ be negative, the contribution we have just
calculated must be cancelled by the Higgs boson bare mass term, and
this cancellation must be made more and more fine to achieve a negative
$m^2$ of the order of $-v^2$ in models where $\Lambda$ is very large. 
This problem is often called the `gauge hierarchy problem'. 
I think of it as
just a special aspect of the fact that the Minimal Standard Model
does not explain why $-\mu^2$ is negative or why electroweak symmetry is
broken.  Once we have left this fundamental question to a mere choice
of a parameter, it is not surprising that the radiative corrections to 
this parameter might drive it in an unwanted direction.

To continue, however, I would like to set this issue aside and think
more carefully about the properties of the massless, chiral particle
multiplets that we find at the TeV energy scale and above.  If these
particles are described by a renormalizable field theory but we can
ignore any mass parameters, the interactions of these particles are
governed by the dimensionless couplings of their renormalizable
interactions.  The scattering amplitudes generated by these couplings
will reflect the maximal parity violation of the field content, with
forward-backward and polarization asymmetries in scattering processes
typically of order 1.

For massless fermions, there is an ambiguity in writing the quantum
numbers in such a chiral situation becuase a left-handed fermion has a
right-handed antifermion, and vice versa.  For reasons that will be
clearer in the next section, I will choose the convention of writing
all species of fermions in terms of their left-handed components,
viewing all right-handed particles as antiparticles.  Thus, I will now
recast  the right-handed $u$ quark in \leqn{eq:x} as the antiparticle
of a left-handed species $\bar u$ which belongs to the $\bar 3$
representation of color $SU(3)$.  The fermions of the Standard Model
thus belong to the left-handed multiplets
\beqa
      L \  : \   I = \frac{1}{2}, \ Y = - \frac{1}{2} & \qquad &
  Q \  : \   I = \frac{1}{2}, \ Y = \frac{1}{6} \CR
    \bar e \  : \   I = 0, \ Y = 1 & \qquad &
  \bar u \  : \   I = 0, \ Y = -\frac{2}{3} \CR
 & \qquad &
  \bar d \  : \   I = 0, \ Y = \frac{1}{3}  \ .
\eeqa{eq:a1}
Here $L$ is the left-handed lepton doublet and $Q$ is the left-handed
quark doublet.  $Q$ is a color $3$, and $\bar u$, $\bar d$ are color
$\bar 3$'s. The right-handed electron is the antiparticle of $\bar e$, 
and there is no right-handed neutrino.
This set of quantum numbers of repeated for each quark and
lepton generation.

If the dimensionless couplings of the theory at TeV energies are small,
these coupling will run according to their renormalization group
equations, but only at a logarithmic rate.  Thus, above the TeV scale,
the description of elementary particles would  change very slowly.  In
this circumstance, it is reasonable to extrapolate many orders of
magnitude above the TeV energy scale and to derive definite physical
conclusions from that extrapolation.  I will now describe two
consequences of this idea.

%%%%%%%%%%%%%%%%%%%%%%%%%%%%%%%%%%%%%%%%%%%%%%%%%%%%%%%%%%%%%%%%%%%%%%
\begin{figure}[t]
\begin{center}
\leavevmode
{\epsfxsize=4in\epsfbox{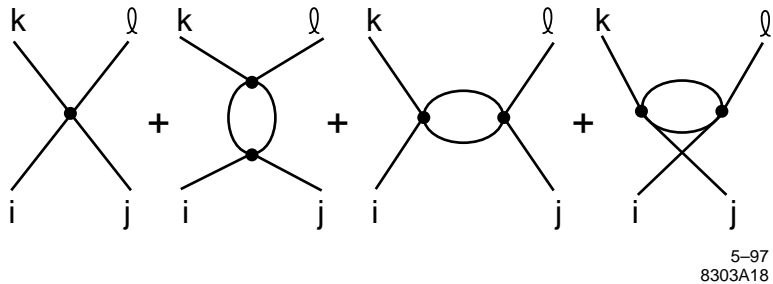}}
\end{center}
 \caption{Diagrams which renormalize the Higgs coupling constant
       in the Minimal Standard Model.}
\label{fig:two}
\end{figure}
%%%%%%%%%%%%%%%%%%%%%%%%%%%%%%%%%%%%%%%%%%%%%%%%%%%%%%%%%%%%%%%%%%%%%%

The first of these concerns the coupling constant of the minimal Higgs
theory.  For this analysis, it is best to write the Higgs multiplet
as four real-valued fields as in \leqn{eq:tt}.  Then the Higgs
Lagrangian (ignoring the mass term) takes the form
\beq
\L = \frac{1}{2} (\del_\mu \phi^i)^2 - \frac{1}{2} \lambda_b \left( (\phi^i)^2
        \right)^2  \ ,
\eeq{eq:b1}
where $i = 1, \ldots, 4$. I have given the coupling a subscript $b$ to
remind us that this is the bare coupling. The value of the first,
tree-level, diagram shown in Figure \ref{fig:two} is
\beq
      - 2i \lambda_b \left( \delta^{ij}\delta^{k\ell} + 
 \delta^{ik}\delta^{j\ell} +  \delta^{i\ell}\delta^{jk} \right) \ .
\eeq{eq:c1}
To compute the three one-loop diagrams in Figure \ref{fig:two}, we
need to contract two of these structures together, using $\delta^{ii} =
4$ where necessary.  The easiest way to do this is to isolate the terms
in each diagram which are proportional to $ \delta^{ij}\delta^{k\ell}$.
Since the set of three diagrams is symmetric under crossing, the other
two index contractions must appear also with equal coefficients.  The
contributions to this term from the three loop diagrams shown in Figure
\ref{fig:two} have the form
\beq
  {(-2i\lambda_b)^2\over 2} \int {d^4k\over (2\pi)^4} {i\over k^2}{i\over k^2}
      \left( [8 + 2 + 2]  \delta^{ij}\delta^{k\ell} + \cdots \right) \ ,
\eeq{eq:d1}
where I have ignored the external momentum, and the numbers in the
bracket give the contribution from each diagram. In a scattering
process, this expression is a good approximation when $k$ lies in the
range from the momentum transfer $Q$ up to the scale $\Lambda$ at which
the Minimal Standard Model breaks down.  Then the sum of the diagrams
in  Figure \ref{fig:two} is
\beq
      - 2i \lambda_b \left(1 - {12 \lambda_b^2\over (4\pi)^2} \log{\Lambda^2
\over Q^2} \right) \cdot
\left[ \delta^{ij}\delta^{k\ell} + \cdots \right] \ .
\eeq{eq:e1}
The coefficient in this expression can be thought of as the effective
value of the Higgs coupling constant for scattering processes at the
momentum transfer $Q$.  Often, we trade the bare coupling $\lambda_b$
for the value of the effective coupling at a low-energy scale (for
example, $v$), which we call the renormalized coupling $\lambda_r$.  In
terms of $\lambda_r$, \leqn{eq:e1} takes the form
\beq
      - 2i \lambda_r \left(1 +  {12 \lambda_r^2\over (4\pi)^2} \log{Q^2
\over v^2} \right) \cdot
\left[ \delta^{ij}\delta^{k\ell} + \cdots \right] \ .
\eeq{eq:f1}

Whichever description we choose, the effective coupling $\lambda(Q)$
has a logarithmically slow variation with $Q$.  The most convenient way
to describe this variation is by writing a differential equation,
called the {\em renormalization group equation} \cite{PS}
\beq
        {d\over d \log Q} \lambda(Q)  = {3\over 2\pi^2} \lambda^2(Q) \ .
\eeq{eq:g1}
If the coupling is not so weak, we should add further terms to the
right-hand side which arise from higher orders of perturbation theory.

The solution of \leqn{eq:g1} is
\beq
       \lambda(Q) =   {\lambda_r \over 1 - (3\lambda/2\pi^2)\log Q/v}
\eeq{eq:h1}
It is interesting that the effective coupling is predicted to become 
strong at high energy, specifically, at the scale
\beq
        Q_* = v \exp\left[{2\pi^2\over 3 \lambda}\right] \ .
\eeq{eq:i1}
Either the minimal Higgs Lagrangian is a consequence of
strong-interaction behavior at the scale $Q_*$, or, at some energy
scale below $Q_*$ the simple Higgs theory must become a part of some
more complex set of interactions.

Making use of \leqn{eq:ee}, we can relate this bound on the validity of
the simple Higgs theory to the value of the Higgs mass, be rewriting
\leqn{eq:i1} as
\beq
        Q_* = v \exp\left[{4\pi^2 v^2 \over 3 m_h^2}\right] \ .
\eeq{eq:i1p}
This is a remarkable formula, because the mass of the Higgs boson sits
in the denominator of an exponential.  Thus, for small $m_h$ or a small
value of  $\lambda$ at $v$, the energy scale $Q_*$ up to which the
minimal Higgs theory can be valid is very high.  On the other hand, as
$m_h$ increases above $v$, the value of $Q_*$ decreases
catastrophically.  Here is a table of the values predicted by
\leqn{eq:i1p}:
\beq
\begin{tabular}{rcr}
  $m_h$  &  \qquad & $Q_*$ \\ \cline{1-1}  \cline{3-3} \\
  150 GeV & &   $6\times 10^{17}$  GeV \\ 
  200 GeV & &   $1 \times 10^{11}$ GeV \\ 
  300 GeV &  &  $2\times  10^6$ GeV \\ 
  500 GeV &  &  $6\times  10^3$ GeV \\ 
  700 GeV &  &  $1\times  10^3$ GeV \\ 
\end{tabular}
\eeq{eq:ii1}
Notice that, as the mass of the Higgs boson goes above 700 GeV, the
scale $Q_*$ comes down to $m_h$.  Larger values of the Higgs boson mass
in the minimal model are self-contradictory.

%%%%%%%%%%%%%%%%%%%%%%%%%%%%%%%%%%%%%%%%%%%%%%%%%%%%%%%%%%%%%%%%%%%%%%
\begin{figure}[t]
\begin{center}
\leavevmode
{\epsfxsize=3.5in\epsfysize=4in\epsfbox{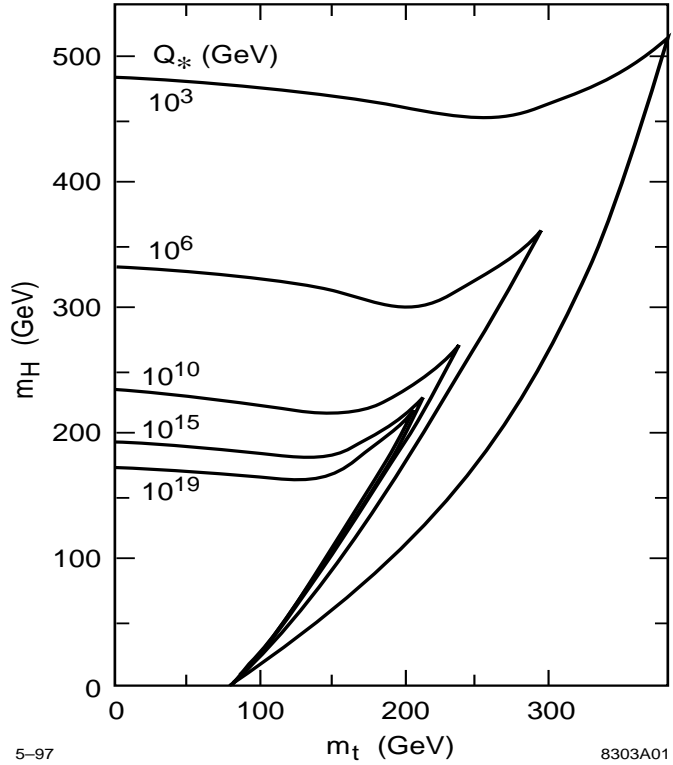}}
\end{center}
 \caption{Region of validity of the minimal Higgs model in the 
$(m_h, m_t)$ plane, including two-loop quantum corrections to the
Higgs potential,
from \protect\cite{Lindner}.}
\label{fig:three}
\end{figure}
%%%%%%%%%%%%%%%%%%%%%%%%%%%%%%%%%%%%%%%%%%%%%%%%%%%%%%%%%%%%%%%%%%%%%%

A more accurate evaluation of the limit $Q_*$ in the Standard Model,
including the full field content of the model and terms in perturbation
theory beyond the leading logarithms, is shown in 
Figure~\ref{fig:three} \cite{Lindner}.
 Note that, in this more sophisticated calculation, the
limit $Q_*$ depends on the value of the top quark mass when $m_t$
becomes large.  The calculation I have just described explains the top
boundary of the regions indicated in the figure; I will describe the
physics that leads to the right-hand boundary in Section 2.6.

%%%%%%%%%%%%%%%%%%%%%%%%%%%%%%%%%%%%%%%%%%%%%%%%%%%%%%%%%%%%%%%%%%%%%%
\begin{figure}[t]
\begin{center}
\leavevmode
{\epsfxsize=2.0in\epsfbox{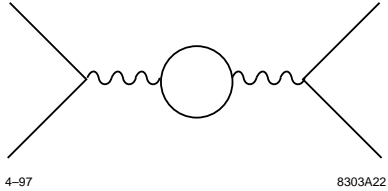}}
\end{center}
 \caption{A one-loop diagram contributing to the renormalization-group
        evolution of a gauge coupling constant.}
\label{fig:four}
\end{figure}
%%%%%%%%%%%%%%%%%%%%%%%%%%%%%%%%%%%%%%%%%%%%%%%%%%%%%%%%%%%%%%%%%%%%%%
 
The same idea, that the basic coupling constants can evolve slowly on a
logarithmic scale in $Q$ due to loop corrections from quantum field
theory, can be applied to the $SU(3)\times SU(2)\times U(1)$ gauge
couplings. The renormalization group equation for the gauge coupling
$g_i$ which includes the effects of one-loop diagrams such as that
shown in Figure~\ref{fig:four}  has the form
\beq
        {d\over d \log Q} g_i(Q)  = - {b_i\over (4\pi)^2} g_i^3 \ .
\eeq{eq:j1}
That is, the rate of change of $g_i^2$ with $\log Q$ is proportional to 
$g_i^4$, as the diagram indicates.

The $b_i$ are constants which depend on the gauge group and on the
matter multiplets to which the gauge bosons couple. For $SU(N)$ gauge
theories with  matter in the fundamental representation,
\beq
    b_N = \left( \frac{11}{3} N - \frac{1}{3} n_f - \frac{1}{6} n_s \right) \ ,
\eeq{eq:k1}
where $n_f$ is the number of chiral (left-handed) fermions and $n_s$ is
the number of complex scalars which couple to the gauge bosons.  For a
$U(1)$ gauge theory in which the matter particles have charges $t$, the
corresponding formula is
\beq
    b_1 =  - \frac{2}{3} \sum_f t_f^2 - \frac{1}{3}  \sum_s t_s^2 \ .
\eeq{eq:kk1}
I will not derive these formulae here; you can find their derivation in
any textbook of quantum field theory (for example, \cite{PS}).  In the
$SU(N)$ case, when $n_f$ and $n_s$ are sufficiently small, $b_N$ is
positive, leading to a decrease of the effective coupling as $Q$
increases.  This is the remarkable phenomenon of {\em asymptotic freedom}.

It is especially interesting that the effect of asymptotic freedom is
stronger for $SU(3)$ than for $SU(2)$ while the $SU(3)$ gauge coupling
is larger at the energy of $Z$ boson mass.  This suggests that, if we
extrapolate to very high energy, the strong- and weak-interaction
coupling constants should become equal, and perhaps the three different
interactions that make up the Standard Model may become unified \cite{GQW}.
  In
the remainder of this section, I will investigate this question
quantitatively.

In order to discuss the unification of gauge couplings,  there
is one small technical point that we must address first.  For a non-Abelian
group, we conventionally normalize the generators $t^a$ so that, in the
fundamental representation,
\beq
                \tr [ t^a t^b ] = \frac{1}{2} \delta^{ab} \ .
\eeq{eq:l1}
Also, for any simple non-Abelian group, $\tr[t^a] = 0$. For example,
the matrices $\tau^a = \sigma^a/2$ which we used to represent the
$SU(2)$ generators below \leqn{eq:b} obey these conditions. However,
for a $U(1)$ group there is no similar natural way to normalize the
charges.  In principle, we could hypothesize that the $SU(2)$ and
$SU(3)$ charges are unified with a charge proportional to the
hypercharge,
\beq
                    t_Y =  c \cdot Y
\eeq{eq:m1}
for any value of the scale factor $c$.

In building a theory of unified strong, weak, and electromagnetic
interactions, we might not want to assume that all fermion species
necessarily
belong to the fundamental representation of some $SU(N)$ group; thus,
we would not wish to impose the condition \leqn{eq:l1} on $t_Y$. But it
is not so unreasonable to insist that there is a single large
non-Abelian group for which $t_Y$ and the $SU(2)$ and $SU(3)$ charges
are all generators, and that the quarks and leptons of the Standard
Model form a representation of this group.  This leads to the
normalization condition for $t_Y$,
\beq
        \tr (t_Y)^2 = \tr (t)^2 \ ,
\eeq{eq:n1}
where $t$ is a generator of $SU(2)$ or $SU(3)$.  Any such generator gives the 
same constraint.  For convenience, I will choose to implement this condition
using $t = t^3$,
the third component of weak-interaction isospin.  The
trace could be taken over three  or over one Standard Model
generations.  Before evaluating $c$, it is interesting to sum over the
fermions with quantum numbers in the table \leqn{eq:a1}, to check that
$t_Y$ has zero trace.  Indeed, including each species in \leqn{eq:a1}
with its $SU(2)$
and color multiplicity, we find
\beqa
 \tr[t_Y] &=& c \tr[Y] \CR
         &=& c \left[ -\frac{1}{2} \cdot 2 + 1\cdot 1 + 
 \frac{1}{6}\cdot 6  - \frac{2}{3} \cdot 3 + \frac{1}{3} \cdot 3 \right]\CR
      &=& 0
\eeqa{eq:o1}
Then we can compute
\beq
 \tr (t^3)^2 = \left(\frac{1}{2}\right)^2 \cdot 2 \cdot 4 = 2 \ , 
\eeq{eq:p1}
and
\beq
 \tr (t_Y)^2 =  c^2 \left[ \left(\frac{1}{2}\right)^2 \cdot 2 + 1\cdot 1 + 
 \left(\frac{1}{6}\right)^2\cdot 6  +\left( \frac{2}{3}\right)^2 \cdot 3 
 +\left( \frac{1}{3}\right)^2 \cdot 3 
\right] = c^2 \cdot \frac {10}{3} \ .
\eeq{eq:q1}
Equating these expressions, we find $c = \sqrt{3/5}$; that is, 
\beq
          t_Y = \sqrt{\frac{3}{5}} Y \ ,
\eeq{eq:r1}
or, writing the $U(1)$ gauge coupling $g'Y = g_1 t_Y$, 
\beq
          g_1 =  \sqrt{\frac{5}{3}} g' \ .
\eeq{eq:s1}
These formulae give the normalization of the $U(1)$ coupling which
unifies with $SU(2)$ and $SU(3)$ in the $SU(5)$ and $SO(10)$ grand
unfied theories, and in many more complicated schemes of unification.

\begin{figure}[t]
\begin{center}
\leavevmode
{\epsfxsize=4in\epsfbox{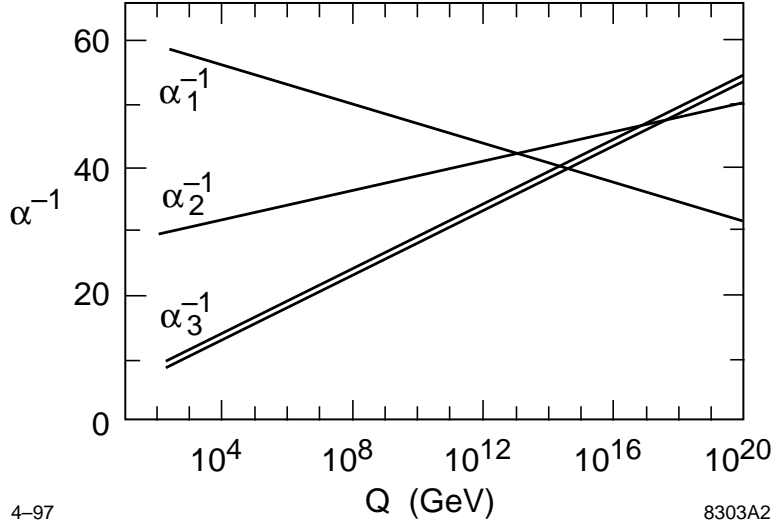}}
\end{center}
 \caption[*]{Evolution of the $SU(3)\times SU(2)\times U(1)$ gauge
couplings to high energy scales, using the one-loop renormalization group
equations of the Standard Model.  The double line for $\alpha_3$ indicates
the current experimental error in  this quantity; the errors in $\alpha_1$
and $\alpha_2$ are too small to be visible.}
\label{fig:five}
\end{figure}

In the Standard Model, the $U(1)$  coupling constant $g_1$ and the
$SU(2)$ and $SU(3)$ couplings $g_2$ and $g_3$ evolve with $Q$ according
to the renormalization group equation \leqn{eq:j1} with
\beqa
    b_3 &=&   11 - \frac{4}{3} n_g \CR
    b_2 &=&   \frac{22}{3} - \frac{4}{3} n_g - \frac{1}{6} n_h\CR 
    b_1 &=&   \phantom{11} -  \frac{4}{3} n_g - \frac{1}{10} n_h \ .
\eeqa{eq:t1}
In this formula, $n_g$ is the number of quark and lepton generations
and $n_h$ is the number of Higgs doublet fields.  Note that a complete
generation of quarks and leptons has the same effect on all three gauge
couplings, so that (at the level of one-loop corrections), the validity
of unification is independent of the number of generations.  The
solution to \leqn{eq:j1} can be written, in terms of the measured
coupling constants at $Q=\mz$, as
\beq
           g_i^2(Q) = {g_i^2(\mz) \over 1 + (b_i/8\pi^2)\log Q/\mz} \ .
\eeq{eq:u1}
Alternatively, if we let $\alpha_i = g_i^2/4\pi$, 
\beq
   \alpha_i^{-1}(Q) =   \alpha_i^{-1}(\mz) + {b_i\over 2\pi}\log{Q\over\mz} \ .
\eeq{eq:v1}
The evolution of coupling constants predicted by \leqn{eq:t1} and
\leqn{eq:v1}, with $n_h = 1$, is shown in Figure \ref{fig:five}.  It is
disappointing that, although the values of the coupling constants do
converge, they do not come to a common value at any scale.

%%%%%%%%%%%%%%%%%%%%%%%%%%%%%%%%%%%%%%%%%%%%%%%%%%%%%%%%%%%%%%%%%%%%%%
%%%%%%%%%%%%%%%%%%%%%%%%%%%%%%%%%%%%%%%%%%%%%%%%%%%%%%%%%%%%%%%%%%%%%%
 
We can be a bit more definite about this test of the unification of
couplings as follows:  I will work in the $\bar{MS}$ scheme for
defining coupling constants. The precisely known values of $\alpha$,
$\mz$, and $G_F$ imply $\alpha^{-1}(\mz) = 127.90\pm .09$, $\sstw(\mz)
= 0.2314\pm.003$ \cite{LEPDG}; combining this with the value of the
strong interaction coupling $\alpha_s(\mz) = 0.118 \pm .003$
\cite{HPDG}, we find
for the $\bar{MS}$ couplings at $Q= \mz$:
\beqa
           \alpha_1^{-1} &=& 58.98 \pm .08 \CR
       \alpha_2^{-1} &=& 29.60 \pm .04 \CR
       \alpha_3^{-1} &=&  8.47 \pm .22 
\eeqa{eq:w1}
On the other hand, if we assume that the three couplings come to a
common value at a scale $m_U$,  we can put $Q= m_U$ into the three
equations \leqn{eq:v1}, eliminate the unknowns $\alpha^{-1}(m_U)$ and
$\log(m_U/m_Z)$, and find one relation among the measured coupling
constants at $\mz$.  This relation is
\beq
       \alpha_3^{-1} = (1+ B)    \alpha_2^{-1} -  B   \alpha_1^{-1} \ , 
\eeq{eq:x1}
where
\beq
          B = {b_3-b_2\over b_2 - b_1}  \ .
\eeq{eq:y1}
From the data, we find
\beq
         B = 0.719 \pm .008 \pm .03 \ ,
\eeq{eq:z1}
where the second error reflects the omission of higher order
corrections, that is, finite radiative corrections at the thresholds and
two-loop corrections in the renormalization group equations.

On the other hand, the Standard Model gives
\beq 
         B = \frac{1}{2} + \frac{3}{110} n_h \ .
\eeq{eq:a2}
This is inconsistent with the unification hypothesis by a large margin.
But perhaps an interesting scheme for physics beyond the Standard Model
could fill this gap and allow a unification of the known gauge
couplings.

\subsection{The special role of the top quark}

In the previous section, we discussed the role of the quarks and
leptons in the energy region above 1 TeV.  However, we ought to give
additional consideration to the role of the top quark.  This quark is
sufficiently heavy that its coupling to the Higgs boson is an important
perturbative coupling at very high energies.  Thus, even in the
simplest models, the top quark plays an important special role in the
renormalization group evolution of couplings. It is possible that the
top quark has an even more central role in electroweak symmetry
breaking, and, in fact, that electroweak symmetry breaking may be {\em
caused} by the strong interactions of the top quark.  I will discuss
this connection of the top quark to electroweak symmetry breaking
later, in the context of specific models.  In this section, I would
like to prepare for that discussion by analyzing the effects of the
large top quark-Higgs boson coupling which is already present in the
Minimal Standard Model.

In the minimal Higgs model, the masses of quarks and leptons arise from
the perturbative couplings to the Higgs boson written in the third line
of \leqn{eq:a}.  These couplings are most often called the `Higgs
Yukawa couplings'. The top quark mass comes from a Yukawa coupling
\beq
       \Delta\L = -\lambda_t \bar t_R \phi\cdot Q_L + \hc \ ,
\eeq{eq:b2}
where $Q_L = (t_L,b_L)$.
When the Higgs field acquires a vacuum expectation value of the form 
\leqn{eq:d}, this term becomes
\beq
       \Delta\L = -{\lambda_tv\over \sqrt{2}}\ \bar t t , 
\eeq{eq:b2plus}
and we can read off the relation $m_t = \lambda_t v/\sqrt{2}$.  The
value of the top quark mass measured at Fermilab is $176 \pm 6$ GeV 
for the on-shell mass \cite{mt}, which corresponds to 
\beq
          (m_t)_{\bar{MS}} = 166 \pm 6 \ \mbox{GeV} \ .
\eeq{eq:c2}
With  the
value of $v$ in \leqn{eq:g}, this implies
\beq
     \lambda_t = 1  \quad \mbox{or}\quad \alpha_t = {\lambda^2\over 4\pi}
            = \left( 14.0 \pm 0.7 \right)^{-1} \ .
\eeq{eq:d2} 
In this simplest model, the top quark Yukawa coupling is weak
at high energies but still is large enough to compete with QCD.

The large value of $\lambda_t$ gives rise to two interesting effects. 
The first of these is an essential modification of the renormalization
group equation for the Higgs boson coupling $\lambda$ given in
\leqn{eq:g1}. Let me now rewrite this equation including the one-loop
corrections due to $\lambda_t$ and also to the weak-interaction
couplings \cite{Sher}: 
\beq
        {d\over d \log Q} \lambda  = {3\over 2\pi^2 }
\left[ \lambda^2 - {1\over 32}\lambda_t^4 + {g^2\over 512}(3 + 2s^2 + s^4)
\right]\ ,
\eeq{eq:e2}
where I have abbreviated $s^2 = \sstw$.

A remarkable property of the formula \leqn{eq:e2} is that the top quark
Yukawa coupling enters the renormalization group equation with a
negative sign (which essentially comes from the factor (-1) for the top
quark fermion loop).  This sign implies that, if the top quark mass is
sufficiently large that that $\lambda^4$ term dominates, the Higgs
coupling $\lambda$ is driven negative at large $Q$.  This is a
dangerous instability which would push the expectation value $v$ of the
Higgs field to arbitrarily high values. The presence of this
instability gives an upper bound on the top quark mass for fixed $m_h$,
or, equivalently, a lower bound on the Higgs mass for fixed $m_t$.  If
we replace $\lambda$, $\lambda_t$, and $g$ in \leqn{eq:e2} with the
masses of $h$, $t$, and $W$, we find the condition 
\beq
    m_h^2 >  \frac{1}{2} \left[ m_t^2 - \frac{3}{4} \mw^2 \right] \ .
\eeq{eq:f2}
I should note that finite perturbative corrections shift this bound in
a way that is important quantitatively.  This effect accounts for the
right-hand boundary of the regions shown in Figure \ref{fig:three}.

The implications of Figure \ref{fig:three} for the Higgs boson mass are
quite interesting.  For the correct value of the top quark mass
\leqn{eq:c2}, the Minimal Standard Model description of the Higgs boson
can be valid only if the mass of the Higgs is larger than about 60 GeV.
But for values of the $m_h$ below 100 GeV or above 200 GeV, the Higgs
coupling must be sufficiently large that this coupling becomes strong
well below the Planck scale.  Curiously, the fit of
current precision electroweak data to the Minimal Standard Model (for
example, to the more precise version of \leqn{eq:w}) gives the value
\cite{LEPDG} 
\beq
           m_h = 124^{+125}_{-71} \ \mbox{GeV}\ ,
\eeq{eq:g2}
which actually lies in the region for which the Minimal Standard Model
is good to extremely high energies.  It is also important to point out
that the regions of Figure~\ref{fig:three} apply only to the Minimal
version of the Standard Model.  In models with additional Higgs
doublets, with the boundaries giving limits on the lightest Higgs
boson, the upper boundary remains qualitatively correct, but the
boundary associated with the heavy top quark is usually pushed far to
the right.

The second perturbative effect of the top quark Yukawa coupling is its
influence back on its own renormalization group evolution.  In the same
simple one-loop approximation as \leqn{eq:e2}, the renormalization
group equation for the top quark Yukawa coupling takes the form 
\beq
        {d\over d \log Q} \lambda_t  = {\lambda_t\over (4\pi)^2 }
\left[ \frac{9}{2}\lambda_t^2 - 8g_3^2 - \frac{9}{4}g^2 \left(1 + 
\frac{17}{24} s^2 \right) \right] \ .
\eeq{eq:h2}
The signs in this equation are not hard to understand.  A theory with
$\lambda_t$ and no gauge couplings cannot be asymptotically free, and
so $\lambda_t$ must drive itself to zero at large distances or small
$Q$. On the other hand, the effect of the QCD coupling $g_3$ is to
increase quark masses and also $\lambda_t$ as $Q$ becomes small.

\begin{figure}[t]
\begin{center}
\leavevmode
{\epsfxsize=2.75in\epsfbox{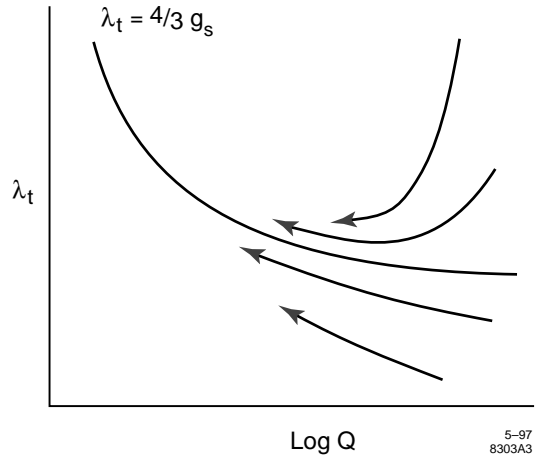}}
\end{center}
 \caption{Renormalization-group evolution of the top quark
 Yukawa coupling $\lambda_t$ and 
the strong interaction coupling $\alpha_s$, from large $Q$ to small $Q$.}
\label{fig:six}
\end{figure}

\begin{figure}[htb]
\begin{center}
\leavevmode
{\epsfxsize=3in\epsfbox{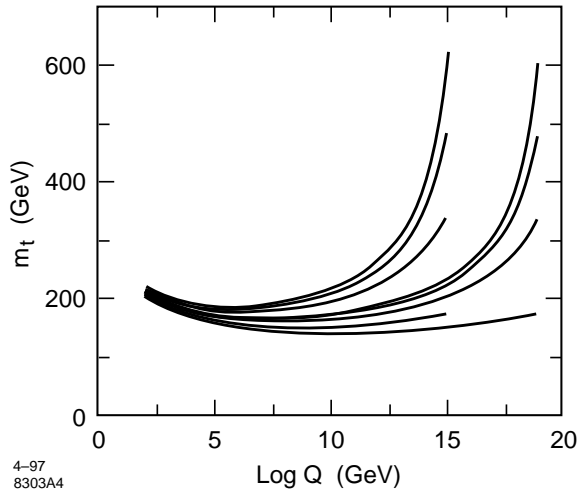}}
\end{center}
 \caption{Convergence of predictions for the top quark mass
   in the Minimal Standard Model, due to 
            renormalization-group evolution, from
     \protect\cite{Hill}.}
\label{fig:seven}
\end{figure}

The two  effects of the $\lambda_t$ and QCD renormalization of
$\lambda_t$ balance at the point 
\beq
          \lambda_t = \frac{4}{3} (4\pi \alpha_s)^{1/2} \sim 1.5\ ,
\eeq{eq:i2}
corresponding to $m_t \sim 250$ GeV.  This condition was referred to by
Hill \cite{Hill} as the `quasi-infrared fixed point' for the top quark
mass. This `fixed point' is in fact a line in the $(\lambda_t,
\alpha_s)$ plane. The renormalization group evolution from large $Q$ to
small $Q$ carries a general initial condition into this line, as shown
in Figure \ref{fig:six}; then the parameters flow along the line, with
$\alpha_s$ increasing in the familiar way as $Q$ decreases, until we
reach $Q\sim m_t$.  The effect of this evolution is that theories with
a wide range of values for $\lambda_t$ at a very high unification scale
all predict the physical value of $m_t$ to lie close to the fixed-point
value \leqn{eq:i2}.  This convergence is shown in Figure
\ref{fig:seven}. The fixed point attracts initial conditions
corresponding to arbitrarily large values of $\lambda_t$ at high
energy.  However, if the initial condition at high energy is
sufficiently small, the value of $\lambda_t$ or $m_t$ might not be able
to go up to the fixed point before $Q$ comes down to the value $m_t$. 
Thus, there are two possible cases, the first in which the physical
value of $m_t$ is very close to the fixed point value, the second in
which the physical value of $m_t$ lies at an arbitrary point below the
fixed-point value.

%%%%%%%%%%%%%%%%%%%%%%%%%%%%%%%%%%%%%%%%%%%%%%%%%%%%%%%%%%%%%%%%%%%%%%
%%%%%%%%%%%%%%%%%%%%%%%%%%%%%%%%%%%%%%%%%%%%%%%%%%%%%%%%%%%%%%%%%%%%%%

%%%%%%%%%%%%%%%%%%%%%%%%%%%%%%%%%%%%%%%%%%%%%%%%%%%%%%%%%%%%%%%%%%%%%%
%%%%%%%%%%%%%%%%%%%%%%%%%%%%%%%%%%%%%%%%%%%%%%%%%%%%%%%%%%%%%%%%%%%%%%

In the Minimal Standard Model, the observed top quark mass \leqn{eq:c2}
must correspond to the second possibility.
  However, in models with two Higgs doublet fields, the
quantity which is constrainted to a fixed point is $m_t/\cos\beta$,
where $\beta$ is the mixing angle defined in \leqn{eq:k}.  The fixed
point location also depends on the full field content of the model.  In
the supersymmetric models to be discussed in the next section, the
fixed-point relation is 
\beq
             {m_t \over \cos\beta} \sim 190\ \mbox{GeV}
\eeq{eq:j2}
for values of $\tan\beta$ that are not too large.  In such theories it
is quite reasonable that the physical value of the top quark mass could
be determined by a fixed point of the renormalization group equation
for $\lambda_t$.

Now that we understand the implications of the large top quark mass in
the simplest Higgs models, we can return to the question of the
implications of the large top quark mass in more general models.  We
have seen that the observed value of $m_t$ can consistently be
generated solely by perturbative interactions.  We have also seen that,
in this case, the coupling $\lambda_t$ can have important effects on
the renormalization group evolution of couplings.  But this observation
shows that the observed value of $m_t$ is not sufficiently large that
it must lead to nonperturbative effects or that it can by itself drive
electroweak symmetry breaking. In fact, we now see that $m_t$ or
$\lambda_t$ can be the cause of electroweak symmetry breaking only if
we combine these parameters with additional new dynamics that lies
outside the Standard Model.  I will discuss some  ideas which follow
this line in Section 5.

\subsection{Recapitulation}

In this section, I have introduced the major questions for physics
beyond the Standard Model by reviewing issues that arise when the
Standard Model is extrapolated to very high energy.  I have highlighted the
issue  of electroweak symmetry breaking, which poses an important
question for the Standard Model which must be solved at energies close
to those of our current accelerators.   There are many possibilities, however,
for the form of this solution.
The new physics responsible for electroweak symmetry breaking
might be a new set of strong interactions which
changes the laws of particle physics fundamentally at some nearby
energy scale.  But the analysis we have done tells us that the 
solution might be constructed in a completely different way, in which
the new interactions are weakly coupled
for many orders of magnitude above the weak interaction scale but 
undergoes qualitative changes 
through the renormalization
group evolution of couplings.

The questions we have asked in Section 2.4 and this dichotomy of
strong-coupling versus weak-coupling solutions to these questions
provide a framework for examining theories of physics beyond the
Standard Model. In the next sections, I will consider some explicit
examples of such models, and we can see how they illustrate the  different
possible answers.

\section{Supersymmetry:  Formalism}

The first class of models that I would like to discuss are
supersymmetric extensions of the Standard Model.  {\em Supersymmetry} is
defined to be  a symmetry of Nature that links bosons and fermions.  As we
will see later in this section, the introduction of supersymmetry into
Nature requires a profound generalization of our fundamental theories,
including a revision of the theory of gravity and a rethinking of our
basic notions of space-time. For many theorists, the beauty of this new
geometrical theory is enough to make it compelling.  For myself, I think
this is quite a reasonable attitude. However, I do not expect you to
share this attitude in order to appreciate my discussion.

For the skeptical experimenter, there are other reasons to study
supersymmetry. The most important is that supersymmetry is a concrete
worked example of physics beyond the Standard Model.  One of the
virtues of extending the Standard Model using supersymmetry is that the
phenomena that we hope to discover at the next energy scale---the  new
spectrum of particles, and the mechanism of electroweak symmetry
breaking---occur in supersymmetric models at the level of perturbation
theory, without the need for any new strong interactions.  
Supersymmetry naturally
predicts are large and complex spectrum of new particles.
These particles have signatures which are interesting, and which test
the capabilities of experiments. Because the theory has weak couplings,
these signatures can be worked out
directly in a rather straightforward way.
  On the other hand, supersymmetric models have a
large number of undetermined parameters, so they can exhibit an
interesting variety of physical effects.  Thus, the study of
supersymmetric models can give you very specific pictures of what it
will be like to experiment on physics beyond the Standard Model and,
through this, should aid you in preparing for these experiments.  For
this reason, I will devote a large segment of these lectures to a detailed
discussion of supersymmetry. However, as a necessary corrective, I will
devote Section 5 of this article to a review of a model of electroweak
symmetry breaking that runs by strong-coupling effects.

This discussion immediately raises a question: Why is supersymmetry
relevant to the major issue that we are focusing on in these lectures,
that of the mechanism of electroweak symmetry breaking?  A quick answer
to this question is that supersymmetry legitimizes the introduction of
Higgs scalar fields, because it connects spin-0 and spin-$\half$ fields
and thus puts the Higgs scalars and the quarks and leptons on the same
epistemological footing.  A better answer to this question is that
supersymmetry naturally  gives rise to a mechanism of electroweak
symmetry breaking associated with the heavy top quark,
 and to many other properties that are attractive
features of the fundamental interactions.  These consequences of the 
theory arise from renormalization group evolution, by arguments similar
to those we used to explain
the features of the Standard Model that we derived in Sections 2.5 and 2.6.
The spectrum of new particles predicted by supersymmetry will also
be shaped strongly by renormalization-group effects. 

 In order to explain
these effects, I must unfortunately subject you to a certain amount of
theoretical formalism.  I will therefore devote this section to describing 
construction of supersymmetric Lagrangians and the analysis of their
couplings.  I will conclude this discussion in Section 3.7 by explaining
the supersymmetric mechanism of electroweak symmetry breaking.  This 
analysis will be lengthy, but it will give us the tools we need to build
a theory of the mass spectrum of supersymmetric particles.  With this 
understanding, we will be ready in Section 4 
to discuss the experimental issues raised
by supersymmetry, and the specific experiments that should resolve them.

\subsection{A little about fermions}

In order to write Lagrangians which are symmetric between boson and
fermion fields, we must first understand the properties of
these fields separately.  Bosons are simple, one component objects. 
But for fermions, I would like to emphasize a few features which are
not part of the standard presentation of the Dirac equation.

The Lagrangian of a massive Dirac field is
\beq
       \L = \bar\psi i \dslash\psi - m \bar\psi \psi \ ,
\eeq{eq:k2}
where $\psi$ is a 4-component complex field, the Dirac spinor.  I would
like to write this equation more explicitly by introducing a particular
representation of the Dirac matrices 
\beq
     \gamma^\mu = \pmatrix{ 0 & \sigma^\mu\cr \bar \sigma^\mu & 0 \cr}\ ,
\eeq{eq:l2}
where the entries are $2\times 2$ matrices with 
\beq
\sigma^\mu = (1, \vec \sigma)\ , \qquad \bar\sigma^\mu = (1, - \vec\sigma)\ .
\eeq{eq:ll2}
We may then write $\psi$ as a pair of 2-component complex fields
\beq
             \psi = \pmatrix{ \psi_L \cr \psi_R \cr} \ .
\eeq{eq:m2}
The subscripts indicate left- and right-handed fermion components, and
this is justified because, in this representation, 
\beq
             \gamma^5 = \pmatrix{-1 & 0 \cr 0 & 1 \cr} \ .
\eeq{eq:n2}
This is a handy representation for calculations involving high-energy
fermions which include  chiral interactions or polarization effects,
even within the Standard Model~\cite{PS}.

In the notation of \leqn{eq:l2}, \leqn{eq:m2}, the Lagrangian
\leqn{eq:k2} takes the form 
\beq
 \L =   \psi^\dagger_L i \bar\sigma^\mu \del_\mu \psi_L + 
 \psi^\dagger_R i \sigma^\mu \del_\mu \psi_R 
     - m \left( \psi^\dagger_R \psi_L +  \psi^\dagger_L \psi_R \right) \ .
\eeq{eq:o2}
The kinetic energy terms do not couple $\psi_L$ and $\psi_R$ but rather
treat them as distinct species.  The mass term is precisely the
coupling between these components.

I pointed out above \leqn{eq:a1} that, since the antiparticle of a
masssless left-handed particle is a right-handed particle, there is an
ambiguity in assigning quantum numbers to fermions.  I chose to resolve
this ambiguity by considering all left-handed states as particles and
all right-handed states as antiparticles. With this philosophy, we
would like to trade $\psi_R$ for a left-handed field.  To do this,
define the $2\times 2$ matrix 
\beq
                c = -i\sigma^2 = \pmatrix{0 & -1 \cr 1 & 0} \ .
\eeq{eq:p2}
and let
\beq
       \chi_L = c \psi^*_R \ , \qquad \chi_L^* = c\psi_R
\eeq{eq:q2}
Note that $c^{-1} = c^T = -c$, $c^* = c$, so \leqn{eq:q2} implies
\beq
        \psi_R = -c \chi^*_L \ , \qquad \psi_R^\dagger = \chi_L^T c \ .
\eeq{eq:r2}
Also note, by multiplying out the matrices, that
\beq
        c \sigma^\mu c^{-1} = (\bar \sigma^\mu)^T \ , \qquad
    c \bar \sigma^\mu c^{-1}  = (\sigma^\mu)^T \ .
\eeq{eq:s2}
Using these relations, we can rewrite
\beqa
  \psi^\dagger_R i \sigma^\mu \del_\mu \psi_R &=&
             \chi_L^T c i\sigma^\mu \del_\mu (-c) \chi^*_L \CR
    &=&  \chi_L^T  i(\bar\sigma^\mu)^T \del_\mu \chi^*_L \CR
  &=&  -\del_\mu \chi_L^\dagger  i(\bar\sigma^\mu) \chi_L \CR
 &=&  \chi^\dagger_L i \bar \sigma^\mu \del_\mu \chi_L \ .
\eeqa{eq:t2}
The minus sign in the third line came from fermion interchange; it was
eliminated in the fourth line by an integration by parts.  After this
rewriting, the two pieces of the Dirac kinetic energy term have
precisely the same form, and we may consider $\psi_L$ and $\chi_L$ as
two species of the same type of particle.

If we replace $\psi_R$ by $\chi_L$, the mass term in \leqn{eq:k2}
becomes 
\beq
     - m \left( \psi^\dagger_R \psi_L +  \psi^\dagger_L \psi_R \right)
    =  - m \left(\chi^T_L c \psi_L - \psi^\dagger_L c \chi^*_L\right)\ .
\eeq{eq:u2}
Note that
\beq
       \chi^T_L c \psi_L = \psi^T_L c \chi_L \ ,
\eeq{eq:v2}
with one minus sign from fermion interchange and a second from taking
the transpose of $c$.  Thus, this mass term is symmetric between the
two species.  It is interesting to know that the most general possible
mass term for spin-$\half$ fermions can be written in terms of
left-handed fields $\psi_L^a$ in the form 
\beq
    - \half m^{ab} \psi^{aT}_L c \psi^b_L + \hc \ ,
\eeq{eq:w2}
where $m^{ab}$ is a symmetric matrix.  For example, this form for the
mass term incorporates all possible different forms of the neutrino
mass matrix, both Dirac and Majorana.

From here on, through the end of Section 4, 
all of the fermions that appear in these lectures will be
2-component left-handed fermion fields.  For this reason, there will be
no ambiguity if I now drop the subscript $L$ in my equations.

\subsection{Supersymmetry transformations}

Now that we have a clearer understanding of fermion fields, I would
like to explore the possible symmetries that could connect fermions to
bosons. To begin, let us try to connect a free massless fermion field
to a free massless boson field.  Because the scalar product
\leqn{eq:v2} of two chiral fermion fields is complex, this connection
will not work unless we take the boson field to be complex-valued. 
Thus, we should look for symmetries of the Lagrangian 
\beq
        \L = \del_\mu \phi^* \del^\mu\phi + 
\psi^\dagger i \bar\sigma \cdot \del \psi 
\eeq{eq:x2}
which mix $\phi$ and $\psi$. 

To build this transformation, we must introduce a symmetry parameter
with spin-$\half$ to combine with the spinor index of $\psi$. I will
introduce a parameter $\xi$ which also transforms as a  left-handed
chiral spinor. Then a reasonable transformation law for $\phi$ is 
\beq
          \delta_\xi \phi = \sqrt{2} \xi^T c \psi \ . 
\eeq{eq:y2}
A fermion field has the dimensions of (mass)$^{3/2}$, while a boson
field has the dimensions of (mass)$^{1}$;  thus, $xi$ must carry the
dimensions (mass)$^{-1/2}$ or (length)$^{1/2}$.  This means that, in
order to form a dimensionally correct transformation law for $\psi$, we
must include a derivative.  A sensible formula is 
\beq
        \delta_\xi \psi = \sqrt{2}i \sigma\cdot \del \phi c \xi^* \ .
\eeq{eq:z2}

It is not difficult to show that the transformation \leqn{eq:y2},
\leqn{eq:z2} is  a symmetry of \leqn{eq:x2}.  Inserting these
transformations, we find 
\beq 
 \delta_\xi \L = \del_\mu\phi^* \del^\mu(
 \sqrt{2}\xi^T c \psi) + (\sqrt{2} i \xi^T c \sigma\cdot 
 \del\phi) i \bar\sigma\cdot \del \psi + (\xi^*)\ .
\eeq{eq:a3}
The term in the first set of parentheses is the right-hand side of
\leqn{eq:y2}.  The term in the second set of parentheses is the
Hermitian conjugate of the right-hand side of \leqn{eq:z2}.  The last
term refers to terms proportional to $\xi^*$ arising from the variation
of $\phi^*$ and $\psi$.  To manipulate \leqn{eq:a3}, integrate both
terms by parts and use the identity 
\beq
                 \sigma\cdot \del \bar\sigma\cdot \del = \del^2
\eeq{eq:b3}
which can be verified directly from \leqn{eq:ll2}.  This gives
\beq
\delta_\xi \L = - \phi^* \del^2( \sqrt{2}\xi^T c \psi) -
    \sqrt{2} i \xi^T c \cdot i  \del^2 \psi 
        + (\xi^*)\ .
\eeq{eq:c3}
The two terms shown now cancel, and the $\xi^*$ terms cancel similarly.
Thus, $\delta_\xi\L = 0$ and we have a symmetry.

The  transformation \leqn{eq:z2} appears rather strange at first sight.
However, this formula takes on a bit more sense when we work out the
algebra of supersymmetry transformations.  Consider the commutator 
\beqa
 (\delta_\eta \delta_\xi - \delta_\xi \delta_\eta) \phi        
   &=&   \delta_{\eta} (\sqrt{2} \xi^Tc\psi) - (\eta \leftrightarrow \xi)\CR
   &=&   \sqrt{2}\xi^T c (\sqrt{2}i \sigma^\mu \del_\mu \phi c \eta^*)
 - (\eta \leftrightarrow \xi)\CR
   &=&   2i\xi^T c \sigma^\mu  c \eta^* \del_\mu \phi
 - (\eta \leftrightarrow \xi)\CR
   &=&   -2i\xi^T (\bar\sigma^\mu )^T \eta^* \del_\mu \phi
 - (\eta \leftrightarrow \xi)\CR
   &=&   2i[\eta^\dagger \bar\sigma^\mu\xi - \xi^\dagger \bar\sigma^\mu\eta]
    \,   \del_\mu\phi \ .
\eeqa{eq:d3} 
To obtain the fourth line, I have used \leqn{eq:s2}; in the passage to
the next line, a minus sign appears due to fermion interchange. In
general, supersymmetry transformations have the commutation relation
\beq
 (\delta_\eta \delta_\xi - \delta_\xi \delta_\eta) A        
  =   2i[\eta^\dagger \bar\sigma^\mu\xi - \xi^\dagger \bar\sigma^\mu\eta]
    \,   \del_\mu A 
\eeq{eq:e3} 
on every field $A$ of the theory.

To clarify the significance of this commutation relation, let me
rewrite the transformations $\delta_\xi$ as the action of a set of
operators, the supersymmetry charges $Q$.   These charges must also be
spin-$\half$.  To generate the supersymmetry transformation, we 
contract them with the spinor parameter $\xi$;  thus 
\beq
   \delta_\xi = \xi^T c Q -  Q^\dagger c \xi^* \ .
\eeq{eq:f3}
At the same time, we may  replace $(i\del_\mu)$ in \leqn{eq:e3} by
the operator which generates spatial translations, the energy-momentum
four-vector $P^\mu$.  Then \leqn{eq:e3} becomes the operator relation
\beq
  \left\{ Q^\dagger_a \ , \ Q_b \right\} =  (\bar\sigma^\mu)_{ab} P_\mu
\eeq{eq:g3}
which defines  the {\em supersymmetry algebra}. This anticommutation
relation has a two-fold interpretation.  First, it says that the square
of the supersymmetry charge $Q$ is the energy-momentum.  Second, it
says that the square of a supersymmetry transformation is a spatial
translation.  The idea of a square appears here in the same sense as we
use when we say that the Dirac equation is the square root of the
Klein-Gordon equation.

We started this discussion by looking for symmetries of the trivial
theory \leqn{eq:x2}, but at this stage we have encountered a
structure with deep connections.
  So it is worth looking back to see whether we
were forced to come to  high level or whether we could have
taken another route.  It turns out that, 
given our premises, we could not have
ended in any other place~\cite{HLS}. We set out to look for an operator
$Q$ that was a symmetry of Nature which carried spin-$\half$.  From
this property, the quantity on the left-hand side of \leqn{eq:g3} is a
Lorentz four-vector which commutes with the Hamiltonian.  In principle,
we could have written a more general formula 
\beq
  \left\{ Q^\dagger_a \ , \ Q_b \right\} =  (\bar\sigma^\mu)_{ab} R_\mu \ ,
\eeq{eq:h3}
where $R^\mu$ is a conserved four-vector charge different from $P^\mu$.
But energy-momentum conservation is already a very strong restriction
on particle scattering processes, since it implies that the only degree
of freedom in a two-particle reaction is the scattering angle in the
center-of-mass system.  A second vector conservation law, to the extent
that it differs from energy-momentum conservation, places new
requirements that contradict these restrictions except at particular,
discrete scattering angles.  Thus, it is not possible to have an
interacting
relativistic field theory with an additional conserved spin-1 charge,
or with any higher-spin charge, beyond standard momentum and angular
momentum conservation \cite{CMthm}.  For this reason, \leqn{eq:g3} is
actually the most general commutation relation that can be obeyed by
supersymmetry charges.

The implications of the supersymmetry algebra \leqn{eq:g3} are indeed
profound.  If the square of a supersymmetry charge is the total
energy-momentum of everything, then {\em supersymmetry must act on
every particle and field in Nature}. We  can exhibit this action
explicitly by writing out the $a=1,b=1$ component of \leqn{eq:g3}, 
\beq
  \left\{ Q^\dagger_1 \ , \ Q_1 \right\} =  P^0 + P^3 = P^+ \ .
\eeq{eq:i3}
On states with $P^+ \neq 0$ (which we can arrange for any  particle
state by a rotation), define 
\beq
         a = {Q_1 \over \sqrt{P^+}}\ , \qquad
  a^\dagger = {Q_1^\dagger \over \sqrt{P^+}} \ .
\eeq{eq:j3}
These operators obey the algebra
\beq
         \{ a^\dagger\ , \ a \} = 1
\eeq{eq:k3}
of fermion raising  and lowering  operators.  They raise and
lower $J^3$ by $\half$ unit.  Thus, in a supersymmetric theory, every
state of nonzero energy has a partner of opposite statistics differing
in angular momentum by $\Delta J^3 = \pm \half$.

On the other hand, for any operator $Q$, the quantity $\{Q^\dagger,
Q\}$ is a Hermitian matrix with eigenvalues that are either positive or
zero.  This matrix has zero eigenvalues for those states that satisfy
\beq
            Q \ket{0} = Q^\dagger \ket{0} = 0 \ ,
\eeq{eq:l3}
that is, for supersymmetric states.  In particular, if supersymmetry is
not spontaneously broken, the vacuum state is supersymmetric and
satisfies \leqn{eq:l3}.  Since the vacuum also has zero three-momentum,
we deduce 
\beq
            \bra{0} H \ket{0} = 0 
\eeq{eq:m3}
as a consequence of supersymmetry.  Typically in a quantum field
theory, the value of the vacuum energy density is given by a
complicated sum of vacuum diagrams.  In a supersymmetric theory, these
diagrams must magically cancel \cite{Zvac}.  This is the first of a
number of magical cancellations of radiative corrections that we will
find in supersymmetric field theories.

\subsection{Supersymmetric Lagrangians}

At this point, we have determined the general formal properties of
supersymmetric field theories. Now it is time to be much more concrete
about the form of the Lagrangians which respect supersymmetry. In this
section, I will discuss the particle content of supersymmetric theories
and present the most general renormalizable supersymmetric Lagrangians
for spin-0 and spin-$\half$ fields.

We argued from \leqn{eq:i3} that all supersymmetric states of nonzero
energy are paired.  In particular, this applies to single-particle
states, and it implies that supersymmetric models contain boson and
fermion fields which are paired in such a way that the particle degrees
of freedom are in one-to-one correspondence.  In the simple example
\leqn{eq:x2}, I introduced a complex scalar field and a left-handed
fermion field.  Each leads to two sets of single-particle states, the
particle and the antiparticle.  I will refer to this set of states---a
left-handed fermion, its right-handed antiparticle, a complex boson,
and its conjugate---as a {\em chiral supermultiplet}.

Another possible pairing is a a massless vector field and a left-handed
fermion, which gives a {\em vector supermultiplet}---two transversely
polarized vector boson states, plus the left-handed fermion and its
antiparticle.  In conventional field theory, a vector boson obtains
mass from the Higgs mechanism by absorbing one degree of freedom from a
scalar field.  In supersymmetry, the Higgs mechanism works by coupling
a vector supermultiplet to a chiral supermultiplet. This coupling
results in a massive vector particle, with three polarization states,
plus an extra scalar.  At the same time, the left-handed fermions in
the two multiplets combine through a mass term of the form \leqn{eq:u2}
to give a massive Dirac fermion, with two particle and two antiparticle
states.  All eight states are degenerate if supersymmetry is unbroken.

More complicated pairings are possible.  One of particular importance
involves the graviton.  Like every other particle in the theory, the
graviton must be paired by supersymmetry.  Its natural partner is a
spin-$\thalf$ field called the {\em gravitino}.  In general relativity,
the graviton is the gauge field of local coordinate invariance.  The
gravitino field can also be considered as a gauge field.  Since it
carries a vector index plus the spinor index carried by $\xi$ or $Q$,
it can have the  transformation law 
\beq
     \delta_\xi \psi_\mu =  {1\over 2\pi G_N}\del_\mu \xi(x) + \cdots
\eeq{eq:n3}
which makes it the gauge field of local supersymmetry.  This gives a
natural relation between supersymmetry and space-time geometry and 
emphasizes the profound character of this generalization of field
theory.

I will now present the most general Lagrangian for chiral
supermultiplets. As a first step, we might ask whether we can give a
mass to the fields in \leqn{eq:x2} consistently with supersymmetry. 
This is accomplished by the Lagrangian 
\beqa
        \L &=& \del_\mu \phi^* \del^\mu\phi + 
\psi^\dagger i \bar\sigma \cdot \del \psi + F^\dagger F  \CR
   & & \hskip 0.5in  + m (\phi F -  \half \psi^T c \psi) + \hc   \ .
\eeqa{eq:o3}
In this expression, I have introduced a new complex field $F$. 
However, $F$ has no kinetic energy and does not lead to any new
particles. Such an object  is called an {\em auxiliary field}.  If we
vary the Lagrangian \leqn{eq:o3} with respect to $F$, we find the field
equations 
\beq
    F^\dagger = - m\phi \ , \qquad   F = - m \phi^* \ .
\eeq{eq:p3} 
Thus $F$ carries only the degrees of freedom that are already present
in $\phi$.  We can substitute this solution back into \leqn{eq:o3} and
find the Lagrangian 
\beq
        \L = \del_\mu \phi^* \del^\mu\phi - m^2 \phi^* \phi + 
\psi^\dagger i \bar\sigma \cdot \del \psi - \half m(\psi^T c \psi - 
      \psi^\dagger c \psi^*) \ , 
\eeq{eq:q3}
which has equal, supersymmetric masses for the bosons and fermions.

It is not difficult to show that the Lagrangian \leqn{eq:o3} is
invariant to the supersymmetry transformation 
\beqa
      \delta_\xi \phi &=& \sqrt{2} \xi^T c \psi\CR
      \delta_\xi \psi &=& \sqrt{2}i \sigma\cdot \del \phi c \xi^* + \xi F\CR
     \delta_\xi F &=& -\sqrt{2}i \xi^\dagger \bar\sigma\cdot \del \psi \ .
\eeqa{eq:r3}
The two lines of \leqn{eq:o3} are invariant separately.  For the first
line, the proof of invariance is a straightforward generalization of
\leqn{eq:c3}. For the second line, we need 
\beqa
    \delta_\xi (\phi F - \half \psi^T c \psi)
 &=& (\sqrt{2}\xi^T c \psi)F + \phi 
      ( -\sqrt{2}i \xi^\dagger \bar\sigma\cdot \del \psi )
       - \psi^T c ( \sqrt{2}i \sigma\cdot \del \phi c \xi^* + \xi F) \CR
  &=&  \sqrt{2}i \xi^\dagger \bar\sigma\cdot \del \psi 
   -\sqrt{2} i \psi^T c \sigma\cdot \del \phi c \xi^* \CR
   &=&  0 \ .
\eeqa{eq:s3}
The first and last terms in the second line cancel by the use of
\leqn{eq:v2}; the terms in the third line cancel after  an integration
by parts and  a rearrangement similar to that in \leqn{eq:d3}  in the
second term.  Thus, \leqn{eq:r3} is an invariance of \leqn{eq:o3}. 
With some effort, one can show that this transformation obeys the
supersymmetry algebra, in the sense that the commutators of
transformations acting on $\phi$, $\psi$, and $F$ follow precisely the
relation \leqn{eq:e3}.

The introduction of the auxiliary field $F$ allows us to write a much
more general class of supersymmetric Lagrangians.  Let $\phi_j$,
$\psi_j$, $F_j$ be the fields of a number of chiral supermultiplets
indexed by $j$. Assign each multiplet the supersymmetry transformation
laws \leqn{eq:r3}.  Then it can be shown by a simple generalization of
the  discussion just given that the supersymmetry transformation leaves
invariant Lagrangians of the general form 
\beqa
        \L &=& \del_\mu \phi_j^* \del^\mu\phi_j + 
\psi_j^\dagger i \bar\sigma \cdot \del \psi_j + F_j^\dagger F_j  \CR
   & & \hskip 0.3in  + (F_j {\del W \over \del \phi_j} 
-  \half \psi^T_j c \psi_k {\del^2 W \over  \del \phi_j \del \phi_k}) + \hc 
  \ , 
\eeqa{eq:t3}
where $W(\phi)$ is an analytic function of the complex fields $\phi_j$
which is called the {\em superpotential}.  It is important to repeat
that $W(\phi)$ can have arbitrary dependence on the $\phi_j$, but it
must not depend on the $\phi_j^*$.  The auxiliary fields  $F_j$ obey
the equations 
\beq
    F^\dagger = -  {\del W \over \del \phi_j} \ . 
\eeq{eq:u3} 
If $W$ is a polynomial in the $\phi_j$, the elimination of the $F_j$ 
by substituting \leqn{eq:u3} into \leqn{eq:t3} produces polynomial
interactions for the scalar fields.

%%%%%%%%%%%%%%%%%%%%%%%%%%%%%%%%%%%%%%%%%%%%%%%%%%%%%%%%%%%%%%%%%%%%%%
\begin{figure}[t]
\begin{center}
\leavevmode
{\epsfxsize=3in\epsfbox{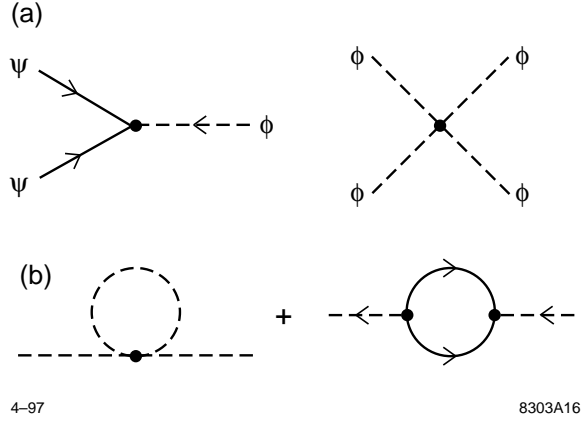}}
\end{center}
 \caption{(a) Yukawa and four-scalar couplings arising from the
supersymmetric Lagrangian with superpotential \protect\leqn{eq:w3}; (b)
Diagrams which give the leading radiative corrections to the scalar
field mass term.}
\label{fig:eight}
\end{figure}
%%%%%%%%%%%%%%%%%%%%%%%%%%%%%%%%%%%%%%%%%%%%%%%%%%%%%%%%%%%%%%%%%%%%%%
 
The free massive Lagrangian \leqn{eq:o3} is a special case of
\leqn{eq:t3} for one supermultiplet with the superpotential 
\beq
        W = \half m \phi^2 \ .
\eeq{eq:v3}
A more interesting model is obtained by setting
\beq 
        W = \frac{1}{3} \lambda \phi^3 \ .
\eeq{eq:w3}
In this case, $W$ leads directly to a Yukawa coupling proportional to
$\lambda$, while  substituting for $F$ from \leqn{eq:u3}
yields a four scalar coupling
proportional to $\lambda^2$:
\beq
        \L = \del_\mu \phi^* \del^\mu\phi + 
\psi^\dagger i \bar\sigma \cdot \del \psi - \lambda^2 |\phi^2|^2
          -\lambda\bigl[ \psi^T c \psi \phi - \phi^* \psi^\dagger c \psi^*
         \bigr] \ .
\eeq{eq:ww3}
  These two vertices are shown in
Figure~\ref{fig:eight}(a).  Their sizes are such that the two leading
diagrams which contribute to the scalar field mass renormalization,
shown in Figure~\ref{fig:eight}(b), are of the same order of magnitude.
In fact, it is not difficult to compute these diagrams for external
momentum $p = 0$.  The first diagram has the value
\beq
 - 4\lambda^2 i \cdot \int {d^4k\over (2\pi)^4} \, {i\over k^2} = 4 \lambda^2  
 \int {d^4k\over (2\pi)^4} \, {1\over k^2} \ .
\eeq{eq:x3} 
To  compute the second diagram, note that the standard form of the 
fermion propagator is
 $\VEV{\psi \psi^\dagger}$, and be careful to include all minus signs
resulting from fermion reordering.  Then you will find
\beqa
 & & \half (-2i\lambda)(2i\lambda)\int {d^4k\over (2\pi)^4} \,
   \tr \left[ {i\sigma\cdot k\over k^2} c \bigl( {-i\sigma\cdot k\over k^2}
      \bigr)^T c \right] \CR
& & \hskip 0.4in = -2\lambda^2 \int {d^4k\over (2\pi)^4} \,
 {  \tr[ \sigma\cdot k \bar \sigma \cdot k]\over k^4 } \ .
\eeqa{eq:xx3}
Using \leqn{eq:b3}, the trace gives $2k^2$, and the two diagrams 
cancel precisely. Thus, the choice \leqn{eq:w3} presents us with
an interacting quantum field theory, but one with exceptional
cancellations in the scalar field mass term.

In this simple model, it is not difficult to see that the scalar field
mass corrections must vanish as a matter of principle.
  The theory with superpotential
\leqn{eq:w3} is invariant under the symmetry 
\beq
      \phi \to e^{2i\alpha} \phi \ , \qquad \psi \to e^{-i\alpha} \psi\ .
\eeq{eq:y3}
This symmetry is inconsistent with the appearance of a fermion mass
term $m\psi^T c \psi$, as in \leqn{eq:q3}.  The symmetry does not
prohibit the appearance of a scalar mass term, but if the theory is to
remain supersymmetric, the scalar cannot have a different mass from the
fermion. However, the cancellation of radiative corrections in models
of the form \leqn{eq:t3} is actually much more profound.  It can be
shown that, in a general model of this type, the only nonvanishing
radiative corrections to the potential terms are field rescalings.  If
a particular coupling---the mass term, a cubic interaction, or any
other---is omitted from the original superpotential, it cannot be
generated by radiative corrections \cite{SPnr,GSRk}.

For later reference, I will write the potential energy associated
with the most general system with a Lagrangian of the form
\leqn{eq:t3}.  This is 
\beq
   V = -  F_j^\dagger F_j  - F_j {\del W \over \del \phi_j} - F^\dagger_j
\left( {\del W \over \del \phi_j}\right)^* \ .
\eeq{eq:z3}
Substituting for $F_j$ from \leqn{eq:u3}, we find
\beq
  V = \sum_j \left|  {\del W \over \del \phi_j} \right|^2 \ .
\eeq{eq:a4}
This simple result is called the {\em F-term} potential.  It is
minimized by setting all of the $F_j$ equal to zero.  If this is
possible, we obtain a vacuum state with $\vev{H} = 0$ which is also
invariant to supersymmetry, in accord with the discussion of
\leqn{eq:m3}.  On the other hand, supersymmetry is  spontaneously
broken if for some reason it is not possible to satisfy all of the
conditions $F_j= 0$ simultaneously.  In that case, we obtain a vacuum
state with $\vev{H} >0$.

\subsection{Coupling constant unification}

At this point, we have not yet completed our discussion of the
structure of supersymmetric Lagrangians.  In particular, we have not
yet written the supersymmetric Lagrangians of vector fields, beyond
simply noting that a vector field combines with a chiral fermion to
form a vector supermultiplet. Nevertheless, it is not too soon to try
to write a supersymmetric generalization of the Standard Model.

I will first list the ingredients needed for this generalization.  For
each of the $SU(3)\times SU(2)\times U(1)$ gauge bosons, we need a
chiral fermion $\lambda^a$ to form a vector supermultiplet.  These
new fermions are called {\em gauginos}.  I will refer  the specific
partners of specific gauge bosons  with  a tilde.
 For example, the fermionic partner of the 
gluon will be called $\widetilde g$, the {\em gluino}, and the fermionic 
partners of the $W^+$ will be called $\widetilde w^+$, the {\em wino}.

  None of
these fermions have the quantum numbers of quarks and leptons.  So we
need to add a complex scalar for each chiral fermion species to put
the quarks and leptons into chiral supermultiplets.  I will use the labels for
left-handed fermion multiplets in \leqn{eq:a1} also to denote the quark
and lepton supermultiplets. Hopefully, it will be clear from context
whether I am talking about the supermultiplet or the fermion.  The
scalar partners of quarks and leptons are called {\em squarks} and {\em 
sleptons}.  I will denote these with a tilde.  For example, the partner
of $e^-_L = L^-$ is the selectron $\widetilde e^-_L$ or $\widetilde L^-$.
The partner of $\bar e^* = e^-_R$ is a distinct selectron which I will call
$\widetilde e^-_R$.  The Higgs fields must
also belong to chiral supermultiplets.   I will denote the scalar
components as $h_i$ and the left-handed fermions as $\widetilde h_i$.  We
will see in a moment that at least two different Higgs multiplets are
required.

Although we need a bit more formalism to write the supersymmetric
generalization of the Standard Model gauge couplings, it is already
completely straightforward to write the supersymmetric generalization
of the Yukawa couplings linking quarks and leptons to the Higgs sector.
The generalization of the third line of \leqn{eq:a} is given by writing
the superpotential 
\beq 
 W =  \lambda^{ij}_u \bar u^i h_2 \cdot  Q^j
 + \lambda^{ij}_d \bar d^i  h_1 \cdot Q^j + 
 \lambda^{ij}_\ell \bar e^i h_1 \cdot L^j
\eeq{eq:b4}
Note that, where in \leqn{eq:a} I wrote $\phi$ and $\phi^*$, I am
forced here to introduce two different Higgs fields $h_1$ and $h_2$. 
The hypercharge assignments of $\bar u$ and $Q$ require for the first
term a Higgs field with $Y = +\half$; for the next two terms, we need a
Higgs field with $Y = - \half$.  Since $W$ must be an analytic function
of supermultiplet fields, as I explained below \leqn{eq:t3}, replacing
$h_1$ by $(h_2)^*$ gives a Lagrangian which is not supersymmetric. 
There is another, more subtle, argument for a second Higgs doublet. 
Just as in the Standard Model, triangle loop diagrams involving the
chiral fermions of the theory contain terms which potentially violate
gauge invariance.  These anomalous terms cancel when one sums over the
chiral fermions of each quark and lepton generation.  However, the
chiral fermion $\widetilde h_2$ leads to a new anomaly term which
violates the conservation of hypercharge.  This contribution is
naturally cancelled by the contribution from $\widetilde h_1$.

We  still need several more ingredients to construct the full 
supersymmetric generalization of the Standard Model, but we have now
made 
a good start.  We have introduced the minimum number of new particles
(unfortunately, this is not a small number), and we have generated new
couplings for them without yet introducing new parameters beyond those
of the Standard Model.

In addition, we already have enough information to study the
unification of forces using the formalism of Section 2.5.  To begin, we
must extend the formulae \leqn{eq:k1}, \leqn{eq:kk1} to supersymmetric
models.  For $SU(N)$ gauge theories, the gauginos
give a contribution $(-\frac{2}{3}N)$ to the right-hand side
of \leqn{eq:k1}. In \leqn{eq:kk1}, there is no contribution either from
the gauge bosons or from their fermionic partners.  We should also
group together the contributions from matter fermions and scalars. 
Then we can write the renormalization group coefficient $b_N$ for
$SU(N)$ gauge theories with $n_f$ chiral supermultiplets in the
fundamental representation as 
\beq
    b_N = 3 N - \frac{1}{2} n_f \ .
\eeq{eq:c4}
Similarly, the renormalization group coefficient for $U(1)$ gauge
theories is now 
\beq
   b_1 =  -  \sum_f t_f^2 \ ,
\eeq{eq:d4}
where the sum runs over chiral supermultiplets.

\begin{figure}[t]
\begin{center}
\leavevmode
{\epsfxsize=4in\epsfbox{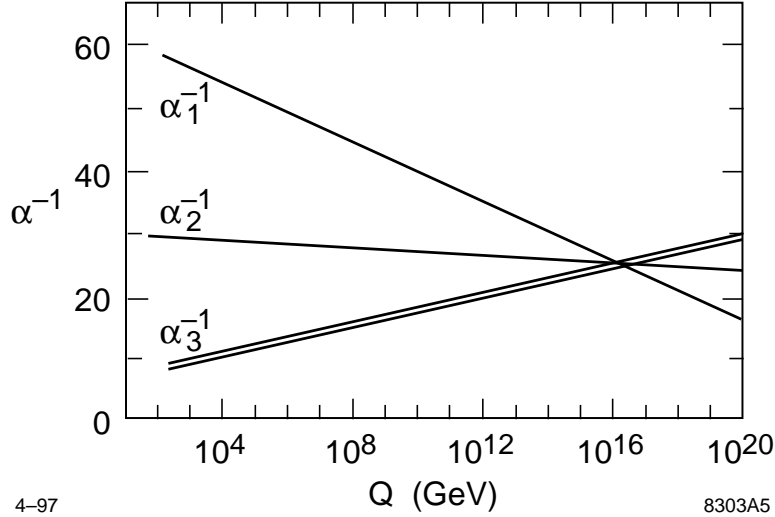}}
\end{center}
 \caption{Evolution of the $SU(3)\times SU(2)\times U(1)$ gauge
couplings to high energy scales, using the one-loop renormalization group
equations of the supersymmetric generalization of the Standard Model.}
\label{fig:nine}
\end{figure}

Evaluating these expressions for $SU(3)\times SU(2)\times U(1)$ gauge
theories with $n_g$ quark and lepton generations and $n_h$ Higgs
fields, we find 
\beqa
    b_3 &=&   9 - 2 n_g \CR
    b_2 &=&   6 - 2 n_g - \frac{1}{2} n_h\CR 
    b_1 &=&   \phantom{9} -  2 n_g - \frac{3}{10} n_h \ .
\eeqa{eq:e4}
Now insert these expressions into \leqn{eq:y1}; for $n_h =2$, we find
\beq
B = \frac{5}{7} = 0.714 \ ,
\eeq{eq:f4}
in excellent agreement with the experimental value \leqn{eq:z1}. 
Apparently,  supersymmetry repairs the difficulty that the
Standard Model has in linking in a simple way to grand unification. 
The running coupling constants extrapolated from the experimental
values \leqn{eq:w1} using the supersymmetric renormalization group
equations are shown in Figure~\ref{fig:nine}.

%%%%%%%%%%%%%%%%%%%%%%%%%%%%%%%%%%%%%%%%%%%%%%%%%%%%%%%%%%%%%%%%%%%%%%
%%%%%%%%%%%%%%%%%%%%%%%%%%%%%%%%%%%%%%%%%%%%%%%%%%%%%%%%%%%%%%%%%%%%%%

Of course it is not difficult to simply make up a model that agrees with 
any previously given value of $B$.
 I hope to have convinced you that the value
\leqn{eq:f4} arises naturally in grand unified theories based on
supersymmetry.  By comparing this agreement to the error bars for $B$
quoted in \leqn{eq:z1}, you can decide for yourself whether this
agreement is fortuitous.

\subsection{The rest of the supersymmetric Standard Model}

I will now complete the Lagrangian of the supersymmetric
generalization of the Standard Model.  First, I must write the
Lagrangian for the vector supermultiplet and then I must  show how to
couple that multiplet
 to matter fields.  After this, I will discuss some general
properties of the resulting system.

The vector multiplet $( A_\mu^a, \lambda^a)$ containing the gauge
bosons of a Yang-Mills theory and their partners has the supersymmetric
Lagrangian 
\beq
 \L = -\frac{1}{4} \left( F_{\mu\nu}^a\right)^2 + \lambda^{\dagger a}
i \bar\sigma^\mu D_\mu \lambda^a + \half (D^a)^2 \ ,
\eeq{eq:g4}
where $D_\mu= (\del_\mu - i g A_\mu^a t^a)$ is the gauge-covariant
derivative, with $t^a$ the gauge group generator. In order to write the
interactions of this multiplet in the simplest form, I have introduced
a set of auxiliary real scalar fields, called  $D^a$.  (The name is
conventional; please do not confuse them with the covariant
derivatives.) The gauge interactions of a chiral multiplet are then
described by generalizing the first line of \leqn{eq:t3} to 
\beqa 
 \L
 &=&  D_\mu \phi_j^* D^\mu\phi_j + \psi_j^\dagger i \bar\sigma^\mu D_\mu
 \psi_j + F_j^\dagger F_j  \CR
  &  & -\sqrt{2}i g\left( \phi_j \lambda^{Ta} t^a c \psi_j - 
         \psi^\dagger t^a c \lambda^{*a} \phi_j\right) + g D^a \phi^\dagger
         t^a \phi \ .
\eeqa{eq:h4}
Eliminating the auxiliary fields using their field equation
\beq
          D^a = - g \sum_j  \phi^\dagger  t^a \phi 
\eeq{eq:i4}
gives a second contribution to the scalar potential, which should be
added to the F-term \leqn{eq:a4}.  This is the {\em D-term} 
\beq
         V =  \frac{g^2}{2} \left( \sum_j  \phi^\dagger  t^a \phi \right)^2\ .
\eeq{eq:j4}
As with the F-term, the ground state of this potential is obtained by
setting all of the $D^a$ equal to zero, if it is possible.  In that
case, one obtains a supersymmetric vacuum state with $\vev{H} = 0$.

The  full supersymmetric generalization of the Standard Model can be
written in the form 
\beq 
 \L =  \L_{\rm gauge} + \L_{\rm kin} + \L_{\rm Yukawa} + \L_\mu \ . 
\eeq{eq:k4} 
The first term is the kinetic energy term for the gauge multiplets of
$SU(3)\times SU(2)\times U(1)$.  The second term is the kinetic energy
term for quark, lepton, and Higgs chiral multiplets, including gauge
couplings of the form \leqn{eq:h4}.  The third term is the Yukawa and
scalar interactions given by the second line of \leqn{eq:t3} using the
superpotential \leqn{eq:b4}. The last term is that following from an
 additional gauge-invariant term that we could add to the superpotential,
\beq
   \Delta W = \mu h_1 \cdot h_2 \ .
\eeq{eq:l4}
This term contributes a supersymmetric mass term to the Higgs fields
and to their fermions partners.  This term is needed on
phenomenological grounds, as I will discuss in Section 4.4.  The
parameter $\mu$ is the only new parameter that we have added so far to
the Standard Model.

This Lagrangian does not yet describe a realistic theory.  It has exact
supersymmetry.  Thus, it predicts charged scalars degenerate with the
electron and massless fermionic partners for the photon and gluons. On
the other hand, it has some very attrative properties.  For the reasons
explained below \leqn{eq:y3}, there is no quadratically divergent
renormalization of the Higgs boson masses, or of any other mass in the
theory.  Thus, the radiative correction \leqn{eq:z}, which was such a
problem for the Standard Model, is absent in this generalization.  In
fact, the only renormalizations in the theory are renormalizations of
the $SU(3)\times SU(2)\times U(1)$ gauge couplings and rescalings of
the various quark, lepton, and Higgs fields.  In the next section, I
will show that we can modify \leqn{eq:k4} to maintain this property
while making the mass spectrum of the theory more realistic.

The Lagrangian \leqn{eq:k4} conserves the discrete quantum number
\beq
         R = (-1)^{L + Q + 2J} \ ,
\eeq{eq:m4}
where $L$ is the lepton number, $Q= 3B$ is the quark number, and $J$ is
the spin.  This quantity is called {\em R-parity}, and it is
constructed precisely so that $R = +1$ for the conventional gauge
boson, quark, lepton, and Higgs states while $R = -1$ for their
supersymmetry partners. If $R$ is exactly conserved, supersymmetric
particles can only be produced in pairs, and the lightest
supersymmetric partner must be absolutely stable.  On the other hand,
$R$-parity can be violated only by adding terms to $\L$ which violate
baryon- or lepton-number conservation.

It is in fact straightforward to write a consistent
$R$-parity-violating supersymmetric theory.  The following terms which
can be added to the superpotential are invariant under $SU(3)\times
SU(2)\times U(1)$ but violate baryon or lepton number: 
\beq 
 \Delta W = 
 \lambda^{ijk}_B \bar u^i \bar d^j \bar d^k
 + \lambda^{ijk}_L  Q^i \cdot L^j \bar d^k + 
 \lambda^{ijk}_e  L^i \cdot L^j \bar e^k + \mu_L^{i} L^i \cdot h_2 \ .
\eeq{eq:n4}
A different phenomenology is produced if one adds the baryon-number
violating couplings $\lambda_B$, or if one adds the other couplings\
written in \leqn{eq:n4},
which violate lepton number.  If one were to add both types of
couplings at once, that would be a disaster, leading to rapid proton
decay.

For a full exploration of the phenomenology of supersymmetric theories,
we should investigate both models in which $R$-parity is conserved, in
which the lightest superpartner is stable, and models in which
$R$-parity is violated, in which the lightest superpartner decays
through $B$- or $L$- violating interactions. In these lectures, since
my plan is to present illustrative examples rather than a systematic
survey, I will restrict my attention to models with conserved
$R$-parity.

\subsection{How to describe supersymmetry breaking}

Now we must address the question of how to modify the Lagrangian
\leqn{eq:k4} to obtain a model that could be realistic.  Our problem is
that the supersymmetry on which the model is based is not manifest in
the spectrum of particles we see in Nature.  So now we must add new
particles or interactions which cause supersymmetry to be spontaneously
broken.

It would be very attractive if there were a simple model of
supersymmetry breaking that we could connect to the supersymmetric
Standard Model. Unfortunately, models of supersymmetry breaking are
generally not simple. So most studies of supersymmetry do not invoke
the supersymmetry breaking mechanism directly but instead try to treat
its consequences phenomenologically. This can be done by adding to
\leqn{eq:k4} terms which violate supersymmetry but become unimportant
at high energy.  Some time ago, Grisaru and Girardello \cite{GG} listed
the terms that one can add to a supersymmetric Lagrangian without
disturbing the cancellation of quadratic divergences in the scalar mass
terms.  These terms are 
\beq
  \L_{\rm soft} = - M^2_j \left| \phi_j \right|^2 
  -  m_a \lambda^{Ta} c \lambda^a
          + B \mu h_1 \cdot h_2 + A W(\phi) \ ,
\eeq{eq:o4}
where $W$ is the superpotential \leqn{eq:b4}, plus other possible
analytic terms cubic in the scalar fields $\phi_j$. These terms give
mass to the squarks and sleptons and to the gauginos, moving
the unobserved superpartners to higher energy. Note that terms of the
structure $\phi^* \phi\phi$ and the mass term $\psi^T c \psi$ do
not appear in \leqn{eq:o4} because they can 
regenerate the divergences of the nonsupersymmetric theory.
All of the coefficients in \leqn{eq:o4} have the dimensions of (mass)
or (mass)$^2$. These new terms in \leqn{eq:o4} are called {\em soft
supersymmetry-breaking terms}.  We can build a phenomenological model
of supersymmetry by adding to \leqn{eq:k4} the various terms in
$\L_{\rm soft}$ with coefficients to be determined by experiment.

It is not difficult to understand that it is the new, rather 
than the familiar, half of the spectrum of the supersymmetric model
that obtains mass from  \leqn{eq:o4}.  In
Section 2.5, I argued that the particles we see in high-energy
experiments are visible only because they are protected
from acquiring very large masses  by some
symmetry principle .  In that
discussion, I invoked only the Standard Model gauge symmetries.  In
supersymmetric models, we have a more complex situation.  In each
supermultiplet, one particle is protected from acquiring mass, as
before, by $SU(2) \times U(1)$.  However, their superpartners---the squarks,
sleptons, and gauginos---are protected from 
obtaining mass only by the  supersymmetry relation to their partner. 
Thus, if supersymmetry is spontaneously broken, all
that is necessary to generate masses for these partners is a coupling
of the supersymmetry-breaking expectation values 
to the Standard Model supermultiplets.

This idea suggests a general structure for  a realistic supersymmetric model.
All of the phenomena of the model are driven by supersymmetry breaking.
First, supersymmetry is broken spontaneously in some new sector of
particles at high energy.  Then, the coupling between these particles
and the quarks, leptons, and gauge bosons leads to soft
supersymmetry-breaking terms for those supermultiplets.  It is very
tempting to speculate further that those terms might then  give rise to
the spontaneous breaking of $SU(2)\times U(1)$ and so to the masses for the 
$W$ and $Z$ and for the quarks and leptons.  I will explain in the next
section how this might happen.

The size of the mass terms in \leqn{eq:o4} depends on two factors.  The
first of these is the mass scale at which supersymmetry is broken. 
Saying for definiteness that supersymmetry breaking is due to the
nonzero value of an $F$ auxiliary field, we can denote this scale by
writing $\vev{F}$, which has the dimensions of (mass)$^2$.  The second
factor is the mass of the bosons or fermions which couple the
high-energy sector to the particles of the Standard Model and thus
communicate the supersymmetry breaking. I will call this mass $\M$, the
{\em messenger scale}.  Then the mass parameters that appear in
\leqn{eq:o4} should be of the order of 
\beq
      m_S = {\vev{F}\over \M} \ .
\eeq{eq:p4}
If supersymmetry indeed gives the mechanism of electroweak symmetry
breaking, then $m_S$ should be of the order of 1 TeV. A case that is
often discussed in the literature is that in which the messenger is
supergravity.  In that case, $\M$ is the Planck mass $m_\Pl$, equal
to  $10^{19}$ GeV, and 
$\vev{F} \sim 10^{11}$ (GeV)$^2$.  Alternatively, both $\vev{F}$ and
$\M$ could be of the order of a few TeV.

The detailed form of the soft supersymmetry-breaking terms depends on
the underlying model that has generated them.  If one allows these
terms to have their most general form (including arbitrary flavor- and
CP-violating interactions, they contain about 120 new parameters. 
However, any particular model of supersymmetry breaking generates a
specific set of these soft terms with some observable regularities. 
One of our goals in Section 4 of these lectures will be to understand
how to determine the soft parameters experimentally and thus uncover
the patterns which govern their construction.

\subsection{Electroweak symmetry breaking from  supersymmetry}

There is a subtlety in trying to determine the pattern of the soft
parameters experimentally.  Like all other coupling constants in a
supersymmetric theory, these parameters run under the influence of the
renormalization group equations.  Thus, the true underlying pattern
might not be seen directly at the TeV
energy scale.  Rather, it might be necessary to extrapolate the
measured values of parameters to higher energy to  look for regularities.

The situation here is very similar to that of the Standard Model
coupling constants.  The underlying picture which leads to the values
of the $SU(3)\times SU(2)\times U(1)$ coupling constants is not very
obvious from the data \leqn{eq:w1}.  Only when these data are
extrapolated to very high energy using the renormalization group do we
see evidence for their unification. Obviously, such evidence must be
 indirect.  On the other hand, the discovery of supersymmetric
particles, and the discovery that these particles showed other
unification relations---with the same unification mass scale---would
give powerful support to this picture.

I will discuss general systematics of the renormalization-group running
of the soft parameters in Section 4.2.  But there is one set of
renormalization group equations that I would like to call your
attention to right away.  These are the equations for the soft mass of
the Higgs boson and the squarks  which are most strongly coupled
to it.  We saw in Section 2.6 that the top quark Yukawa coupling was
sufficiently large that it could have an important effect in
renormalization group evolution.  Let us consider, then, the evolution
equations for the three scalars that interact through this coupling,
the Higgs boson $h_2$, the scalar top $\widetilde Q_{t} = \widetilde t_L$,
 and the
scalar top $\widetilde t_R$.  The most important terms in
these equations are the following: 
\beqa
  {d\over d \log Q} M^2_{h} &=& {1\over (4\pi)^2} \left\{ 3 \lambda_t^2 
          (M^2_h + M^2_Q + M^2_t) + \cdots \right\}\CR
  {d\over d \log Q} M^2_{Q} &=& {1\over (4\pi)^2} \left\{ 2 \lambda_t^2 
          (M^2_h + M^2_Q + M^2_t) - {32\over 3}g_3^2 m_3^2\cdots \right\}\CR
  {d\over d \log Q} M^2_{t} &=& {1\over (4\pi)^2} \left\{ 3 \lambda_t^2 
          (M^2_h + M^2_Q + M^2_t) - {32\over 3}g_3^2 m_3^2 + \cdots \right\}
         \ ,           \CR
\eeqa{eq:r4}
where $g_3$ is the QCD coupling, $m_3$ is the mass of the gluino, 
and the omitted terms are of electroweak
strength.  The last two equations exhibit the competition between the
top quark Yukawa coupling and QCD renormalizations which we saw earlier
in \leqn{eq:e2} and \leqn{eq:h2}.  The supersymmetric QCD couplings
cause the masses of the $\widetilde Q_{t}$ and $\widetilde t_R$ to
increase at low energies, while the effect of $\lambda_t$ causes all
three masses to decrease.

%%%%%%%%%%%%%%%%%%%%%%%%%%%%%%%%%%%%%%%%%%%%%%%%%%%%%%%%%%%%%%%%%%%%%%
\begin{figure}[t]
\begin{center}
\leavevmode
{\epsfxsize=4in\epsfbox{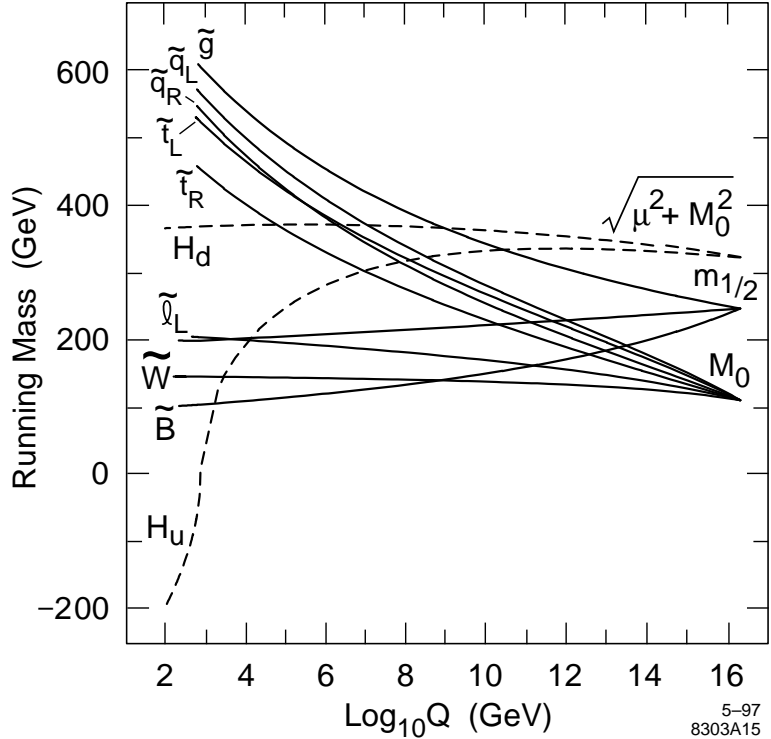}}
\end{center}
 \caption[*]{Example of the 
evolution of the soft supersymmetry-breaking mass terms from 
       the grand unification scale to the weak interaction scale, 
   from \protect\cite{Gordycrew}. The initial conditions for the 
evolution equations at the grand unification scale are taken to be the 
 universal among species, in a simple pattern presented in Section 4.2.}
\label{fig:ten}
\end{figure}
%%%%%%%%%%%%%%%%%%%%%%%%%%%%%%%%%%%%%%%%%%%%%%%%%%%%%%%%%%%%%%%%%%%%%%
 
Indeed, if the  $\widetilde Q_{t}$ and $\widetilde t_R$ masses
stay large, the equations \leqn{eq:r4} predict that $M^2_h$ should go
down through zero and become negative \cite{RI}.  Thus, if all scalar mass
parameters are initially positive at high energy scales, these
equations imply that the Higgs boson $h_2$ will acquire a negative
parameter and thus an instability to electroweak symmetry breaking.  An
example of the solution to the full set of renormalization group
equations, exhibiting the instability in $M^2_h$, is shown in
Figure~\ref{fig:ten} \cite{Gordycrew}.

At first sight, it might  have been any of the scalar fields in the
theory whose potential would be  unstable by
renormalization group evolution.  But the Higgs scalar $h_2$ has
the strongest instability if the top quark is heavy.  In this way, the
supersymmetric extension of the Standard Model naturally contains the
essential  feature that we set out to find, a physical  mechanism for 
electroweak symmetry breaking.  As a bonus, we find that this mechanism
is closely associated with the heaviness of the top quark.

If you have been patient through all of the formalism I have presented
in this section, you now see that your patience has paid off.  It was
not obvious when we started that supersymmetry would give the essential
ingredients of a theory of electroweak symmetry breaking.  But it
turned out to be so.  In the next section, I will present more details
of the physics of supersymmetric models and present a program for their
experimental exploration.

\section{Supersymmetry: Experiments}

In the previous section, I have presented the basic formalism of
supersymmetry. I have also explained that supersymmetric models have
several features that naturally answer questions posed by the Standard
Model.  At the beginning of Section 3, I told you that supersymmetry
might be considered a worked example of physics beyond the Standard
Model.  Though I doubt you are persuaded  by now that physics beyond
the Standard Model must be supersymmetric,  I hope you see these models
as reasonable alternatives that can be understood in very concrete
terms.

Now I would like to analyze the next step along this line of reasoning.
What if, at LEP 2 or at some higher-energy machine, the superpartners
appear?  This discovery would change the course of experimental
high-energy physics and shape it along a certain direction.  We should
then ask, what will be the important issues in high-energy physics, and
how will we resolve these issues experimentally? In this section, I
will give a rather detailed answer to this question.

I emphasize again that I am not asking you to become a believer in
supersymmetry. A different discovery about physics beyond the Standard
Model would change the focus of high-energy physics in a different
direction.  But we will learn more by choosing a particular direction
and studying its far-reaching implications than by trying to reach
vague but  general conclusions. I will strike off in a different
direction in Section 5.

On the other hand, I hope you are not put off by the complexity of the
supersymmetric Standard Model.  It is true that this model has many
ingredients and a very large content of new undiscovered particles.  On
the other hand, the model develops naturally from a single physical
idea.   I argued in Section 2.2 that this
structure, a complex phenomenology built up around a definite 
principle of physics, is seen often in Nature.  It
 leads to a more attractive solution to the problems of the
Standard Model than a model whose only virtue is minimality.

It is true that, in models with complex consequences, it may not be
easy to see the underlying structure in the experimental data.  This
is the challenge that experimenters will face.  I will now discuss how
we can meet this challenge for the particular case in which the physics
beyond the Standard Model is supersymmetric.

\subsection{More about soft supersymmetry breaking}

As we discussed in Section 3.6, a realistic supersymmetric theory has a
Lagrangian of the form 
\beq
 \L =  \L_{\rm gauge} +  \L_{\rm kin} + \L_{\rm Yukawa} + \L_\mu 
 + \L_{\rm soft}\ .
\eeq{eq:s4}
Of the various terms listed here, the first three contain only
couplings that are already present in the Lagrangian of the Standard
Model.  The fourth term contains one new parameter $\mu$.  The last
term, however, contains a very large number of new parameters.

I have already explained that one should not be afraid of seeing a
large number of undetermined parameters here.  The same proliferation 
of parameters occurs in any theory with a certain level of complexity
when viewed from below.  The low-energy scattering amplitudes of QCD,
for example, contain many parameters  which turn out to be the masses
and decay constants of hadronic resonances.  If it is possible to
measure these parameters, we will obtain a large amount of new
information.

In thinking about the values of the soft supersymmetry-breaking
parameters, there are two features that we should take into account. 
The first is that the soft parameters obey renormalization group
equations.  Thus, they potentially change significantly from their
underlying values at the messenger scale defined in \leqn{eq:p4} to
their physical values observable at the TeV scale.  We have seen in
Section 3.7 that these changes can have important physical
consequences.  In the next section, I will describe the renormalization
group evolution of the supersymmetry-breaking mass terms in more
detail, and we will use our understanding of this evolution to work out
some general predictions for the superparticle spectrum.

%%%%%%%%%%%%%%%%%%%%%%%%%%%%%%%%%%%%%%%%%%%%%%%%%%%%%%%%%%%%%%%%%%%%%%%%%
\begin{figure}[t]
\begin{center}
\leavevmode
{\epsfxsize=2.0in\epsfbox{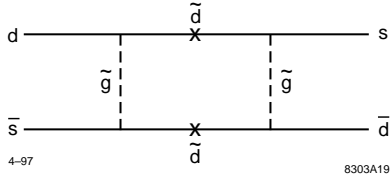}}
\end{center}
 \caption{A potentially 
dangerous contribution of supersymmetric particles to 
             flavor-changing neutral current processes.}
\label{fig:eleven}
\end{figure}
%%%%%%%%%%%%%%%%%%%%%%%%%%%%%%%%%%%%%%%%%%%%%%%%%%%%%%%%%%%%%%%%%%%%%%

The second feature is that there are strong constraints on the flavor
structure of soft supersymmetry breaking terms which come from
constraints on flavor-changing neutral current processes.  In
\leqn{eq:o4}, I have written independent mass terms for each of the
scalar fields.  In principle, I could also have written mass terms that
mixed these fields.  However, if we write the scalars in the basis in
which the quark masses are diagonalized, we must not find substantial
off-diagonal terms.  A mixing 
\beq
            \Delta\L =  \Delta M^2_d  \bar s^\dagger \bar d \ ,
\eeq{eq:t4}
for example, would induce an excessive contribution to the $K_L$--$K_S$
mass difference through the diagram shown in Figure~\ref{fig:eleven}
unless 
\beq
         {   \Delta M^2_d \over M^2_d}< 10^{-2}\left( {M_d \over
 300 \ \mbox{GeV}}\right)^2 \ .
\eeq{eq:u4}
Similar constraints arise from $D$--$\bar D$ mixing, $B$--$\bar B$
mixing, $\mu\to e \gamma$ \cite{SUSYFCNC}.

The strength of the constraint \leqn{eq:u4} suggests that the physical
mechanism that generates the soft supersymmetry breaking terms 
contains a natural feature that suppresses such off-diagonal terms. 
One possibility is that equal  soft masses are generated for all
scalars with the same $SU(2)\times U(1)$ quantum numbers. Then the
scalar mass matrix is proportional to the matrix {\bf 1} and so is
diagonal in any basis \cite{DGg,Sakai}.
 Another possibility is that, by virtue of
discrete flavor symmetries, the scalar mass matrices are approximately
diagonal in the same basis in which the quark mass matrix is diagonal
\cite{NLS}. These two solutions to the potential problem of
supersymmetric flavor violation are called, respectively, 
`universality' and `alignment'. A problem with the alignment scenario
is that the bases which diagonalize the $u$ and $d$ quark mass matrices
differ by the weak mixing angles, so it is not possible to completely
suppress the mixing both for the $u$ and $d$ partners.  This scenario
then leads to a prediction of $D$--$\bar D$ mixing near the current
experimental bound.

\subsection{The spectrum of superparticles---concepts}

We are now ready to discuss the expectations for the mass spectrum of
supersymmetric partners.  Any theory of this spectrum must have two
parts giving , first, the generation of the underlying soft parameters
at the messenger scale and , second, the modification of these
parameters through renormalization group evolution.  In this section, I
will make the simplest assumptions about the underlying soft parameters
and concentrate on the question of how these parameters are modified by
the renormalization group.  In the next section, we will confront the
question of how these simple assumptions can be tested.

Let us begin by considering the fermionic partners of gauge bosons,
the gauginos.
  If the messenger scale lies above the scale of grand
unification, the gauginos associated with the $SU(3)\times SU(2)\times
U(1)$ gauge bosons will be organized into a single representation of
the grand unification group and thus will have a common soft mass term.
This gives a very simple initial condition for renormalization group
evolution.

The renormalization group equation for a gaugino mass $m_i$ is
\beq
     {d \over d \log Q} m_i =  - {1\over (4\pi)^2} \cdot 2 b_i \cdot m_i\ ,\
\eeq{eq:v4}
where $i= 3,2,1$ for the gauginos of $SU(3)\times SU(2)\times U(1)$ and
$b_i$ is the coefficient in the equation \leqn{eq:j1} for the coupling
constant renormalization.  Comparing these two equations, we find that
$m_i(Q)$ and $\alpha_i(Q)$ have the same renormalization group
evolution, and so their ratio is constant as a function of $Q$.  This
relation is often written 
\beq
              {m_i(Q)\over \alpha_i(Q)} =   {m_{1/2}\over \alpha_U}\ ,
\eeq{eq:w4}
where $\alpha_U$ is the unification value of the coupling constant
($\alpha_U^{-1} = 24$), and $m_{1/2}$ is the underlying soft mass
parameter. As the $\alpha_i$ flow from their unified value at very
large scales to their observed values at $Q= \mz$, the gaugino masses
flow along with them.  The result is that the grand unification of
gaugino masses implies the following relation among the observable
gaugino masses: 
\beq
    {m_1 \over  \alpha_1} =  {m_2 \over  \alpha_2} =  {m_3 \over  \alpha_3}\ . 
\eeq{eq:x4}
I will refer to this relation as {\em gaugino unification}.  It implies
that, for the values at the weak scale, 
\beq
   {m_1 \over  m_2} = 0.5 \ , \qquad  {m_3 \over  m_2} = 3.5 \ .
\eeq{eq:y4}
I caution you that these equations apply to a perturbative (for
example, $\bar{MS}$) definition of the masses.  For the gluino mass
$m_3$, the physical, on-shell, mass may be larger than the $\bar{MS}$
mass by 10--20\%, due to a radiative correction which depends on the
ratio of the masses of the gluon and quark partners \cite{Damien}.

%%%%%%%%%%%%%%%%%%%%%%%%%%%%%%%%%%%%%%%%%%%%%%%%%%%%%%%%%%%%%%%%%%%%%%
\begin{figure}[t]
\begin{center}
\leavevmode
\epsfbox{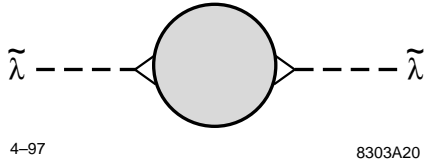}
\end{center}
 \caption{A simple radiative correction giving gaugino masses
 in the pattern of `gaugino unification'.}
\label{fig:elevenplus}
\end{figure}
%%%%%%%%%%%%%%%%%%%%%%%%%%%%%%%%%%%%%%%%%%%%%%%%%%%%%%%%%%%%%%%%%%%%%%

Though gaugino unification is a consequence of the grand unification of
gaugino masses, it does not follow uniquely from this source. On the
contrary, this result can also follow from models  in which gaugino
masses arise from radiative corrections at lower energy. For example,
in a model of Dine, Nelson, Nir, and Shirman \cite{DNNS}, gaugino
masses are induced by the diagram shown in Figure~\ref{fig:elevenplus},
in which a supersymmetry-breaking expectation value of $F$ couples to
some new supermultiplets of mass roughly 100 TeV, and this influence is
then tranferred to the gauginos through their Standard Model gauge
couplings.  As long as the mass pattern of the heavy particles is
sufficiently simple, we obtain gaugino masses $m_i$ proportional to the
corresponding $\alpha_i$, which reproduces \leqn{eq:x4}.

Now consider the masses of the squarks and sleptons, the 
scalar partners of quarks and leptons.
We saw in Section 3.4 that, 
 since  the left- and right-handed quarks belong to different
supermultiplets $Q$, $\bar u$, $\bar d$,  each has its own scalar
partners. The same situation applies for the leptons.  In this section,
I will assume for maximum 
simplicity that the underlying values of the squark
and slepton mass parameters are completely universal, with the value
$M_0$.  This is a stronger assumption than the prediction of grand
unification, and one which does not necessarily have a 
fundamental justification.
  Nevertheless, there are two effects that distort this
universal mass prediction into a complex particle spectrum.

The first of these effects comes from the $D$-term potential
\leqn{eq:j4}. Consider the contributions to this potential from the
Higgs fields $h_1$, $h_2$ and from a squark or slepton field
$\widetilde f$. Terms contributing to the $\widetilde f$ mass comes
from the $D^a$ terms associated with the $U(1)$ and the neutral $SU(2)$
gauge bosons, 
\beqa
  V &=& {g^{\prime 2}\over 2} \left( h^*_1 (-\half) h_1 + 
            h^*_2 (\half) h_2 + \widetilde f^* Y \widetilde f\right)^2 \CR
    & &  + {g^{2}\over 2} \left( h^*_1 \tau^3 h_1 + 
            h^*_2 \tau^3 h_2 + \widetilde f^* I^3 \widetilde f\right)^2 \ .
\eeqa{eq:z4}
The factors in the first line are the hypercharges of the fields $h_1$,
$h_2$. Now replace these Higgs fields by their vacuum expectation
values 
\beq
       \VEV{h_1} = {1\over \sqrt{2}}\pmatrix{v \cos \beta\cr 0 \cr} \qquad
   \VEV{h_2} = {1\over \sqrt{2}}\pmatrix{0 \cr v \sin \beta\cr}  
\eeq{eq:a5}
and keep only the cross term in each square. This gives
\beqa
 V &=&  2 {g^{\prime 2}\over 2}{v^2\over 2} \left(-\half \cos^2\beta + \half
 \sin^2\beta\right) \widetilde f^* Y \widetilde f + 
 2 {g^{2}\over 2}{v^2\over 2} \left(+\half \cos^2\beta - \half
 \sin^2\beta\right) \widetilde f^* I^3 \widetilde f \CR
&=& - c^2 \mz^2 (\sin^2\beta - \cos^2\beta)  \widetilde f^*
(I^3 - {s^2\over c^2} Y)\widetilde f \CR
&=& - \mz^2  (\sin^2\beta - \cos^2\beta)  \widetilde f^*
(I^3 - s^2 Q)\widetilde f \ .
\eeqa{eq:b5}
Thus, this term gives a contribution to the scalar mass
\beq 
 \Delta M_f^2 =  - \mz^2 {\tan^2\beta-1\over \tan^2\beta+1}  (I^3 - s^2 Q)\ .
\eeq{eq:c5}

%%%%%%%%%%%%%%%%%%%%%%%%%%%%%%%%%%%%%%%%%%%%%%%%%%%%%%%%%%%%%%%%%%%%%%
\begin{figure}[htb]
\begin{center}
\leavevmode
\epsfbox{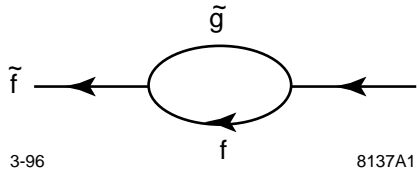}
\end{center}
 \caption{Renormalization of the soft scalar mass due to the gaugino
mass.}
\label{fig:elf}
\end{figure}
%%%%%%%%%%%%%%%%%%%%%%%%%%%%%%%%%%%%%%%%%%%%%%%%%%%%%%%%%%%%%%%%%%%%%%

The second effect is the renormalization group running of the scalar
mass induced by the gluino mass through the diagram shown in
Figure~\ref{fig:elf}. The renormalization group equation for the scalar
mass $M_f$ is 
\beq
     {d \over d \log Q} M_f =  - {1\over (4\pi)^2} \cdot 8\cdot 
 \sum_i C_2(r_i) g_i^2 m_i^2 \ , 
\eeq{eq:d5}
where 
\beq
C_2(r_i) = \cases{ \frac{3}{5} Y^2 & $U(1)$ \cr
                   0, \ \frac{3}{4} & singlets, doublets of $SU(2)$\cr
                   0, \ \frac{4}{3} & singlets, triplets of $SU(3)$\cr} \  .
\eeq{eq:e5}
In writing this equation, I have ignored the Yukawa couplings of the
flavor $f$. This is a good approximation for light flavors, but we have
already seen that it is not a good approximation for the top squarks,
and it may fail also for the $b$ and $\tau$ partners if $\tan\beta$ is
large.  In those cases, one must add further terms
 to the renormalization group equations, such as those given
in \leqn{eq:r4}.

To integrate the equation \leqn{eq:d5}, we need to know the behavior of
the gaugino masses as a function of $Q$.  Let me assume that this is
given by gaugino unification according to \leqn{eq:w4}. Then 
\beq
 {g_i^2 m_i^2\over 4\pi} = \alpha_i(Q)\cdot {\alpha_i^2(Q)
 m_i(\M)\over \alpha_{i\M}^2} = \alpha_i^3(Q) \cdot 
 {m_2\over \alpha_2^2}\ .
\eeq{eq:f5}
where $\alpha_{i\M}$ is the value of $\alpha_i$ at the messenger scale,
and the quantities at the extreme right are to be evaluated at the weak
interaction scale.  If we inserting this expression into \leqn{eq:d5}
and taking the evolution of $\alpha_i(Q)$ to be given by \leqn{eq:v1},
the right-hand side of \leqn{eq:d5} is given as an explicit function of
$Q$. To integrate the equation from messenger scale to the weak scale,
we only need to evaluate 
\beqa
 \int^\M_{\mz} d\,\log Q \, \alpha_i^3(Q) &=&
 \int^\M_{\mz} d\,\log Q  {\alpha^3_{i \M} \over (1 +
                       (b_i/2\pi)\alpha_{i\M}\log(Q/\M))^3}\CR
 &=&
{2\pi\over b_i\alpha_{i \M}}   {\alpha^3_{i \M} \over (1 +
                       (b_i/2\pi)\alpha_{i\M}\log(Q/\M))^2}\bigg|^\M_{\mz} \CR
&=& {2\pi\over b_i}(\alpha_i^2 - \alpha_{i \M}^2)
\eeqa{eq:g5}
Then, assembling the renormalization group and D-term contributions, 
the physical scalar mass at the weak interaction scale is given by 
\beq
M_f^2 = M_0^2 + \sum_i {2\over b_i}C_2(r_i)\,
 {\alpha_i^2 - \alpha_{i \M}^2\over \alpha_2^2}\, m_2^2 + \Delta M_f^2 \ .
\eeq{eq:h5}

The term in \leqn{eq:h5} induced by the renormalization group effect is
not simple, but it is also not so difficult to understand. It is
amusing that it is quite similar in form to the formula one would find
for a one-loop correction from a diagram of the general structure shown
in Figure~\ref{fig:elf}.  Indeed, in the model of Dine, Nelson, Nir,
and Shirman referred to above, for which the messenger scale is quite
close to the weak interaction scale, the computation of radiative
corrections gives the simple result 
\beq
 M_f^2 =  \sum_i 2 C_2(r_i)
 { \alpha_i^2\over \alpha_2^2} m_2^2 + \Delta M_f^2 \ ,
\eeq{eq:i5}
where, in this formula, the quantity $m_2/\alpha_2$ is simply the mass
scale of the messenger particles.  The formulae \leqn{eq:h5} and
\leqn{eq:i5} do differ quantitatively, as we will see in the next
section.

The equations \leqn{eq:w4} and \leqn{eq:h5} give a characteristic
evolution from the large scale $\M$ down to the weak interaction scale.
The colored particles are carried upward in mass by a large factor,
while the masses of color-singlet sleptons and gauginos change by a
smaller amount. The effects of the top Yukawa coupling discussed in
Section 3.7 add to these mass shifts, lowering the masses of the top
squarks and sending the (mass)$^{2}$ of the Higgs field $h_2$ down through
zero.  These observations explain all of the basic qualitative features of
the evolution which we saw illustrated
in Figure~\ref{fig:ten}.

\subsection{The spectrum of superparticles---diagnostics}

Now that we understand the various effects that can contribute to the
superpartner masses, we can try to analyze the inverse problem: Given a
set of masses observed experimentally, how can we read the pattern of
the underlying mass parameters and determine the value of the messenger
scale? In this section, I will present some general methods for
addressing this question.

This question of the form of the underlying soft-supersymmetry breaking
parameters requires careful thought.  If supersymmetric particles are
discovered at LEP 2 or LHC, this will become the most important
question in high-energy physics.  It is therefore important not to
trivialize this question or to address it only in overly restrictive
contexts. In reading the literature on supersymmetry experiments
at colliders, it is important to keep in mind the broadest range of
possibilities for the spectrum of superparticles.  Be especially
vigilant for code-words such as `the minimal SUGRA framework'
\cite{BSnow} or `the Monte Carlo generator described in [93]'
\cite{Atlas} which imply the restriction to the special case in which
$M_0$ is universal and $\M$ is close to the Planck mass.

Nevertheless, in this section, I will make some simplifying
assumptions. If the first supersymmetric partners are not found a LEP
2, the $D$-term contribution \leqn{eq:c5} is a small correction to the
mass formula. In any event,  I will ignore it from here on.  Since this
term is model-independent, it can in principle be computed and
subtracted if the value of $\tan\beta$ is known.  (It is actually not so
easy to measure $\tan\beta$; a collection of methods is given in
\cite{FMor}.)  In addition, I will ignore the effects of weak-scale
radiative corrections.  These are sometimes important and can distort
the overall pattern unless they are subtracted correctly
\cite{DamienII}.

I will also assume, in my description of the spectrum of scalars, that
the spectrum of gauginos is given in terms of $m_2$ by gaugino
unification. As I have explained in the previous section, gaugino
unification is a feature of the simplest schemes for generating the
soft supersymmetry-breaking masses both when $\M$ is very large and
when it is relatively small. However, there are many more complicated
possibilities. The assumption of gaugino unification can be tested
experimentally, as I will explain in Section~4.5.  This is an essential
part of any experimental investigation of the superparticle spectrum.
If the assumption is not valid, that also affects the interpretation of
the spectrum of scalar particles. In particular, the renormalization
effects included in the various curves shown in this section must be
recomputed using the correct mass relations among the three gauginos.

\begin{figure}[tb]
\begin{center}
\leavevmode
{\epsfysize=4.5in\epsfbox{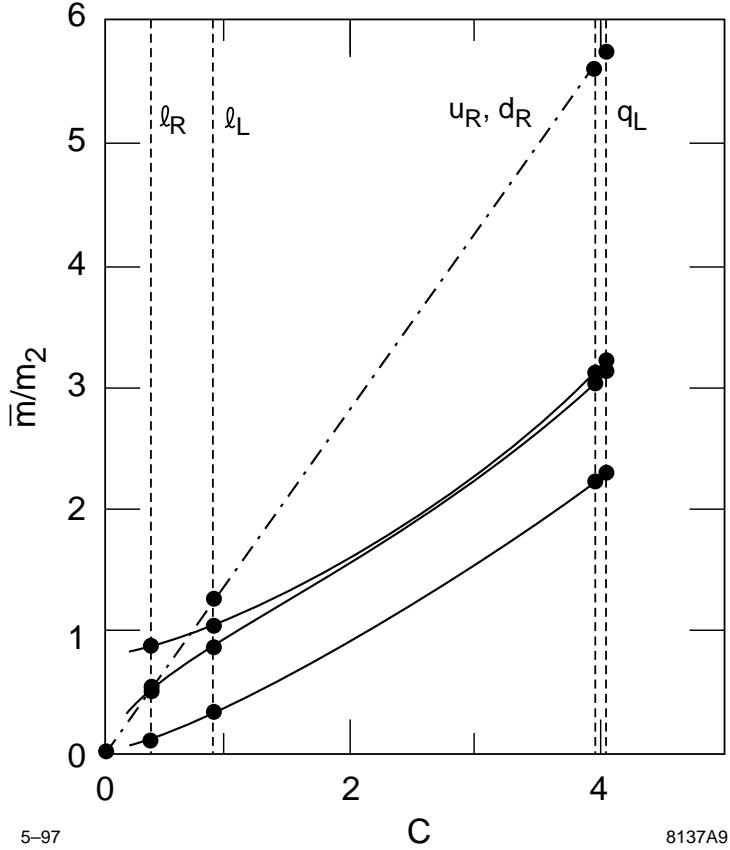}}
\end{center}
 \caption[*]{The simplest predictions for the mass spectrum of 
       squarks and sleptons, expressed on the Dine-Nelson plot 
          \protect\leqn{eq:j5}. The dot-dashed curve is the 
       prediction of \protect\leqn{eq:i5}; the solid curves show
       the effect of renormalization-group evolution with (from 
       bottom to top) $\M = 10^5$ GeV, $2\times 10^{16}$ GeV,
        $10^{18}$ GeV.}
\label{fig:twelve}
\end{figure}

Once the gaugino masses are determined, we can ask about the relation
between the mass spectrum of gauginos and that of scalars.  To analyze
this relation, it is useful to form the `Dine-Nelson plot', that is,
the plot of 
\beq
         {M_f\over m_2} \quad \mbox{against} \quad
 C \equiv \left[\sum_i C_2(r_i){\alpha_i^2\over \alpha_2^2} \right]^{1/2} \ ,
\eeq{eq:j5}
suggested by \leqn{eq:i5}.  Some sample curves on this plot are shown
in Figure~\ref{fig:twelve}.  The quantity $C$ takes on only five
distinct values, given by the $SU(3)\times SU(2)\times U(1)$ quantum
numbers of $\bar e$, $L$, $\bar d$, $\bar u$, and $Q$.  These are
indicated in the figure as vertical dashed lines. (The values of $C$
for $\bar d$ and $\bar u$ are almost identical.  The dot-dash line is
the prediction of \leqn{eq:i5}.  The solid lines are the predictions of
the renormalization group term in \leqn{eq:h5} for $\M = 100$ TeV,
$2\times 10^{16}$ GeV (the grand unification scale), and $10^{18}$ GeV
(the superstring scale).

%%%%%%%%%%%%%%%%%%%%%%%%%%%%%%%%%%%%%%%%%%%%%%%%%%%%%%%%%%%%%%%%%%%%%%
%%%%%%%%%%%%%%%%%%%%%%%%%%%%%%%%%%%%%%%%%%%%%%%%%%%%%%%%%%%%%%%%%%%%%%

\begin{figure}[p]
\begin{center}
\leavevmode
{\epsfysize=6.0in\epsfbox{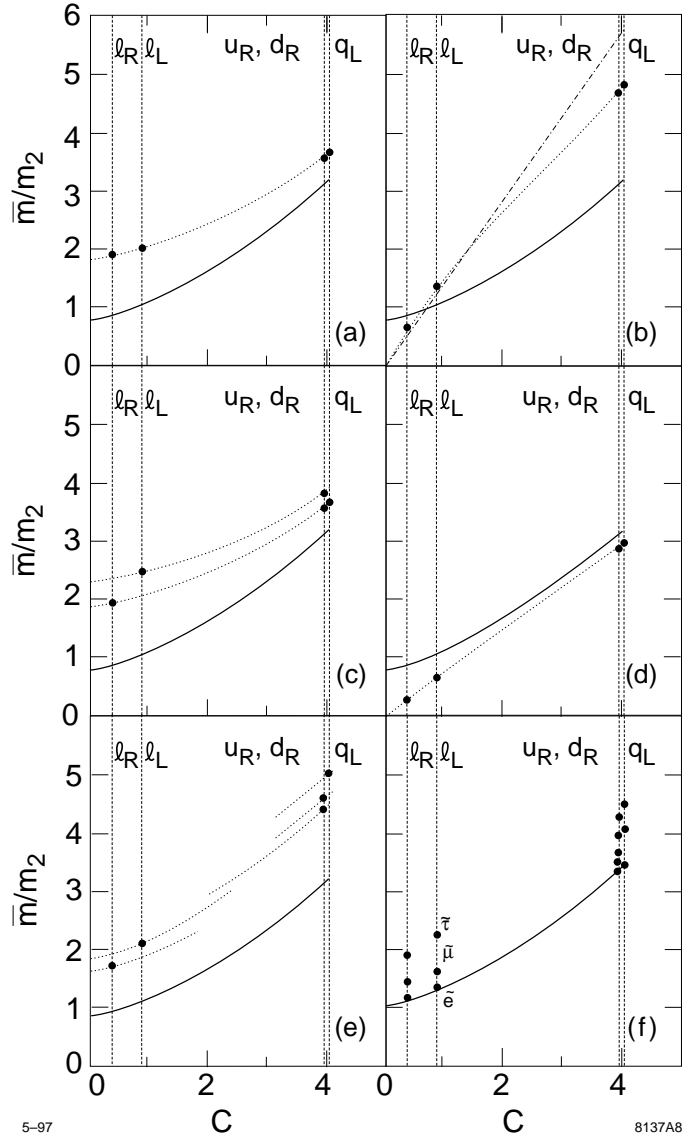}}
\end{center}
 \caption{Scalar spectrum predicted in a number of theoretical models of
  supersymmetry breaking, as displayed on the Dine-Nelson plot, from 
         \protect\cite{mysusy}.}
\label{fig:thirteen}
\end{figure}

With this orientation, it is interesting to ask how a variety of models
of supersymmetry breaking appear in this presentation. In
Figure~\ref{fig:thirteen}, I show the Dine-Nelson plot for a collection
of models from the literature discussed in \cite{mysusy}. The highest
solid curve from Figure~\ref{fig:twelve} has been retained for
reference. The model in the upper left-hand corner is the `minimal
SUGRA' model with a universal $M_0$ at the Planck scale.  In this case,
the dashed curve lies a constant distance  in $m^2$ above the solid
curve.  The model in the upper right-hand corner is that of \cite{DNNS}
with renormalization-group corrections properly included.  The model in
the bottom right-hand corner gives an example of the alignment scenario
of \cite{NLS}.  The plot is drawn in such a way as to suggest that, the
underlying soft scalar masses tend to zero for the first generation of
quarks and leptons.  This behavior could be discovered experimentally
with the analysis I have suggested here.

%%%%%%%%%%%%%%%%%%%%%%%%%%%%%%%%%%%%%%%%%%%%%%%%%%%%%%%%%%%%%%%%%%%%%%
%%%%%%%%%%%%%%%%%%%%%%%%%%%%%%%%%%%%%%%%%%%%%%%%%%%%%%%%%%%%%%%%%%%%%%

It is interesting that the various models collected in
Figure~\ref{fig:thirteen} look quite different to the eye in this
presentation. This fact gives me confidence that, if we could actually
measure the mass parameters needed for this analysis, those data would
provide us with incisive information on the physics of the very large
scales of unification and supersymmetry breaking.

\subsection{The superpartners of $W$ and Higgs}

Now that we have framed the problem of measuring the mass spectrum of
superparticles, we must address the question of how this can be done.
What are the signatures of the presence of supersymmetric particles,
and how can we translate from the characteristics of observable
processes to the values of the parameters of which determine the
supersymmetry spectrum?

I will discuss the signatures and decay schemes for superparticles in
the next section.  First, though, we must discuss a complication which
needs to be taken into account in this phenomenology.

After $SU(2)\times U(1)$ symmetry-breaking, any two particles with the
same color, charge, and spin can mix.  Thus, the
spin-$\half$ supersymmetric partners of the $W$ bosons and the charged
Higgs bosons can mix with one another.  Similarly, the partners of the
$\gamma$, $Z^0$, $h_1^0$, and $h^0_2$ enter into a $4\times 4$ mixing
problem.

Consider first the mixing problem of the charged fermions. The mass
terms for these fermions arise from the gaugino-Higgs coupling in
\leqn{eq:h4}, the soft gaugino mass term in \leqn{eq:o4}, and the
fermion mass term arising from the superpotential \leqn{eq:l4}. The
relevant terms from the Lagrangian are 
\beqa
 \Delta\L &=& - \sqrt{2}i{g\over 2}\left(
 h_2^0 \widetilde w^{-T} c \widetilde h^+_2 
- \widetilde h^{-T}_1 c \widetilde w^+ h^0_1 \right) \CR
 & & \hskip 0.4in
- m_2 \widetilde w^{-T} c \widetilde w^+ +\mu \widetilde h_1^{-T} c
  \widetilde h_2^+\ .
\eeqa{eq:k5}
If we replace $h^0_1$ and $h^0_2$ by their vacuum expectation values
in \leqn{eq:a5}, these terms take the form 
\beq
\Delta\L = - \pmatrix{ \widetilde w^- & i\widetilde h^-_1\cr}^T c\, 
\mbox{\bf m}
                  \pmatrix{ \widetilde w^+ \cr i \widetilde h^+_2\cr}\ ,
\eeq{eq:l5}
where ${\bf m}$ is the mass matrix
\beq
          \mbox{\bf m} =  \pmatrix{ m_2 &   \sqrt{2}\mw \sin\beta \cr
                    \sqrt{2}\mw \cos\beta & \mu\cr }
\eeq{eq:m5}
The physical massive fermions are the eigenstates of this mass matrix.
They are called {\em charginos}, $\widetilde \chi^\pm_{1,2}$, where 1
labels the lighter state.   More precisely, the charginos
$\widetilde\chi_1^+$,  $\widetilde\chi_2^+$ are the linear combinations
that diagonalize the matrix $\mbox{\bf m}^\dagger \mbox{\bf m}$, and
$\widetilde\chi_1^-$,  $\widetilde\chi_2^-$ are the linear combinations
that diagonalize the matrix $\mbox{\bf m}\mbox{\bf m}^\dagger$.

%%%%%%%%%%%%%%%%%%%%%%%%%%%%%%%%%%%%%%%%%%%%%%%%%%%%%%%%%%%%%%%%%%%%%%
\begin{figure}[htb]
\begin{center}
\leavevmode
{\epsfxsize=3.5in\epsfbox{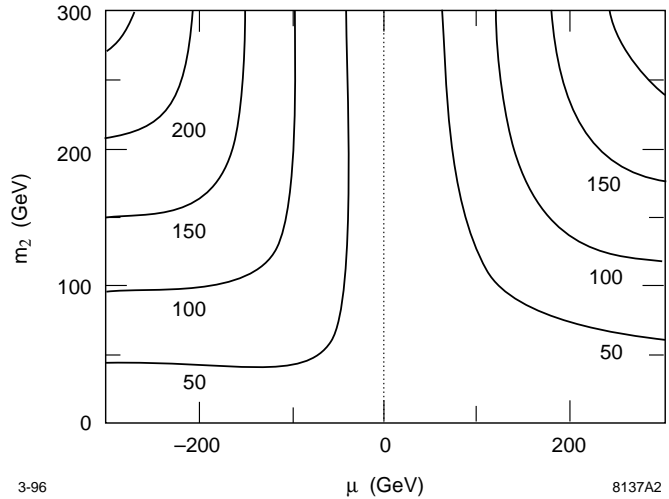}}
\end{center}
 \caption{Contours of fixed chargino mass in the plane of the 
      mass parameters $(\mu,m_2)$, computed for $\tan\beta = 4$.}
\label{fig:fourteen}
\end{figure}
%%%%%%%%%%%%%%%%%%%%%%%%%%%%%%%%%%%%%%%%%%%%%%%%%%%%%%%%%%%%%%%%%%%%%%

The diagonalization of the matrix \leqn{eq:m5} is especially simple in
the limit in which the supersymmetry parameters $m_2$ and $\mu$ are
large compared to $\mw$.  In the region $\mu > m_2 \gg \mw$,
$\widetilde\chi^+_1$ is approximately $\widetilde w^+$, with mass $m_1
\approx m_2$, while $\widetilde\chi^+_2$ is approximately $\widetilde
h_2^+$, with mass $m_2 \approx \mu$. For  $m_2  > \mu \gg \mw$, the
content of $\widetilde\chi^+_1$ and $\widetilde\chi^+_2$ reverses. 
More generally, we refer to the region of parameters in which
$\widetilde\chi^+_1$ is mainly $\widetilde w^+$ as the {\em gaugino
region}, and that in which $\widetilde\chi^+_1$ is mainly $\widetilde
h_2^+$ as the {\em Higgsino region}.  If charginos are found are LEP 2, it
is quite likely that they may be mixtures of gaugino and Higgsino;
however, the region of parameters in which the charginos are
substantially mixed decreases as the mass increases.  The contours of
constant   $\widetilde\chi^+_1$ mass in the $(\mu, m_2)$ plane, for
$\tan\beta =4$ are shown in Figure~\ref{fig:fourteen}.

An analysis similar to that leading to  \leqn{eq:m5} gives the mass
matrix of the neutral fermionic partners.  This is a $4\times 4$ matrix
acting on the vector $(\widetilde b, \widetilde w^3, i\widetilde
h_1^0, i\widetilde h_2^0)$, where $\widetilde b$ and $\widetilde w^3$
are the partners of the $U(1)$ and the neutral $SU(2)$ gauge boson.  In
this basis, the mass matrix takes the form 
\beq
 \mbox{\bf m} = \pmatrix{m_1 & 0 &  - \mz s \cos\beta & \mz s\sin\beta \cr
             0 & m_2 &  \mz c \cos\beta & - \mz c\sin\beta \cr
             - \mz s \cos\beta & \mz c\cos\beta & 0 & -\mu\cr
             \mz s \sin\beta & - \mz c\sin\beta &  -\mu & 0 \cr} \ .
\eeq{eq:n5}
The linear combinations which diagonalize this matrix are called
{\em neutralinos}, $\widetilde\chi^0_1$ through  $\widetilde\chi^0_4$ from
lowest to highest mass.  The properties of these states are similar to
those of the charginos.  For example, in the gaugino region,
$\widetilde \chi^0_1$ is mainly $\widetilde b$ with mass $m_1$, and
$\widetilde \chi^0_2$ is mainly $\widetilde w^3$, with mass $m_2$.

Note that, when $\mu =0$, the neutralino mass matrix \leqn{eq:n5} has
an eigenvector with zero eigenvalue $(0,0,\sin\beta,\cos\beta)$.  In
addition, the vector $(0,0, \cos\beta, -\sin\beta)$ has a relatively
small mass $m_\chi \sim \mz^2/m_2$.  This situation is excluded by the
supersymmetry searches at LEP 1, for example,  \cite{LEPsusy}. 
 Thus, we are required
on phenomenological grounds to include the superpotential \leqn{eq:l4}
with a nonzero value of $\mu$.  It is also important to note that, with
the `minimal SUGRA' assumptions used in many phenomenological studies,
it is easiest to arrange electroweak symmetry breaking through the
renormalization group mechanism discussed in Section 3.7 if $\mu$
is of order
$m_3 \approx  3.5 m_2$. Thus, this set of assumptions typically leads to
the gaugino region of the chargino-neutralino physics.
 
\subsection{Decay schemes of superpartners}

With this information about the mass eigenstates of the superpartners,
we can work out their decay schemes and, from this, their signatures.
As I have explained at the end of Section 3.5, I restrict this
discussion to the situation in which $R$-parity, given by \leqn{eq:m4},
is conserved and so the lightest supersymmetric partner is stable.  In
most of this discussion, I will assume that this stable particle is the
lightest neutralino $\widetilde\chi^0_1$.  The neutralino is a massive
but weakly-interacting particle. It  would  not be  observed directly
in a detector at a high-energy collider but rather would appear as
missing energy and unbalanced momentum.

In this context, we can discuss the decays of specific superpartners.
Clearly, the lighter superpartners will have the simplest decays, while
the heavier superpartners will decay to the lighter ones.  Since heavy
squarks and sleptons often decay to charginos and neutralinos, it is
convenient to begin with these.

\begin{figure}[t]
\begin{center}
\leavevmode
{\epsfxsize=3.75in\epsfbox{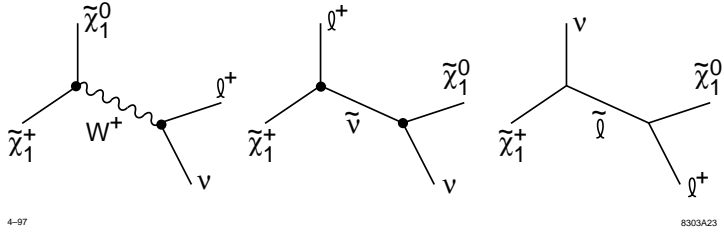}}
\end{center}
 \caption{Diagrams leading to the decay of the chargino $\ch{1}$ to the
3-body final state $\ell^+\nu \ne{1}$. The chargino can decay to $u\bar
d \ne{1}$ by similar processes.}
\label{fig:fifteen}
\end{figure}

The decay pattern of the lighter chargino depends on its field content
and, in particular, on whether its parameters lie in the gaugino region
or the Higgsino  region.  In the gaugino region, the lighter chargino
is mainly $\widetilde w^+$, with mass $m_2$.  The second neutralino is
almost degenerate, but the first neutralino has mass $m_1 = 0.5 m_2$,
assuming gaugino unification.  If $m_2 > 2 m_W$, the decay $\ch{1}\to
W^+ \ne{1}$ typically dominates. If $m_2$ is smaller, the chargino
decays to 3-body final states through the diagrams shown in
Figure~\ref{fig:fifteen}, and through the analogous diagrams involving
quarks.  The last two diagrams involve virtual sleptons.  If the
slepton mass is large, the branching ratio to quarks versus leptons is
the usual color factor of 3.  However, if the sleptons are light, the
branching ratio to leptons may be enhanced.

In the Higgsino region, the chargino $\ch{1}$ and the two lightest
neutralinos $\chi{1}$, $\chi{2}$ are all roughly degenerate at the mass
$\mu$.  The first diagram in Figure~\ref{fig:fifteen} dominates in this
case, but leads to only a small visible energy in the $\ell^+ \nu$ or
$u\bar d$ system.

%%%%%%%%%%%%%%%%%%%%%%%%%%%%%%%%%%%%%%%%%%%%%%%%%%%%%%%%%%%%%%%%%%%%%%
%%%%%%%%%%%%%%%%%%%%%%%%%%%%%%%%%%%%%%%%%%%%%%%%%%%%%%%%%%%%%%%%%%%%%%

The decay schemes of the second neutralino $\ne{2}$ are similar to
those of the chargino.  Since supersymmetry models typically have a
light neutral Higgs boson $h^0$, the decay $\ne{2} \to \ne{1} h^0$ may
be important.  If neither this process nor the on-shell decay to $Z^0$
are allowed, the most important decays are the 3-body processes such as
$\ne{2} \to \ne{1} q \bar q$.  The process $\ne{2} \to \ne{1}
\ell^+\ell^-$ is particularly important at hadron colliders, as we will
see in Section~4.8.

Among the squarks and sleptons, we see from Figure~\ref{fig:thirteen}
that the $\widetilde e_R^-$ of each generation is typically the
lightest. This particle couples to $U(1)$ but not $SU(2)$ and so, in
the gaugino region, it decays through  $\widetilde e^-_R \to e
\ne{1}$.  On the other hand, the partners $\tilde L$ of the left-handed
leptons prefer to decay to $\ell \ne{2}$ or $\nu \ch{1}$ if these modes
are open.

It is a typical situation that the squarks are heavier than the gluino.
For example, the renormalization group term in \leqn{eq:h5}, with $\M$
of the order of the unification scale, already  gives a contribution
equal to $3m_2$.  In that case, the squarks decay to the gluino,
$\widetilde q \to q \widetilde g$.  If the gluinos are heavier, then,
in the gaugino region, the superpartners of the right-handed quarks
decay dominantly to $q\ne{1}$, while the partners of the left-handed
quarks prefer to decay to $q \ne{2}$ or $q \ch{1}$.

%%%%%%%%%%%%%%%%%%%%%%%%%%%%%%%%%%%%%%%%%%%%%%%%%%%%%%%%%%%%%%%%%%%%%%
\begin{figure}[t]
\begin{center}
\leavevmode
\epsfbox{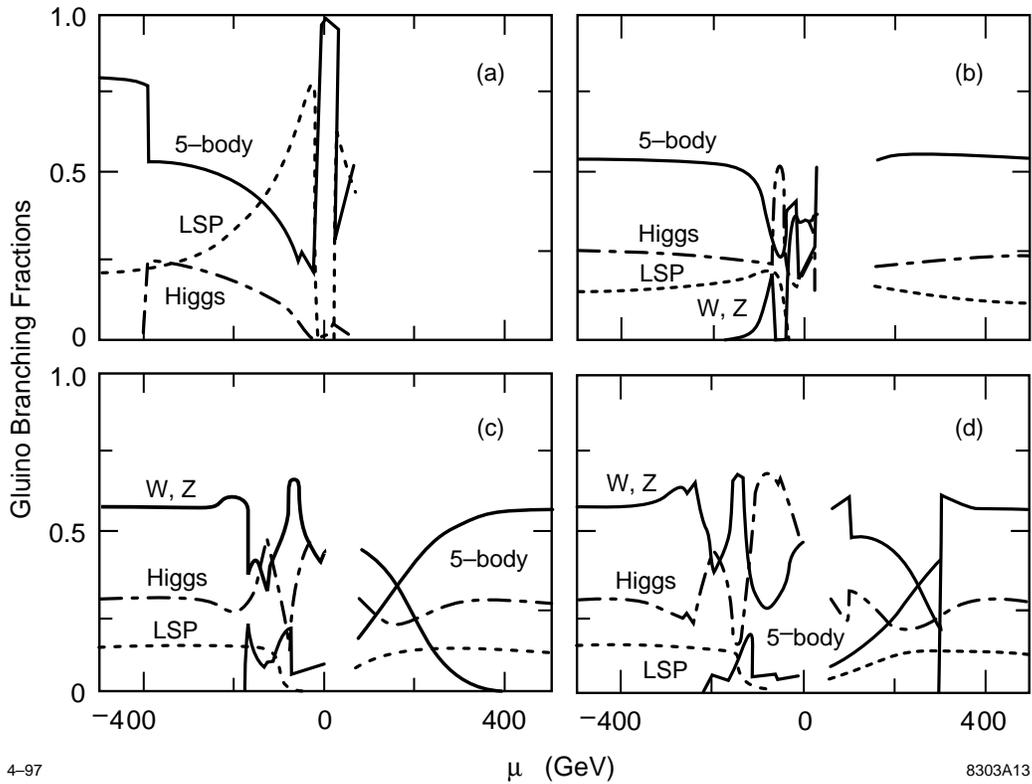}
\end{center}
 \caption{Branching fractions for gluino decay in the various classes 
      of final states possible for $m(\widetilde g) < m(\widetilde q)$,
        from \protect\cite{BGH}.
        The four graphs correspond to the gluino masses (a) 120 GeV,
      (b) 300 GeV, (c) 700 GeV, (d) 1000 GeV. The branching fractions
are given as a function of $\mu$ with $m_2$ determined from the gluino
mass by the gaugino unification relation (133).}
\label{fig:sixteen}
\end{figure}
%%%%%%%%%%%%%%%%%%%%%%%%%%%%%%%%%%%%%%%%%%%%%%%%%%%%%%%%%%%%%%%%%%%%%%

If the squarks and gluinos are much heavier than the color-singlet
superpartners, their decays can be quite complex, including cascades
through heavy charginos, neutralinos, and Higgs bosons \cite{BBTW,BGH,BTW}.
Figure~\ref{fig:sixteen} shows the branching fractions of the gluino as
a function of $\mu$, assuming gaugino unification and the condition
that the squarks are heavier than the gluino.  The boundaries apparent
in the figure correspond to the transition from the gaugino region (at
large $|\mu|$) to the Higgsino region.  The more complex decays
indicated in the figure can be an advantage in hadron collider
experiments, because they lead to characteristic signatures such as
multi-leptons or direct $Z^0$ production in association with missing
transverse momentum.  On the other hand, as the dominant gluino decay
patterns become more complex, the observed inclusive cross sections
depend more indirectly on 
the underlying supersymmetry parameters.

Up to now, I have been assuming that the lightest superpartner is the
$\ne{1}$.  However, there is an alternative possibility that is quite
interesting to consider.  According to Goldstone's theorem, when a
continuous symmetry is spontaneously broken, a massless particle
appears as a result.  In the most familiar examples, the continuous
symmetry transforms the internal quantum numbers of fields, and the
massless particle is a Goldstone boson.  If the spontaneously broken
symmetry is coupled to a gauge boson, the Goldstone boson combines with
the gauge boson to form a massive vector boson; this is the Higgs
mechanism.  Goldstone's theorem also applies to the spontaneous
breaking of supersymmetry, but in this case the massless particle is a
Goldstone fermion or {\em Goldstino}.  Then it would seem that the 
Goldstino should be the
lightest superpartner into which all other superparticles decay?

To analyze this question, we need to know two results from the theory of
the Goldstino.  Both have analogues in the usual theory of Goldstone
bosons. I have already pointed out in \leqn{eq:n3} that the gravitino,
the spin-$\thalf$ supersymmetric partner of the graviton, acts as the
gauge field of local supersymmetry.  This particle can participate in a
supersymmetric version of the Higgs mechanism. If supersymmetry is
spontaneously broken by the expectation value of an $F$ term, the
gravitino and the Goldstino combine to form a massive spin-$\thalf$
particle with mass 
\beq
            m_\psi = {\VEV{F}\over \sqrt{3} m_\Pl}\ ,
\eeq{eq:o5}
where $m_\Pl$ is the Planck mass.  Notice that, if the messenger scale
$\M$ is of the order of $m_\Pl$, this mass scale is of the order of the
scale $m_S$ of soft  supersymmetry-breaking mass terms given in
\leqn{eq:p4}.  In fact, in this case, the massive gravitino is
typically heavier than the $\ne{1}$.  On the other hand, if $\M$ is of
order 100 TeV, with $\VEV{F}$ such that the superparticle masses are at
the weak interaction scale, $m_\psi$ is of order $10^{-2}$ eV and so is
much lighter than any of the superpartners we have discussed above.

The second result bears on the probability for producing Goldstinos.
The methods used to analyze pion physics in QCD generalize to this case
and predict that the Goldstino $\widetilde G$ is produced through the
effective Lagrangian 
\beq
  \Delta\L =   {1\over \VEV{F}} j^T_\mu c \del^\mu \widetilde G \, 
\eeq{eq:p5}
where $\VEV{F}$ is the supersymmetry-breaking vacuum expectation value
in \leqn{eq:o5} and $j_\mu$ is the conserved current associated with
supersymmetry. Integrating by parts, this gives a coupling for the vertex
$\widetilde f \to   f \widetilde G$ proportional to 
\beq
                   {\Delta m\over \VEV{F}} \ ,
\eeq{eq:q5}
where $\Delta m$ is the supersymmetry-breaking mass difference between
$f$ and $\widetilde f$.  If the Goldstino becomes incorporated into a
massive spin-$\thalf$ field, this does not affect the production
amplitude, as long as the Goldstinos are emitted at energies large
compared to their mass.  I will discuss this point for the more 
standard case of a Goldstone boson in Section 5.3.
  This result tells us that, if the messenger scale $\M$ is of
order $m_\Pl$ and $\VEV{F}$ is connected with $\M$ through \leqn{eq:p4},
the rate for the decay of any superpartner to the Goldstino is so slow
that it is irrelevant in accelerator experiments.  On the other hand,
if $\M$ is less than 100 TeV, decays to the Goldstino can become
relevant.

For the case of the coupling of the  $\widetilde b$, the superpartner
of the $U(1)$ gauge boson, to the photon and $Z^0$ fields, the
effective Lagrangian \leqn{eq:p5} takes the more explicit form 
\beq
  \Delta\L =   {m_1\over \VEV{F}} \widetilde b^\dagger \sigma^{\mu\nu}
(c F_{\mu\nu} - s Z_{\mu\nu}) \widetilde G  \ . 
\eeq{eq:r5}
This interaction leads to the decay $ \widetilde b \to  \gamma
\widetilde G$ with lifetime \cite{DDTR} 
\beq
   c\tau = (\mbox{0.1 mm})\left({\mbox{100\ GeV}\over m_1 }\right)^5
 \left({\VEV{F}^{1/2}\over \mbox{100\ TeV}}\right)^4 \ .
\eeq{eq:s5}
It is difficult to estimate whether the value of $c\tau$ resulting from
\leqn{eq:s5} should be meters or microns.  But this argument does
predict that, if the $\ne{1}$ is the lightest superpartner of Standard
Model particles, all decay chains should end with the decay of the
$\ne{1}$ to $\gamma \widetilde G$.  If the lifetime \leqn{eq:s5} is
short, each $\ne{1}$ momentum vector, which we visualized above as
missing energy, should be realized instead as missing energy plus a
direct photon.

It is also  possible in this case of small $\VEV{F}$ that the lightest
sleptons $\widetilde e^-_R$ could be lighter than the $\ne{1}$.  If
these particles are the lightest superparticles, they lead to an
unacceptable cosmological abundance of stable charged matter.  This
problem disappears, however, if they can decay to the Goldstino. In
that case, all supersymmetric decay chains terminate with leptons and
missing energy, for example, 
\beq
   \ne{1} \to \ell^- \widetilde\ell^+_R \to \ell^- \ell^+ \widetilde G \ .
\eeq{eq:t5}

From here on, I will concentrate on the most straightforward case in
which the  $\ne{1}$ is the lightest superparticle and is stable over
the time scales observable in collider experiments.  However, it is
important to keep these alternative phenomenologies in mind when you
are actually looking for superparticle signatures in the data.

\subsection{The mass scale of supersymmetry}

At last, we have all the background we require to discuss the
experiments which will detect and study supersymmetric particles at
colliders.  In this section, I would like to recapitulate the general
ideas that we have formulated for this study.  I will also note the 
implication of the these idea for the mass range of supersymmetric 
particles.  If the picture of supersymmetry that I have constructed here is
correct, the supersymmetric particles should be discovered at planned,
or even at the present, accelerators.

Although the mass scale of supersymmetry depends on many
parameters and is in principle adjustible over a large range, there is
a good reason to expect to find supersymmetric particles relatively
near at hand.  As I have discussed in Section 3.7, supersymmetry
provides a mechanism for electroweak symmetry breaking.  If we assume
that this indeed is the mechanism of supersymmetry breaking, the $W$
and $Z$ masses must be masses characteristic of the scale of soft
supersymmetry-breaking parameters. Alternatively,  $\mw$ can only be 
much less than $m_S$ in \leqn{eq:p4} by virtue of an unnatural
cancellation or fine-tuning of parameters. This possibility has been
studied quantitatively in a number of theoretical papers
\cite{ENT,BG,ACas}, with the conclusion that  the relation between
$\mw$ and $m_S$ is natural (by the authors' definitions) only when 
\beq
m_2 < 3 \mw \ . 
\eeq{eq:tt5} 
Of course, it is possible that the mechanism of electroweak symmetry
breaking does not involve supersymmetry.  In that case, there might
still be supersymmetry at a very high scale (to satisfy aesthetic
arguments or to aid in the quantization of gravity), but in this case
supersymmetry would not be relevant to experimental high-energy physics.

The schemes for the supersymmetric mass spectrum
discussed in Sections 4.2 and 4.3 give a definite expectation for the
ordering of states.  The gaugino unification relation predicts that
the gluino is the heaviest of the gauginos, with the on-shell gluino
mass satisfying 
\beq
            m(\widetilde g) \sim 4 m_2 \ .
\eeq{eq:u5}
Our results were much less definitive about the mass relations of the 
squarks and sleptons.  Roughly, though,
\beq
             m(\widetilde q) \sim (2-6) \cdot m(\widetilde \ell)\ , \quad
 \mbox{and} \quad
          m(\widetilde\ell) \sim
m_2   \ ,
\eeq{eq:v5}
in the models discussed in Section 4.3.

The relations  \leqn{eq:tt5}--\leqn{eq:v5} predict that we should 
find charginos below 250 GeV in mass and gluinos below 1 TeV.  This 
mass region is not very far away.
 The LEP 2 and Tevatron experimental programs will cover
almost half of this parameter space in the next five years.  The LHC
can probe for supersymmetric particles up to masses about a factor 3
beyond the region predicted by the relations above, and an  $\ee$ linear
collider with up to 1.5 TeV in the center of mass would have a roughly
equivalent reach.

Search strategies for supersymmetric particles depend on the 
detailed properties of the model.  But in general,
assuming $R$-parity conservation and the identification
of $\ne{1}$ as the lightest superparticle, the basic signature of
supersymmetry is new particle production associated with missing
energy.  In collider experiments, we would typically be looking for a
multi-jet or multi-lepton final state, together with the characteristic
missing transverse momentum or acoplanarity.

Because I would like to continue in a somewhat different direction, I
will not describe in detail the techniques and strategies for the
discovery of supersymmetry at these colliders.  The search strategies
for various supersymmetric particles at LEP 2 are described in
\cite{LEPtwo}.  Experimental strategies for discovering supersymmetry
at the Tevatron are reviewed in \cite{Tevatron}, together with an
estimation of the reach in the mass spectrum.

%%%%%%%%%%%%%%%%%%%%%%%%%%%%%%%%%%%%%%%%%%%%%%%%%%%%%%%%%%%%%%%%%%%%%%
\begin{figure}[p]
\begin{center}
\leavevmode
{\epsfxsize=3.5in\epsfbox{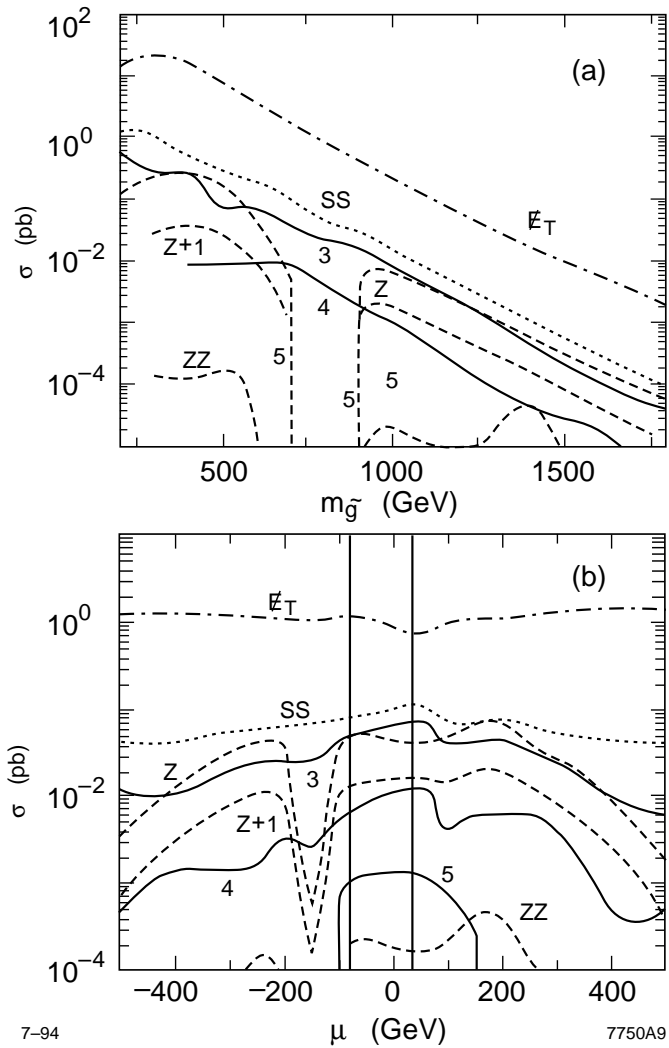}}
\end{center}
 \caption{Cross sections for various signatures of supersymmetric particle
 production at the LHC, from from \protect\cite{BTW}.  The observables
 studied are, from top to bottom, missing $E_T$, like-sign dileptons, 
  multi-leptons, and $Z +$ leptons.  The top graph plots the cross sections
as a function of $m(\widetilde g)$ for $m(\widetilde q) = 2 m(\widetilde g)$,
and $\mu = -150$ GeV, and $m_2$ given by gaugino unification.  The bottom
graph, plotted for $m(\widetilde g) = 750$ GeV as a function of $\mu$, 
shows the model-dependence of the cross sections.}
\label{fig:seventeenxx}
\end{figure}
%%%%%%%%%%%%%%%%%%%%%%%%%%%%%%%%%%%%%%%%%%%%%%%%%%%%%%%%%%%%%%%%%%%%%%

\begin{figure}[tb]
\begin{center}
\leavevmode
{\epsfxsize=4in\epsfysize=4.5in\epsfbox{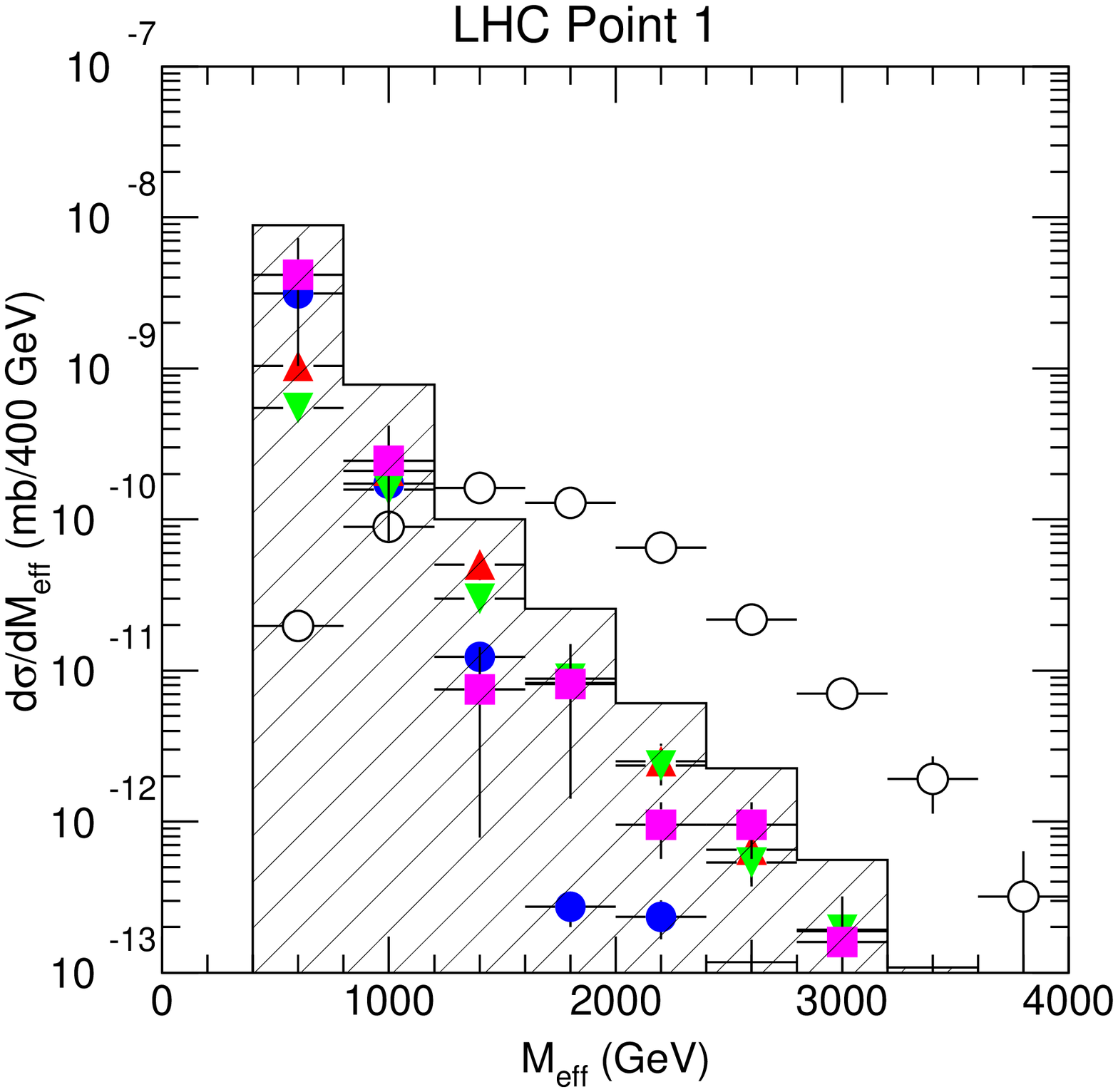}}
\end{center}
 \caption{Simulation of the observation of supersymmetric particle 
production at the LHC, from \protect\cite{Ianscrew}, at a point in 
parameter space with $m(\widetilde g) = 1$ TeV. The observable 
$M_{eff}$ is given by the sum of the missing $E_T$ and the sum of the
$E_T$ values for the four hardest jets.  The supersymmetry signal is 
shown as the open circles.  Among the backgrounds, the squares are
due to QCD processes, and the other points shown are due to $W$,
$Z$, and $t$ production.}
\label{fig:seventeen}
\end{figure}

It is important to point out, though, that if the phenomenology of
supersymmetry follows the general lines I have laid out here, it will
be discovered, at the latest, by the LHC.  The cross sections for LHC
signatures of supersymmetry involving multiple leptons and direct $Z^0$
production associated with missing transverse energy are shown in
Figure~\ref{fig:seventeenxx} \cite{BTW}. These cross sections are very
large, of order 100 fb, for example, for the like-sign dilepton signal,
at a collider that is designed to produce an event sample of 100
fb$^{-1}$ per year per detector.  Supersymmetry can also be seen by
looking for events with large jet activity and missing transverse
momentum.  A sample comparison of signal and background for an
observable that measures the jet activity is shown in
Figure~\ref{fig:seventeen}~\cite{Ianscrew}. The authors of this
analysis conclude that, at the LHC, the major backgrounds to
supersymmetry reactions do not come from Standard Model background
processes but rather from other 
supersymmetry reactions.

%%%%%%%%%%%%%%%%%%%%%%%%%%%%%%%%%%%%%%%%%%%%%%%%%%%%%%%%%%%%%%%%%%%%%%
%%%%%%%%%%%%%%%%%%%%%%%%%%%%%%%%%%%%%%%%%%%%%%%%%%%%%%%%%%%%%%%%%%%%%%

That prospect is enticing, but it is only the beginning of an experimental
research program on supersymmetry.  We have seen that the theory of the 
supersymmetry spectrum is complex
 and subtle.  The investigation of supersymmetry
should allow us to measure this spectrum.  That in turn will give us access
to the soft supersymmetry-breaking parameters, which are generated at
very short distances and which therefore should hold information about
the very deep levels of fundamental physics. So it is important to 
investigate to what extent these experimental measurements
are actually feasible using accelerators that we can foresee.

In discussing this question, I will assume, pessimistically, that the 
scale of supersymmetry is relatively high, and so I will concentrate on 
experiments for the high-energy colliders of the next generation, the LHC
and the $\ee$ linear collider discussed in the introduction.  
As a byproduct, this approach will illustrate the deep analytic 
power that both of these machines can bring to bear on new 
physical phenomena.

\subsection{Superspectroscopy at $\ee$ colliders}

I will start this discussion of supersymmetry measurements from the
side of $\ee$ colliders.  It is intuitively clear that, if we had an
$\ee$ collider operating in the energy region appropriate to
supersymmetric particle production, some precision measurements could
be made.  But I have stressed that the soft supersymmetry-breaking
Lagrangian can contain a very large number of parameters which become
intertwined in the mass spectrum.  Thus, it is important to ask, is
there a set of measurements which extracts and disentangles
these parameters? I will explain now how to do that.

I do not wish to imply, with this approach, that precision supersymmetry
measurements are possible only at $\ee$ colliders.  In fact, the next
section will be devoted to precision information that can be obtained 
from hadron collider experiments.  And, indeed, to justify the 
construction of an $\ee$ linear collider, it is necessary to show that 
the $\ee$ machine adds significantly to the results that will be
available from the LHC.  Nevertheless, it has pedagogical virtue to begin 
from 
the $\ee$ side, because the $\ee$ experiments allow a completely 
systematic approach to the issues of parameter determination.  I will
return to the question of comparing $\ee$ and $pp$ colliders in 
Section 4.9.

To begin, let me review some of the parameters of future $\ee$ colliders.
Cross sections for $\ee$ annihilation decreases with
the center-of-mass energy as $1/E_\CM^2$. Thus, to be effective, a
future collider must provide a data sample of 20-50 fb$^{-1}$/year at
an center of mass energy of 500 GeV, and a data sample increasing from
this value as $E_\CM^2$ at higher energies.  The necessary luminosities
are envisioned in the machine designs \cite{loew}.  Though new sources
of machine-related background appear, the experimental environment is
anticipated to be similar to that of LEP \cite{nlcbook}. An important
feature of the experimental arrangement not available at LEP is an
expected 80--90\% polarization of the electron beam.  We will see in a moment
that this 
polarization provides  a powerful physics analysis tool.

The simplest supersymmetry analyses at $\ee$ colliders involve $\ee$
annihilation to slepton pairs.  Let $\widetilde \mu_R$ denote the
second-generation $\widetilde e^-_R$.  This particle has a simple
decay $\widetilde \mu_R \to \mu \ne{1}$, so  pair-production of
$\widetilde \mu_R$ results in  a final state with $\mu^+\mu^-$ plus
missing energy. The production process is simple $s$-channel
annihilation through a vitual $\gamma$ and $Z^0$; thus, the cross
section and polarization asymmetry are characteristic of the standard
model quantum numbers of the $\widetilde\mu_R$ and are independent of the
soft supersymmetry-breaking parameters.

%%%%%%%%%%%%%%%%%%%%%%%%%%%%%%%%%%%%%%%%%%%%%%%%%%%%%%%%%%%%%%%%%%%%%%
\begin{figure}[t]
\begin{center}
\leavevmode
{\epsfxsize=2.5in\epsfbox{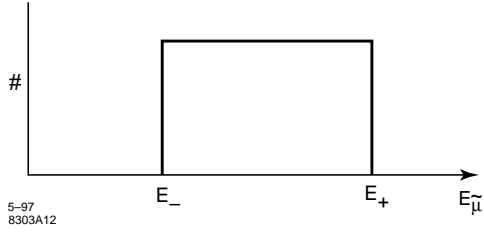}}
\end{center}
 \caption{Schematic  energy distribution in a slepton or squark decay,
   allowing  a precision supersymmetry mass measurement at an
        $\ee$ collider.}
\label{fig:eighteenvv}
\end{figure}
%%%%%%%%%%%%%%%%%%%%%%%%%%%%%%%%%%%%%%%%%%%%%%%%%%%%%%%%%%%%%%%%%%%%%%

It is straightforward to measure the mass of the $\widetilde \mu_R$, and
the method of this analysis can be applied to many other examples. 
Because the $\widetilde \mu_R$ is a scalar, it decays isotropically to
its two decay products. When we transform to the lab frame, the
distribution of $\mu$ energies is flat between the kinematic endpoints,
as indicated in Figure~\ref{fig:eighteenvv}. The endpoints occur at 
\beq
          E_\pm = (1 \pm \beta) \gamma \E \ ,
\eeq{eq:w5}
with $\beta = (1 - 4 m(\widetilde{\mu})^2/E_\CM^2)^{1/2}$, $\gamma = E_\CM/
   2 m(\widetilde{\mu})$, and 
\beq
   \E = {m(\widetilde{\mu})^2 - m(\ne{1})^2\over 2 m(\widetilde{\mu}^2) }
\ .
\eeq{eq:x5}
Given the measured values of $E_\pm$, one can solve algebraically for
the mass of the parent $\widetilde\mu_R$ and the mass of the missing
particle $\ne{1}$.  Since many particles have two-body decays to the
$\ne{1}$, this mass can be determined redundantly.  For heavy
supersymmetric particles, the lower endpoint may sometimes  be obscured
by background from cascade decays through heavier charginos and
neutralinos.  So it is also interesting to note that, once the mass of
the $\ne{1}$ is known, the mass of the parent particle can be
determined from the measurement of the higher endpoint only.

%%%%%%%%%%%%%%%%%%%%%%%%%%%%%%%%%%%%%%%%%%%%%%%%%%%%%%%%%%%%%%%%%%%%%%
\begin{figure}[tb]
\begin{center}
\leavevmode
{\epsfxsize=5in\epsfbox{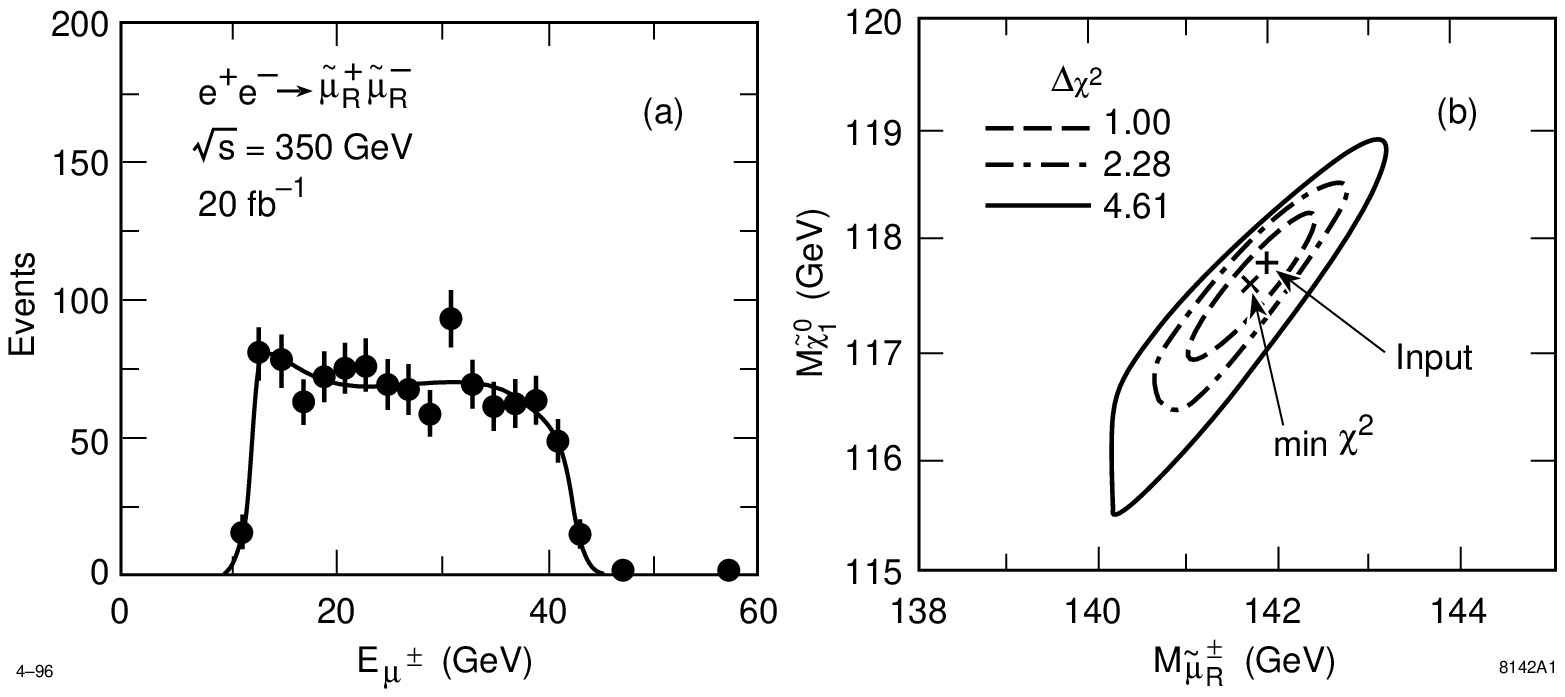}}
\end{center}
 \caption{Simulation of the $\widetilde\mu_R$ mass measurement at an
       $\ee$ linear collider, from \protect\cite{Tsuk}.  The left-hand
      graph gives the event distribution in the decay muon energy.
   The right-hand graph shows the $\chi^2$ contours as a function of the
     masses of the parent $\widetilde\mu_R$ and the daughter $\ne{1}$.}
\label{fig:eighteen}
\end{figure}
%%%%%%%%%%%%%%%%%%%%%%%%%%%%%%%%%%%%%%%%%%%%%%%%%%%%%%%%%%%%%%%%%%%%%%

\begin{figure}[t]
\begin{center}
\leavevmode
{\epsfxsize=2.7in\epsfbox{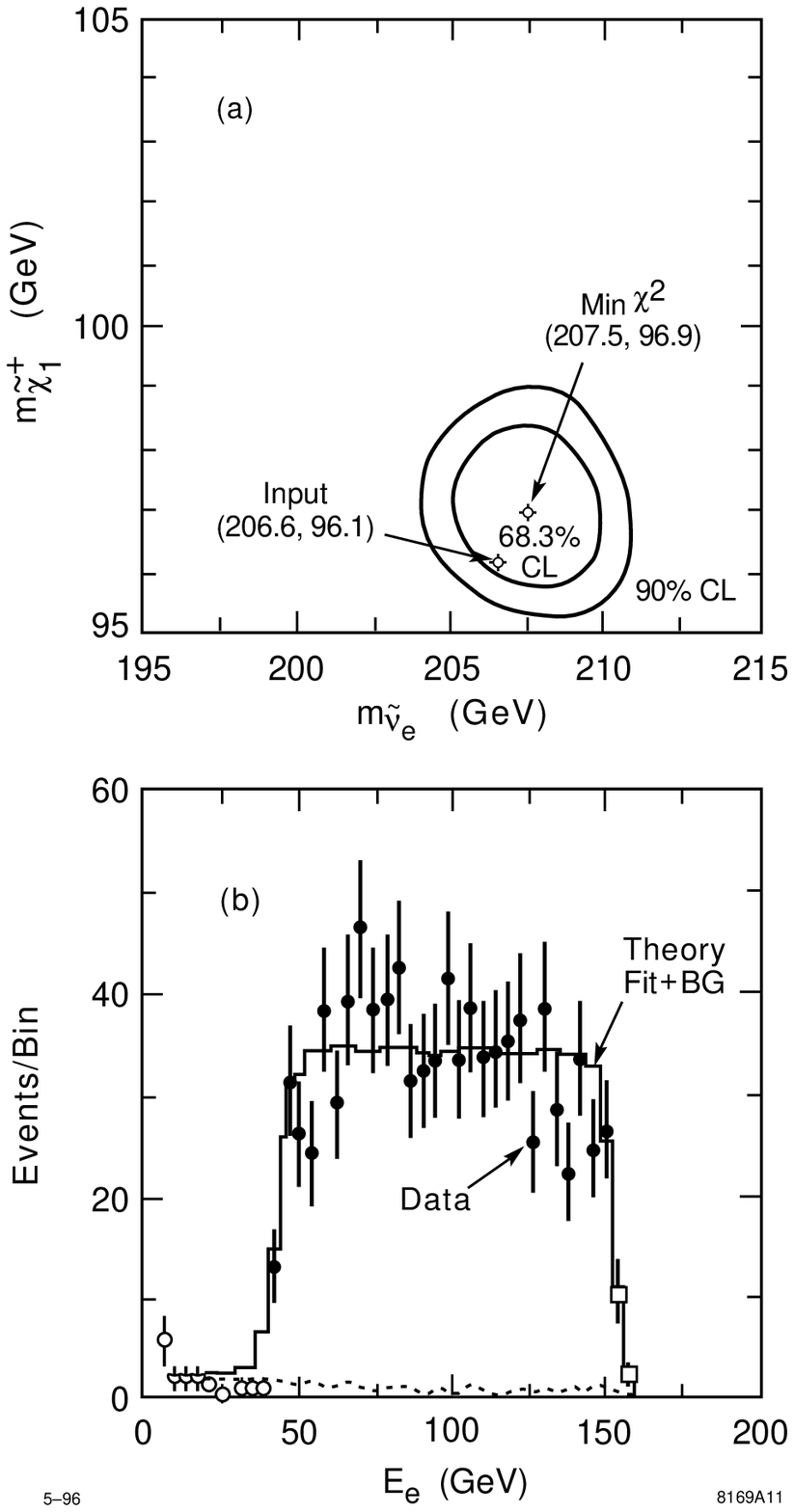}}
\end{center}
 \caption{Simulation of the $\widetilde\nu$ mass measurement at an
       $\ee$ linear collider, from \protect\cite{nlcbook}.  The bottom
      graph gives the event distribution in the decay electron  energy.
   The top graph shows the $\chi^2$ contours as a function of the
     masses of the parent $\widetilde\nu$ and the daughter $\ch{1}$.}
\label{fig:eighteenxx}
\end{figure}

A simulation of the $\widetilde \mu_R$ mass measurement done by the JLC
group \cite{Tsuk} is shown in Figure~\ref{fig:eighteen}.  The
simulation assumes 95\% right-handed
electron polarization, which essentially
eliminates the dominant background  $\ee\to W^+W^-$, but even with 80\%
polarization the endpoint discontinuities are clearly visible. The
measurement gives the masses of $\widetilde{\mu}_R$ and $\ne{1}$ to about
1\% accuracy.  As another example of this technique,
Figure~\ref{fig:eighteenxx} shows a simulation by the NLC group
\cite{nlcbook} of the mass measurement of the $\widetilde \nu$ in
$\widetilde\nu \to e^- \ch{1}$.

%%%%%%%%%%%%%%%%%%%%%%%%%%%%%%%%%%%%%%%%%%%%%%%%%%%%%%%%%%%%%%%%%%%%%%
%%%%%%%%%%%%%%%%%%%%%%%%%%%%%%%%%%%%%%%%%%%%%%%%%%%%%%%%%%%%%%%%%%%%%%

%%%%%%%%%%%%%%%%%%%%%%%%%%%%%%%%%%%%%%%%%%%%%%%%%%%%%%%%%%%%%%%%%%%%%%
\begin{figure}[tb]
\begin{center}
\leavevmode
{\epsfxsize=3.0in\epsfbox{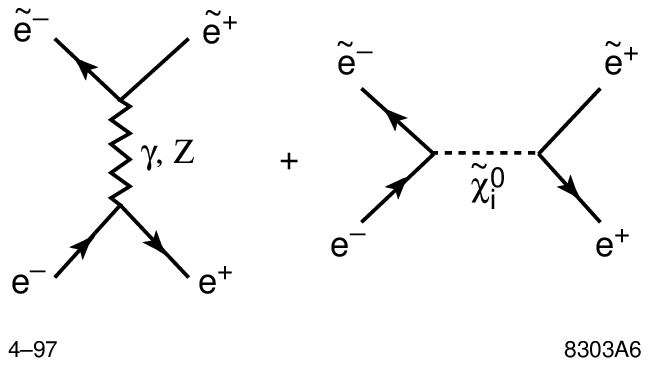}}
\end{center}
 \caption{Feynman diagrams for the process of selectron pair production.}
\label{fig:nineteen}
\end{figure}
%%%%%%%%%%%%%%%%%%%%%%%%%%%%%%%%%%%%%%%%%%%%%%%%%%%%%%%%%%%%%%%%%%%%%%

To go beyond the simple mass determinations, we can look at processes in
which  the production reactions are more complex.
Consider, for example, the pair-production of the first-generation
$\widetilde e^-_R$.   The production
goes through two Feynman diagrams, which are shown in
Figure~\ref{fig:nineteen}.  Because the $\ne{1}$ is typically light
compared to other superparticles, it is the second diagram that is
dominant, especially at small angles.  By measuring the forward peak in
the cross section, we obtain an additional measurement of the lightest
neutralino mass, and a measurement of its coupling to the electron. We
have seen in \leqn{eq:h4} that the coupling of $\widetilde{b}$ to $e^+
\widetilde e^-_R$ is proportional to the standard model $U(1)$ coupling
$g'$.   Thus, this information can be used to determine one of the
neutralino mixing angles. Alternatively, if we have other diagnostics
that indicate that the neutralino parameters are in the gaugino region,
this experiment can check the supersymmmetry relation of couplings.
For a 200 GeV $\widetilde e^-_R$, with 
 a 100 fb$^{-1}$ data sample at 500 GeV, the ratio of
 couplings can be  determined to 1\% accuracy~\cite{NFT}.

Notice that the neutralino exchange diagram in
Figure~\ref{fig:nineteen} is present only for $e^-_R e^+_L \to
\widetilde e^-_R\widetilde e^+_R$, since $\widetilde e^-_R$ is the
superpartner of the right-handed electron. On the other hand, 
with the initial state $e^-_L e^+_R$, we have the analogous diagram
producing the 
superpartner of the left-handed electron $\widetilde L^-$. In the
gaugino region, the process $e^-_L e^+_R \to \widetilde L^- \widetilde
L^+$ has large contributions both from $\ne{1}$ ($\widetilde b$)
exchange and from  $\ne{2}$ ($\widetilde w^3$) exchange.  The reaction
 $e^-_L e^+_L \to \widetilde L^- \widetilde e^+_R$ is also mediated by
neutralino exchange and contains additional useful information.

%%%%%%%%%%%%%%%%%%%%%%%%%%%%%%%%%%%%%%%%%%%%%%%%%%%%%%%%%%%%%%%%%%%%%%
\begin{figure}[t]
\begin{center}
\leavevmode
{\epsfxsize=3.5in\epsfbox{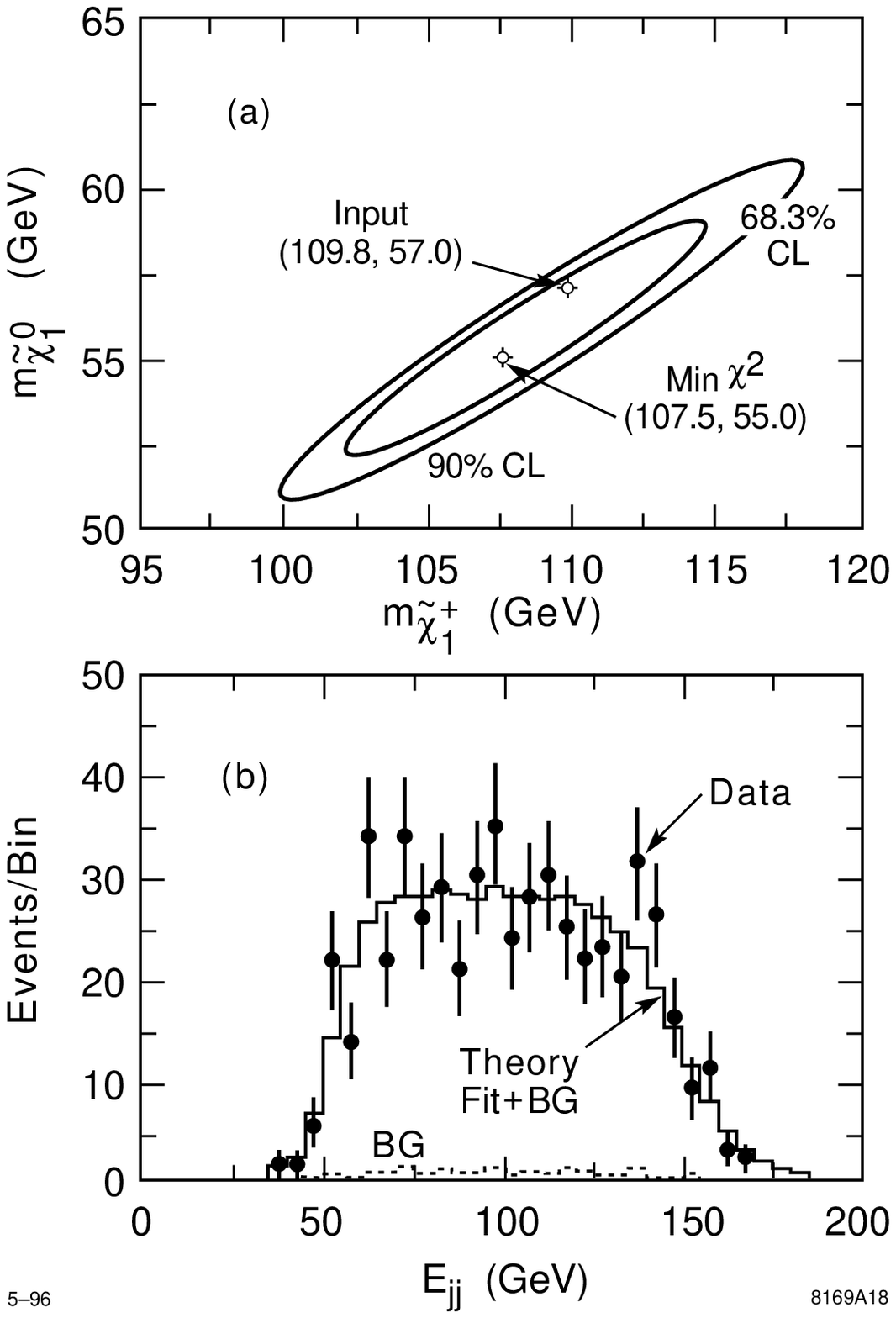}}
\end{center}
 \caption{Simulation of the $\ch{1}$ mass measurement at an
       $\ee$ linear collider, from \protect\cite{nlcbook}.  The bottom 
      graph gives the event distribution in the  energy of the $\bar q q$
  pair emitted in a $\ch{1}$ hadronic decay.  The hadronic system is 
 restricted to a  bin in mass around 30 GeV.
   The bottom  graph shows the $\chi^2$ contours as a function of the
     masses of the parent $\ch{1}$ and the daughter $\ne{1}$.}
\label{fig:twentyandahalf}
\end{figure}
%%%%%%%%%%%%%%%%%%%%%%%%%%%%%%%%%%%%%%%%%%%%%%%%%%%%%%%%%%%%%%%%%%%%%%

Along with the sleptons, the chargino $\ch{1}$ is expected to be a
relatively light particle which is available for precision measurements
at an $\ee$ collider.   The dominant decays of the chargino are $\ch{1}
\to q\bar q \ne{1}$ and $\ch{1}\to \ell^+\nu \ne{1}$, leading to events
with quark jets, leptons, and missing energy.  In mixed hadron-lepton
events, one chargino decay can be analyzed as a two-body decay into the
observed $q\bar q$ system plus the unseen neutral particle $\ne{1}$; then the
mass measurement technique of Figure~\ref{fig:eighteenvv} can be
applied.  The simulation of a sample measurement, using jet pairs
restricted to an interval around 30 GeV in mass, is shown in
Figure~\ref{fig:twentyandahalf}~\cite{nlcbook}.
 The full data sample (50 fb$^{-1}$ at
500 GeV) gives the $\ch{1}$ mass to an accuracy of 1\%~\cite{NLCsusy}.

%%%%%%%%%%%%%%%%%%%%%%%%%%%%%%%%%%%%%%%%%%%%%%%%%%%%%%%%%%%%%%%%%%%%%%
\begin{figure}[tb]
\begin{center}
\leavevmode
{\epsfxsize=3.0in\epsfbox{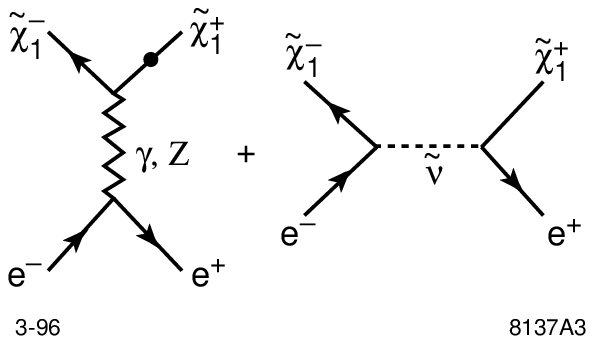}}
\end{center}
 \caption{Feynman diagrams for the process of chargino pair production.}
\label{fig:twenty}
\end{figure}
%%%%%%%%%%%%%%%%%%%%%%%%%%%%%%%%%%%%%%%%%%%%%%%%%%%%%%%%%%%%%%%%%%%%%%

The diagrams for chargino pair production are shown in
Figure~\ref{fig:twenty}.  The cross section depends strongly on the
initial-state polarization.  If the $\widetilde \nu$ is very heavy, it
is permissible to ignore the second diagram; then the first diagram
leads to a cross section roughly ten times larger for $e^-_L$ than for
$e^-_R$.  If the $\widetilde \nu$ is light, this diagram interferes
destructively to lower the cross section.

\begin{figure}[t]
\begin{center}
\leavevmode
{\epsfxsize=4.0in\epsfbox{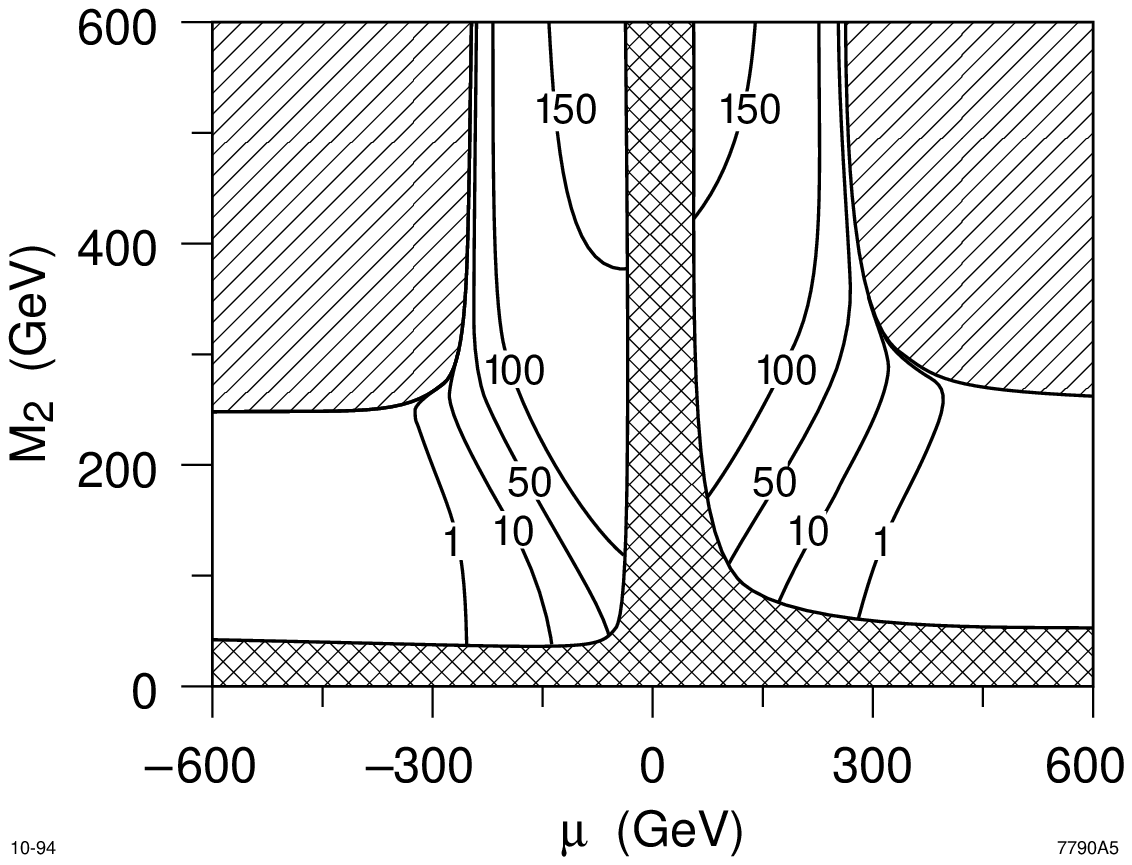}}
\end{center}
 \caption[*]{Contours of constant cross section, in fb,
for the reaction  $e^-_R e^+ \to \ch{1}
\chm{1}$ at $E_\CM = 500$ GeV, from \protect\cite{Feng}.  The plot
shows how the value of this cross section
 maps to the position in the ($\mu, m_2$)
plane.  The boundaries of the indicated regions are the curves on which
the $\ch{1}$ mass equals 50 GeV and 250 GeV.}
\label{fig:twentyone}
\end{figure}

For a right-handed electron beam, the second diagram vanishes.  Then
there is an interesting connection between the chargino production
amplitude and the values of the chargino mixing angles \cite{Tsuk}.
Consider first the limit of very high energy, $E_\CM^2 \gg \mz^2$.   In
this limit, we can ignore the $Z^0$ mass and consider the virtual gauge
bosons in the first diagram to be the $U(1)$ and the neutral $SU(2)$
bosons.  But the $e^-_R$ does not couple to the $SU(2)$ gauge bosons. 
On the other hand, the $W^+$ and $\widetilde w^+$ have zero hypercharge
and so do not couple to the $U(1)$ boson.  Thus, at high energy, the
amplitude for $e^-_R e^+ \to \ch{1} \chm{1}$ is nonzero only if the
charginos have a Higgsino component and is, in fact, proportional to
the chargino mixing angles.  Even if we do not go to asymptotic
energies, this polarized cross section is large in the Higgsino region
and small in the gaugino region, as shown in
Figure~\ref{fig:twentyone}.   This information can be combined with the
measurement of the forward-backward asymmetry  to determine both of the
chargino mixing angles in a manner independent of the other
supersymmetry parameters \cite{Feng}.

%%%%%%%%%%%%%%%%%%%%%%%%%%%%%%%%%%%%%%%%%%%%%%%%%%%%%%%%%%%%%%%%%%%%%%
%%%%%%%%%%%%%%%%%%%%%%%%%%%%%%%%%%%%%%%%%%%%%%%%%%%%%%%%%%%%%%%%%%%%%%

If the study with  $e^-_R$ indicates that the chargino parameters are in
the gaugino region,  measurement of  the differential cross section for
$e^-_L e^+ \to  \ch{1} \chm{1}$ can be used to determine the magnitude
of the second diagram in Figure~\ref{fig:twenty}.  The value of this
diagram can be used to estimate the $\widetilde\nu$ mass or to test
another of the coupling constant relations predicted by supersymmetry.
With a 100 fb$^{-1}$ data sample, the ratio between the $\widetilde{w}^+
\widetilde{\nu} e^-_L$ coupling and the $W^+ \nu e^-_L$ coupling can be
determined to 25\% accuracy if $m(\widetilde\nu)$ must also be
determined by the fit, and to 5\% if $m(\widetilde\nu)$ is known from
another measurement.

These examples demonstrate how the $\ee$ collider experiments can
determine superpartner masses and the mixing angle of the charginos and
neutralinos.  The experimental program is systematic and does not
depend on assumptions about the values of other supersymmetry
parameters.  It only demands the basic requirement that the
color-singlet superpartners are available for study at the energy at
which the collider can run. If squarks can be pair-produced at these
energies, they can also be studied in this systematic way.  Not only
can their masses be measured, but polarization observables can be used
to measure the small mass differences predicted by \leqn{eq:h5} and
\leqn{eq:i5}~\cite{FFin}.

\subsection{Superspectroscopy at hadron colliders}

At the end of Section 4.6, I explained that it should be relatively
straightforward to identify the signatures of supersymmetry at the LHC.
However, it is a challenging problem there to extract precision
information about the underlying supersymmetry parameters.  For a long
time, it was thought that this information would have to come from
cross sections for specific signatures whose origin is complex and
model-dependent.  However, it has been realized more recently that the
LHC can, in certain situations, offer ways to determine supersymmetry
mass parameters kinematically.

Let me briefly describe the parameters of the LHC \cite{LHCrep}.
This is a $pp$ collider with 14 TeV in the center of mass.  The design
luminosity corresponds to a data sample, per experiment, of 
100 fb$^{-1}$ per year.  A simpler experimental environment, without
multiplet hadronic collisions per proton bunch crossing, is obtained
by running at a lower luminosity of 10  fb$^{-1}$ per year, and this 
is probably what will be done initially.  If the supersymmetric partners
of Standard Model particles indeed lie in the region defined by 
our estimates \leqn{eq:tt5}--\leqn{eq:v5}, this low luminosity should
already be sufficient to begin detailed exploration of the supersymmetry
mass spectrum.

%%%%%%%%%%%%%%%%%%%%%%%%%%%%%%%%%%%%%%%%%%%%%%%%%%%%%%%%%%%%%%%%%%%%%%
\begin{figure}[htb]
\begin{center}
\leavevmode
{\epsfxsize=3.5in\epsfbox{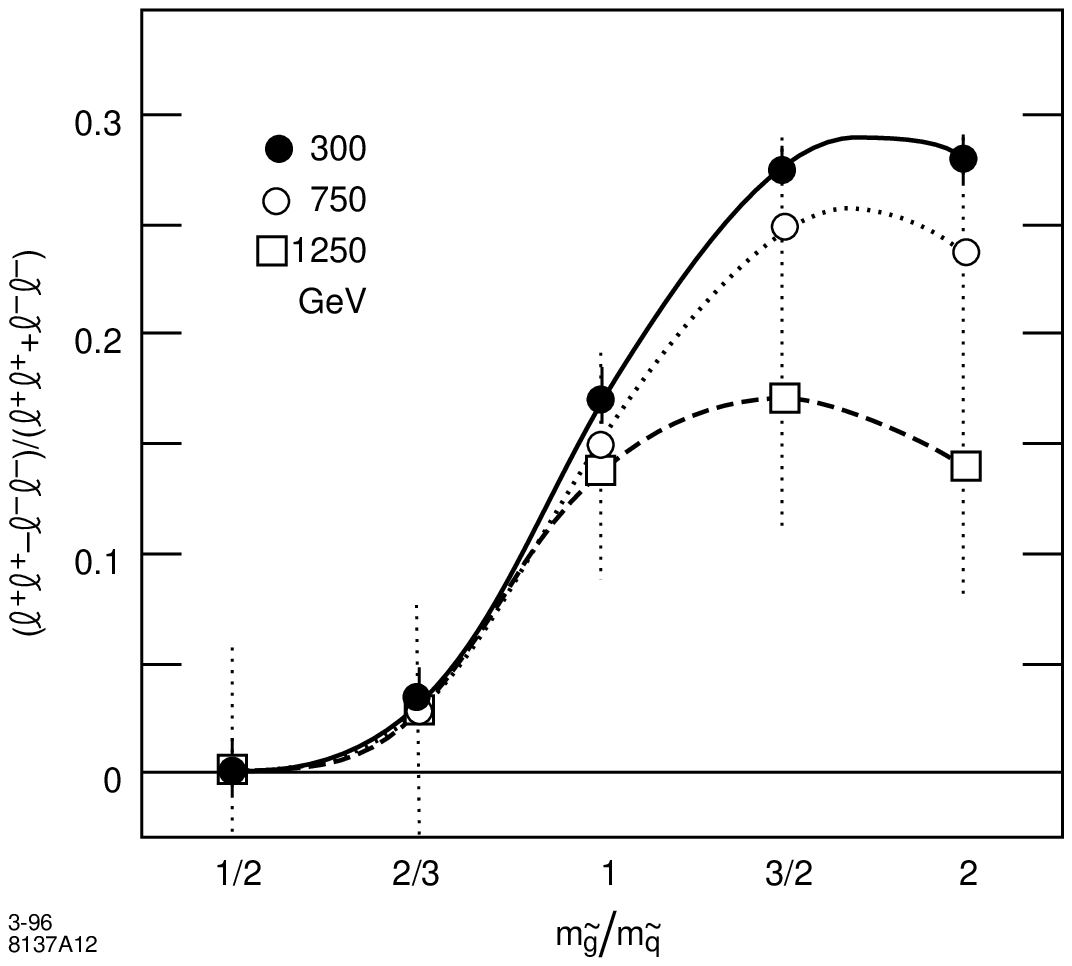}}
\end{center}
 \caption{The asymmetry between the cross sections for dilepton events 
       with $\ell^+\ell^+$ and those with $\ell^-\ell^-$ expected
         at the LHC, plotted as a function of the ratio of the 
        gluino to the squark mass, from \protect\cite{Atlas}.  The 
        three curves refer to the idicated values of the lighter
      of the squark and gluino  masses.} 
\label{fig:twentytwo}
\end{figure}
%%%%%%%%%%%%%%%%%%%%%%%%%%%%%%%%%%%%%%%%%%%%%%%%%%%%%%%%%%%%%%%%%%%%%%

Before we discuss methods for direct mass measurement, I should point
out that the many signatures available at the LHC which do not 
give explicit kinematic reconstructions
do offer a
significant amount of information.  For example, the
ATLAS collaboration \cite{Atlas, Basa} has suggested comparing the
cross-sections for like-sign dilepton events with $\ell^+\ell^+$ versus
$\ell^-\ell^-$.  The excess of events with two positive leptons comes
from the process in which two $u$ quarks exchange a gluino and convert
to $\widetilde u$, making use of the fact that the proton contains more
$u$ than $d$ quarks. The contribution of this process peaks when the
squarks and gluinos have roughly equal masses, as shown in
Figure~\ref{fig:twentytwo}.   Thus, this measurement allows one to 
estimate the ration of the squark and gluino masses.
 Presumably, if the values of $\mu$,
$m_1$, and $m_2$ were known from the $\ee$ collider experiments, it
should be possible to make a precise theory of multi-lepton production
and to use the rates of these processes to determine $m(\widetilde g)$
and $m(\widetilde q)$.

%%%%%%%%%%%%%%%%%%%%%%%%%%%%%%%%%%%%%%%%%%%%%%%%%%%%%%%%%%%%%%%%%%%%%%
\begin{figure}[htb]
\begin{center}
\leavevmode
{\epsfxsize=3.5in\epsfbox{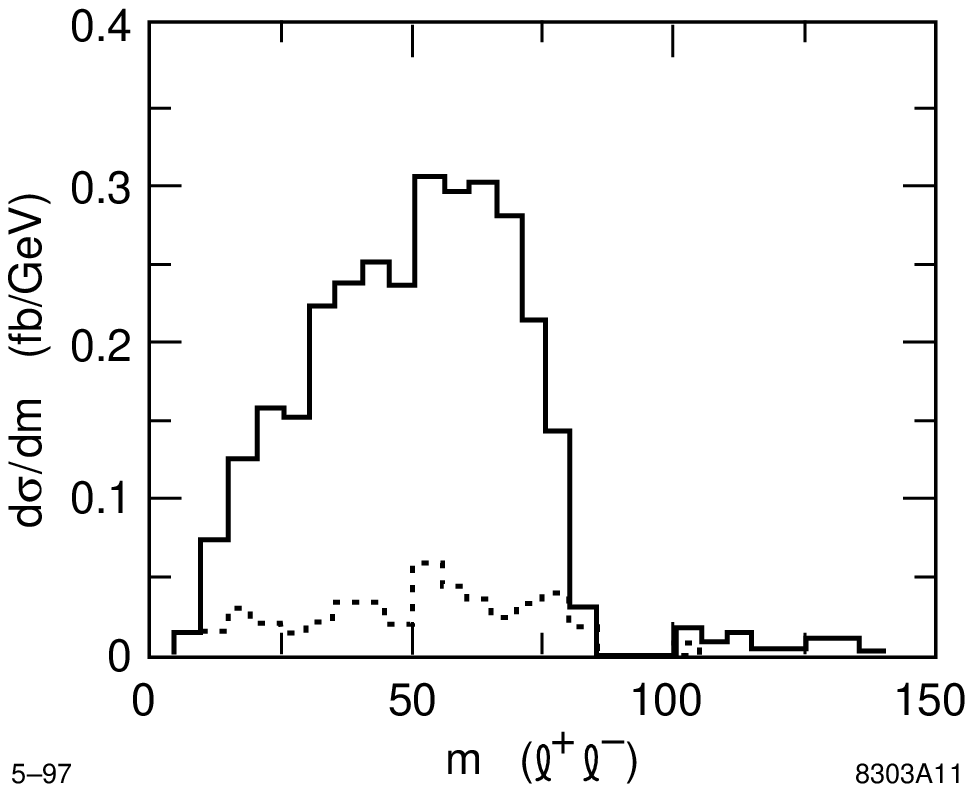}}
\end{center}
 \caption{Distribution of the dilepton mass in the process
   $p\bar p \to \ch{1} \ne{2}+X$, with the $\ne{2}$ decaying to 
    $\ell^+\ell^- \ne{1}$, from \protect\cite{bcpt}.}
\label{fig:twentythree}
\end{figure}
%%%%%%%%%%%%%%%%%%%%%%%%%%%%%%%%%%%%%%%%%%%%%%%%%%%%%%%%%%%%%%%%%%%%%%

In some circumstances, however, the LHC provides direct information on
the superparticle spectrum.  Consider, for example, decay chains which
end with the decay  $\ne{2}\to \ell^+\ell^-\ne{1}$ discussed in
Section~4.5. The dilepton mass distribution has a discontinuity at the
kinematic endpoint where 
\beq
 m(\ell^+\ell^-) = m(\ne{2}) - m(\ne{1})  \ .
\eeq{eq:y5}
The sharpness of this kinematic edge is shown in
Figure~\ref{fig:twentythree}, taken from a study of the process $q\bar
q \to \ch{1}\ne{2}$ \cite{bcpt}.  Under the assumptions of gaugino
unification plus the gaugino region of parameter space, the mass
difference in \leqn{eq:y5} equals $0.5m_2$.  Thus, if we have some
independent evidence for these assumptions, the position of this edge
can be used to give the overall scale of superparticle masses.  Also,
if the gluino mass can be measured, the ratio of that mass to the
mass difference \leqn{eq:y5} provides a test of these assumptions.

At a point in parameter space studied for the ATLAS Collaboration
 in \cite{Ianscrew}, it is possible to go much further. We
need not discuss why this particular point in the `minimal SUGRA'
parameter space was chosen for special study, but it turned out to have
a number of advantageous properties.  The value of the gluino mass was
taken to be 
300 GeV, leading to a very large gluino production cross section, equal to 
1 nb,
at the LHC.  The effect of Yukawa couplings discussed in Section 3.7
lowers the masses of the superpartners of $t_L$ and $b_L$, in
particular, making $\widetilde b_L$ the lightest squark.  Then a major
decay chain for the $\widetilde g$ would be \beq \widetilde g \to
\widetilde b_L \bar b \to b \bar b \ne{2}\ , \eeq{eq:z5} which could be
followed by the dilepton decay of the $\ne{2}$.

%%%%%%%%%%%%%%%%%%%%%%%%%%%%%%%%%%%%%%%%%%%%%%%%%%%%%%%%%%%%%%%%%%%%%%
\begin{figure}[phtb]
\begin{center}
\leavevmode
{\epsfxsize=3in\epsfbox{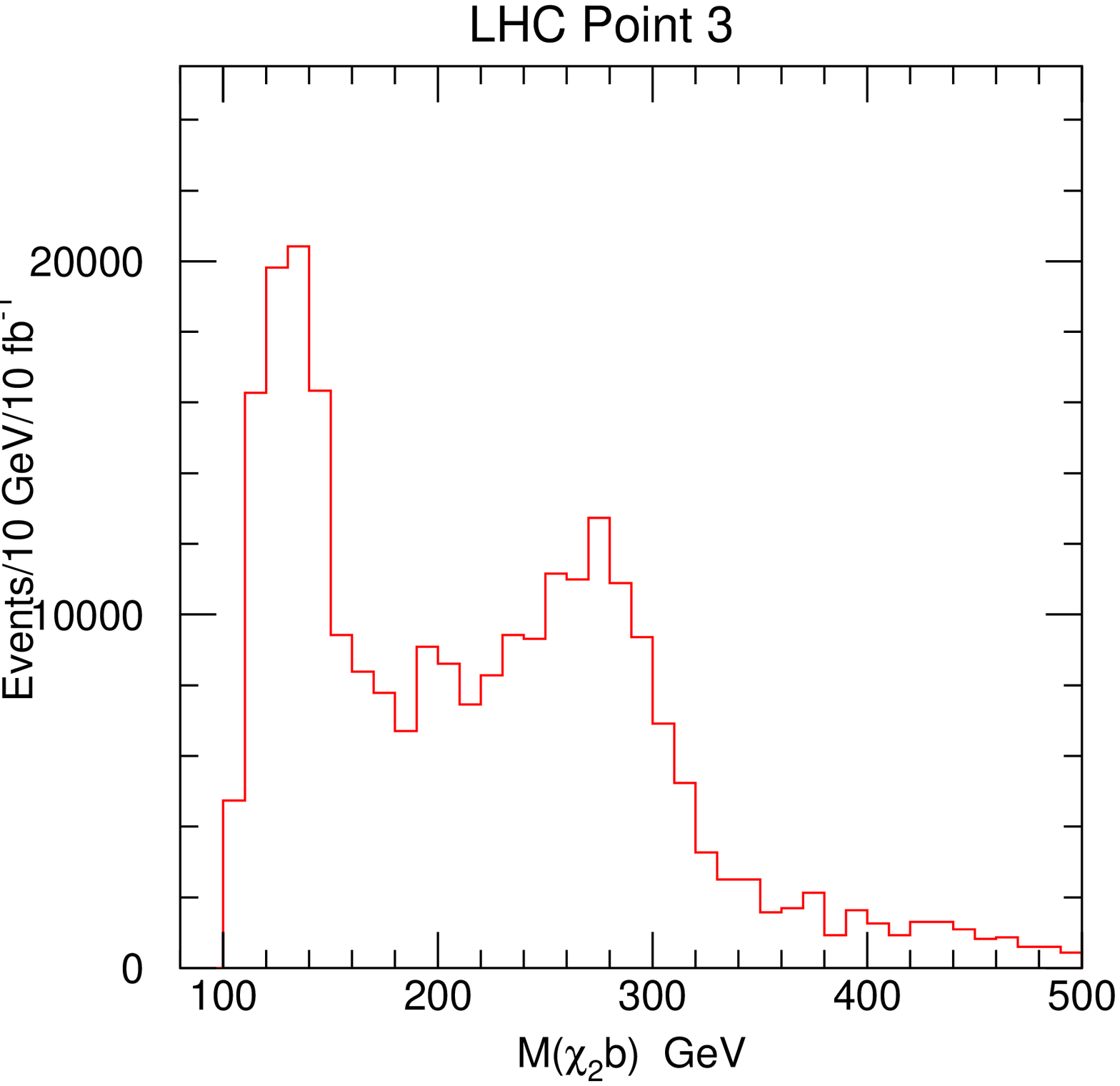}}
\medskip
{\epsfxsize=3in\epsfbox{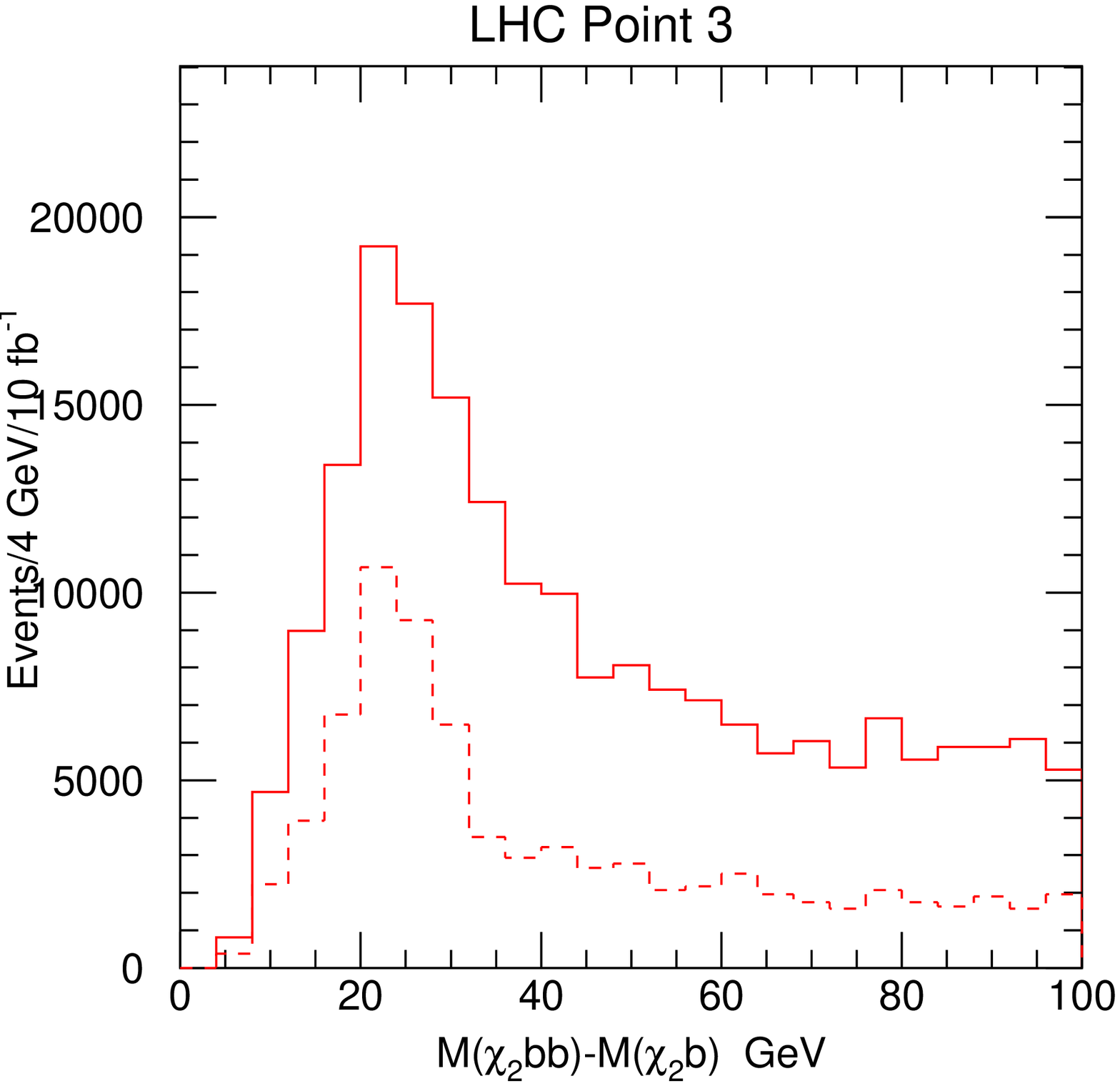}}
\end{center}
 \caption[*]{Reconstruction of the mass of the $\widetilde b$ and the 
 $\widetilde g$ at the LHC, at a point in supersymmetry parameter space 
studied in \protect\cite{Ianscrew}.  In the plot  on the left,
 the peak near 300 GeV shows the reconstructed $\widetilde b$.  The 
 plot  on the right shows the event distribution in the variable
 $m(\widetilde g) - m(\widetilde b)$.  The dashed distribution shows the 
values for the events lying between 230 GeV and 330 GeV in the left-hand
 figure.}
\label{fig:twentyfour}
\end{figure}
%%%%%%%%%%%%%%%%%%%%%%%%%%%%%%%%%%%%%%%%%%%%%%%%%%%%%%%%%%%%%%%%%%%%%%

Since the number of events expected at this point is 
 very large, we can select events in which
the $\ell^+\ell^-$ pair falls close to its kinematic endpoint. For
these events, the dilepton pair and the daughter $\ne{1}$ are both at
rest with respect to the parent $\ne{2}$.  Then, if we are also given
the mass of the $\ne{1}$, the energy-momentum 4-vector of the $\ne{2}$
is determined. This mass might be obtained from the assumptions listed
below \leqn{eq:y5}, from a more general fit of the LHC supersymmetry
data to a model of the supersymmetry mass spectrum, or from a direct
measurement at an $\ee$ collider.  In any event, once the momentum
vector of the $\ne{2}$ is determined, there is no more missing momentum
in the decay chain.  It is now possible to successively add $b$ jets to
reconstruct the  $\widetilde b_L$ and then the $\widetilde g$.  The
mass peaks for these states obtained from the 
 simulation results of \cite{Ianscrew} are
shown in Figure~\ref{fig:twentyfour}. For a fixed $m(\ne{1})$, the
masses of $\widetilde b_L$ and $\widetilde g$ are determined to 1\%
accuracy.

It may seem  that this example uses many special features of the
particular point in parameter space which was chosen for the analysis. 
At another point, the spectrum might be different in a way that would
compromise parts of this analysis.  For example, the $\ne{2}$ might be
allowed to decay to an on-shell $Z^0$, or the gluino might lie below
the $\widetilde b_L$.   On the other hand, the method just described
can be extended to any superpartner with a three-body becay involving
one unobserved neutral.  In \cite{Ianscrew}, other examples are discussed
which 
apply these
ideas to decay chains that end with  $\widetilde q \to \ne{1} h^0 q$
  and $\widetilde t \to \ne{1} W^+ b$.

To properly evaluate the capability of the LHC to perform precision
supersymmetry measurements, we must remember that Nature has chosen (at
most) one point in the supersymmetry parameter space, and that every
point in parameter space is special in its own way.  It is not likely
that we will know, in advance, which  particular trick that will be
most effective.  However, we have now only begun the study of
strategies to determine the superparticle spectrum from the kinematics
of LHC reactions.  There are certainly many more tricks to be
discovered.

\subsection{Recapitulation}

If physics beyond the Standard Model is supersymmetric, I
am optimistic about the future prospects for experimental particle
physics.  At the LHC, if not before, we will discover the superparticle
spectrum.  This spectrum encodes information about physics at the
energy scale of supersymmetry breaking, which  might be as high as the
grand unification or even the superstring scale.  If we can measure
the basic parameters that determine this spectrum,
 we can uncover the patterns that will let us
decode this information and see much more deeply into fundamental
physics.

It is not clear how much of this program can already be done at the LHC
and how much must be left to the experimental program of an $\ee$
linear collider.  For adherents of the linear collider, the worst case
would be that Nature has chosen a minimal parameter set and also some
special mass relations that allow the relevant three or four parameters
to be determined at the LHC.  Even in this case, the linear collider
would have a profoundly interesting experimental program.  In this
simple scenario, the LHC experimenters will be able to fit their data
to a small number of parameters, but the hadron
collider experiments cannot verify that this is the whole story.  To give
one example, it is not known how, at a hadron collider, to measure the 
mass of the $\ne{1}$, the particle that provides the basic quantum of 
missing energy-momentum used to build up the supersymmetry mass spectrum.
The LHC experiments may give indirect determinations of $m(\ne{1})$.
The linear collider can provide a direct precision measurement of this
particle mass.  If the predicted value were found, that would be an
intellectual triumph comparable to the direct discovery of the $W$ boson
in $p\bar p$ collisions.

I must also emphasize that there is an important difference between the
study of the supersymmetry spectrum and that of the spectrum of weak
vector bosons.  In the latter case, the spectrum was predicted by a 
coherent theoretical model, the $SU(2)\times U(1)$ gauge theory.  In the 
case of supersymmetry, as I have emphasized in Section 4.3, the minimal
parametrization is just a guess---and one guess among many.  Thus, 
it is a  more likely
outcome that a simple parametrization of the supersymmetry
spectrum would omit crucial details.  To discover these features, one would
need the model-independent approach to supersymmetry parameter
measurements that the $\ee$ experiments can provide.

In this more general arena for the construction and testing of supersymmetry
model, the most striking feature of the comparison of colliders is how 
much each facility adds to the results obtainable at the other.
  From the $\ee$ side, we
will obtain a precision understanding of the color-singlet portion of
the supersymmetry spectrum. We will measure parameters which determine
what decay chains the colored superparticles will follow.
From the $pp$ side, we will observe some of these decay chains 
directly and obtain precise inclusive cross sections for the decay 
products.  This should allow us to analyze these decay chains back to their
origin and to measure the superspectrum parameters of heavy colored
superparticles.  Thus,
  if the problem that Nature poses for us is supersymmetry,
these two colliders together can solve that problem experimentally.

\section{Technicolor}

In the previous two sections, I have given a lengthy discussion of the
theoretical structure of models of new physics based on supersymmetry.
I have explained how supersymmetry leads to a solution to the problem
of electroweak symmetry breaking.  I have explained that the
ramifications of supersymmetry are quite complex and lead to a rich
variety of phenomena that can be studied experimentally at colliders.

This discussion illustrated one of the major points that I made at the
beginning of these lectures.  In seeking an explanation for electroweak
symmetry breaking, we could just write down the minimal Lagrangian
available. However, for me, it is much more attractive to look
for a theory in which electroweak symmetry breaking emerges from a
definite physical idea.  If the idea is a  profound one, it will 
naturally lead to new phenomena that we can discover in experiments.

Supersymmetry is an idea that illustrates this picture, but it might
not be the right idea.  You might worry that this example was a very
special one. Therefore, if I am to provide an overview of ideas on
physics beyond the Standard Model, I should give at least one more
example of a physical idea that leads to electroweak symmetry breaking,
and one assumptions of a very different kind.
 Therefore, in this section, I will
discuss models of electroweak symmetry breaking based on the postulate of
new strong
interactions at the electroweak scale.
  We will see that this idea leads to a different set of
physical predictions but nevertheless implies a rich and intriguing
experimental program.

\subsection{The structure of technicolor models}

The basic structure of a model of electroweak symmetry breaking by new
strong interactions is that of the Weinberg-Susskind model discussed at
the end of Section 2.2.  This model was based on a strong-interaction
model that was essentially a scaled up version of QCD.  From here on, I
will refer to the new strong interaction gauge symmetry as
`technicolor'. In this section, I will discuss more details of this
model, and also add features that are necessary to provide for quark
and lepton mass generation.

In Section 2.2, I pointed out that the Weinberg-Susskind model leads to
a vacuum expectation value which breaks $SU(2)\times U(1)$.  To
understand this model better, we should first try to compute the $W$
and $Z$ boson mass matrix that comes from this symmetry breaking.

QCD with two massless flavors has the global symmetry $SU(2)\times
SU(2)$; independent $SU(2)$ symmetries can rotate the doublets $q_L =
(u_L,d_L)$ and $q_R = (u_R,d_R)$.  When the operator $\bar q q$ obtains
vacuum expectation values as in \leqn{eq:o}, the two $SU(2)$ groups are
locked together by the pairing of quarks with antiquarks in the vacuum.
Then the overall $SU(2)$ is unbroken; this is  the manifest isospin
symmetry of QCD.  The second $SU(2)$ is that associated with the axial
vector currents 
\beq
             J^{\mu 5 a} = \bar q \gamma^\mu \gamma^5 \tau^a q \ .
\eeq{eq:a6}
This symmetry is spontaneously broken.  By Goldstone's theorem, the
symmetry breaking leads to a massless boson for each spontaneously
broken symmetry, one created or annihilated by each broken symmetry
current.  These three particles are identified with the pions of QCD.
The matrix element between the axial $SU(2)$ currents and the pions can
be parametrized as 
\beq
    \bigl\langle 0 \big| J^{\mu 5 a}\big|\pi^b(p)\bigr\rangle
 = i f_\pi  p^\mu \delta^{ab} \ .
\eeq{eq:b6}
By recognizing that $J^{\mu 5 a}$ is a part of the weak interaction
current, we can identify $f_\pi$ as the pion decay constant, $f_\pi =
93$ MeV. The assumption of Weinberg and Susskind is that the same story
is repeated in technicolor. However, since the technicolor quarks are
assumed to be massless,
the pions remain precisely massless at this stage of the argument.

%%%%%%%%%%%%%%%%%%%%%%%%%%%%%%%%%%%%%%%%%%%%%%%%%%%%%%%%%%%%%%%%%%%%%%
\begin{figure}[tb]
\begin{center}
\leavevmode
{\epsfxsize=3.5in\epsfbox{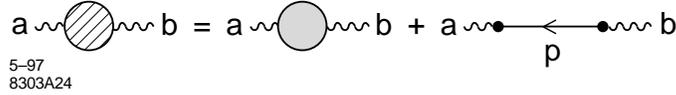}}
\end{center}
 \caption{Contributions to the vacuum polarization of the $W$ boson from 
        technicolor states.}
\label{fig:twentyfive}
\end{figure}
%%%%%%%%%%%%%%%%%%%%%%%%%%%%%%%%%%%%%%%%%%%%%%%%%%%%%%%%%%%%%%%%%%%%%%

If the system with spontaneously broken symmetry and massless pions is
coupled to gauge fields, the gauge boson should obtain mass through the
Higgs mechanism.  To compute the mass term, consider the gauge boson
vacuum polarization diagram shown in Figure~\ref{fig:twentyfive}.

  Let
us assume first that we couple only the weak interaction $SU(2)$ bosons
to the techniquarks.  The coupling is 
\beq
       \Delta \L = g A^a_\mu J^a_{L\mu}\ .
\eeq{eq:c6}
Then the matrix element \leqn{eq:b6} allows a pion to be
annihilated and a gauge boson created, with the amplitude
\beq
     ig  \cdot (-\half) \cdot i f_\pi p_\mu \delta^{ab} \ ;
\eeq{eq:d6}
the second factor comes from $J^a_{L\mu} = \half(J^a_\mu -
J^{a5}_\mu)$. Using this amplitude, we can evaluate the amplitude for a 
process in which a gauge boson converts to a Goldstone boson and then
converts back.  This corresponds to the diagram contributing to the 
vacuum polarization shown as the second term on the right-hand side of
 Figure~\ref{fig:twentyfive}.  The value of this diagram is
\beq
        \left({gf_\pi p_\mu\over 2}\right) {1\over p^2}  
 \left(- {gf_\pi p_\nu\over 2}\right) \ .
\eeq{eq:e6}
The full vacuum polarization amplitude $i\Pi^{ab}_{\mu\nu}(p)$ consists
of this term  plus more
complicated terms with massive particles or multiple particles
exchanged.  These are indicated as the shaded blob 
in Figure~\ref{fig:twentyfive}.
  If there are no massless particles in the symmetry-breaking
sector other than the pions, \leqn{eq:e6} is the only term with a
$1/p^2$ singularity near $p=0$. Now recognize that   the gauge
current $J^a_{L\mu}$ is conserved, and so the vacuum polarization must
satisfy 
\beq
           p^\mu\, \Pi^{ab}_{\mu\nu}(p) = 0  \ .
\eeq{eq:f6}
These two requirements are compatible only if the vacuum polarization 
behaves near $p=0$ as
\beq 
         \Pi^{ab}_{\mu\nu} = \left({gf_\pi\over 2}\right)^2 \left(g_{\mu\nu}
      - {p_\mu p_\nu\over p^2} \right) \delta^{ab} \ .
\eeq{eq:g6}
This is a mass term for the vector boson, giving 
\beq
       \mw = g{v\over 2}\ , \quad \mbox{with} \quad v = f_\pi\ .
\eeq{eq:h6}
This is the result that I promised above \leqn{eq:q}.

Now add to this structure the $U(1)$ gauge boson $B_\mu$ coupling to 
hypercharge.  Repeating the same arguments, we find the mass matrix
\beq
      m^2 =    \left({f_\pi\over 2}\right)^2\pmatrix{g^2  & & & \cr
                                             & g^2 & & \cr
                                            &  & g^2 & -gg'\cr
                                            &   &-gg'& (g')^2 \cr} \ ,
\eeq{eq:i6}
acting on $(A^1_\mu, A^2_\mu, A^3_\mu, B_\mu)$. 
This has just the form of \leqn{eq:t}.   The eigenvalues of
this matrix give the vector boson masses \leqn{eq:f}, with $v = 246\
\mbox{GeV}= f_\pi$.  This is the result promised above \leqn{eq:q}. 
More generally, in a model with $N_D$ technicolor doublets, we require,
\beq
              v^2 = N_D f_\pi^2 \ .
\eeq{eq:j6}
Thus, a larger technicolor sector lies lower in energy and is closer 
to the scale
of present experiments.

In my discussion of \leqn{eq:t}, I pointed out that this equation
calls for  the presence of an unbroken $SU(2)$ global symmetry of the 
new strong interactions, called {\em custodial $SU(2)$},
 in addition to the spontaneously broken weak 
interaction $SU(2)$ symmetry.  This global $SU(2)$ symmetry requires that 
the first three diagonal entries in \leqn{eq:i6} are equal, giving the 
mass relation  $ \mw/ \mz = \cos\theta_w$.  Custodial $SU(2)$ symmetry also 
acts on the heavier states of the new strong interaction theory and will
play an important role  in our analysis of the experimental probes of this
sector.

\begin{figure}[tb]
\begin{center}
\leavevmode
\epsfbox{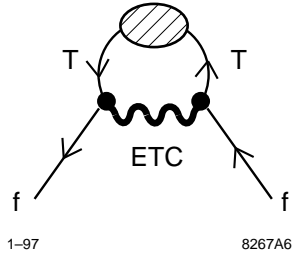}
\end{center}
 \caption{ETC generation of quark and lepton masses.}
\label{fig:twentysix}
\end{figure}

The model I have just described gives mass the the $W$ and $Z$ bosons,
but it does not yet give mass to quarks and leptons.  In order to
accomplish this, we must couple the quarks and leptons to the
techniquarks.  This is done by introducing further gauge bosons called
{\em Extended Technicolor} (ETC) bosons \cite{DimS,EandL}. 
 If we imagine that the
ETC bosons connect light fermions to techniquarks, and that they
are very heavy, a typical coupling induced by these bosons would have
the form 
\beqa
     i \Delta\L &=&  (i g_E \bar u_L\gamma^\mu U_L) {-i\over -m_E^2}
 (i g_E \bar U_R \gamma_\mu u_R) \CR
     &=& -i {g_E^2\over m_E^2} \bar u_L \gamma^\mu  U_L \bar U_R 
               \gamma_\mu u_R
\eeqa{eq:k6}
Now replace $U_L\bar U_R$ by its vacuum expectation value due to
dynamical techniquark mass generation: 
\beq
             \VEV{ U_L \bar U_R} = -{1\over 4} \VEV{\bar U U} = 
                    {1\over 4} \Delta \ ,
\eeq{eq:l6}
where $m_E$ and $g_E$ are the ETC mass and coupling, $\Delta$ is as in
\leqn{eq:o} and the unit matrix is in the space of Dirac indices. 
Inserting \leqn{eq:l6} into \leqn{eq:k6}, we find a fermion mass term
\beq
         m_u = {g^2_E\over m_E^2} \Delta \ .
\eeq{eq:m6}
The origin of this term is shown diagrammatically in
Figure~\ref{fig:twentysix}. In principle, masses could be generated in
this way for all of the quarks and leptons.

%%%%%%%%%%%%%%%%%%%%%%%%%%%%%%%%%%%%%%%%%%%%%%%%%%%%%%%%%%%%%%%%%%%%%%
%%%%%%%%%%%%%%%%%%%%%%%%%%%%%%%%%%%%%%%%%%%%%%%%%%%%%%%%%%%%%%%%%%%%%%

From \leqn{eq:m6}, we can infer the mass scale required for the ETC
interactions.  Estimating with $g_E \approx 1$, and $\Delta \sim 4\pi
f_\pi^3$ (which gives $\VEV{\bar u u} = ($300 MeV$)^3$ in QCD), we find
\beq
     m_E = g_E \left({4 \pi f_\pi^3\over m_f}\right)^{1/2} = 
                 \cases{43 \ \mbox{TeV}& $f= s$ \cr
            1.0 \ \mbox{TeV}& $f = t$ \cr }\ ,
\eeq{eq:n6}
using the $s$ and $t$ quark masses as reference points in the fermion
mass spectrum.

The detailed structure of the ETC exchanges must be paired with a
suitable structure of the techiquark sector.  We might call `minimal
technicolor' the theory with precisely one weak interaction $SU(2)$
doublet of techniquarks.  In this case, all of the flavor structure
must appear in the ETC group.  In particular, some ETC bosons must be
color triplets to give mass to the quarks through the mechanism of
Figure~\ref{fig:twentysix}.  Another possibility is that the
technicolor sector could contain techniquarks with the $SU(3)\times
SU(2)\times U(1)$ quantum numbers of a generation of quarks and leptons
\cite{FS}. Then the ETC bosons could all be color singlets, though they
would still carry generation quantum numbers.  In this case also,
\leqn{eq:j6} would apply with $N_D = 4$, putting $f_\pi = 123$ GeV. 
More complex cases in which ETC bosons can be doublets of $SU(2)$ have
also been discussed in the literature \cite{CST}.

\subsection{Experimental constraints on technicolor}

The model that I have just described makes a number of characteristic
physical predictions that can be checked in experiments at energies
currently available.  Unfortunately, none of these predictions checks
experimentally.  Many theorists view this as a repudiation of the
technicolor program.  However, others point to the fact that we have
built up the technicolor model assuming that the dynamics of the
technicolor interactions exactly copies that of QCD.  By modifying the
pattern or the explicit energy scale of chiral symmetry breaking, it is
possible to evade these difficulties.  Nevertheless, it is important to
be aware of what the problems are.  In this section, I will review the
three major experimental problems with technicolor models and then
briefly examine how they may be avoided through specific assumptions
about the strong interaction dynamics.

The first two problems are not specifically associated with technicolor
but rather with the ETC interactions that couple techniquarks to the
Standard Model quarks and leptons.  If two  matrices of the ETC group
link quarks with techniquarks, the commutator of these
matrices should  link quarks with quarks.  This implies that there should
be ETC bosons which create new four-quark interactions with
coefficients of order $g_E^2/m_E^2$.  In the Standard Model, there are
no flavor-changing neutral current couplings at the tree level.  Such
couplings are generated by weak interaction box diagrams and other loop
effects, but the flavor-changing part of these interactions is
suppressed to the level observed experimentally by the GIM cancellation
among intermediate flavors \cite{GIM}.  This cancellation follows from
the fact that the couplings of the various flavors of quarks and
leptons to the $W$ and $Z$ depend only on their $SU(2)\times U(1)$
quantum numbers.  For ETC, however, either the couplings or the boson
masses must depend strongly on flavor in order to generate the observed
pattern of quark and lepton masses.  Thus, generically, one expects
large flavor-changing neutral current effects.  It is possible to
suppress these couplings to a level at which they do not contribute
excessively to the $K_L$--$K_S$ mass difference, but only by raising
the ETC mass scale to $m_E \geq 1000$ TeV.  In a similar way, ETC
interactions generically give excessive contributions to $K^0 \to \mu^+
e^-$ and to $\mu \to e \gamma$ unless $m_E \geq 100$ TeV 
\cite{ELP,DimEll}.  These estimates contradict the value of the 
ETC boson masses
required in \leqn{eq:n6}.  There are schemes for natural flavor
conservation in technicolor theories, but they require a very large
amount of new structure just above 1~TeV \cite{DimGR,Randall,Georgi}.
 
The second problem comes in the value of the top quark mass.  If ETC is
weakly coupled, the value of any quark mass should be bounded by
approximately 
\beq
                m_f \leq  {g_E^2\over 4\pi} \Delta^{1/2}\ ,
\eeq{eq:o6}
where $\Delta$ is the techniquark bilinear expectation value. 
Estimating as above, this bounds the quark masses at about 70 GeV
\cite{Raby}. To see this problem from another point of view, look back
at the mass of the ETC boson associated with the top quark, as given in
\leqn{eq:n6}. This is comparable to the mass of the technicolor $\rho$
meson, which we would estimate from \leqn{eq:q} to have a value of
about 2 TeV.  So apparently the top quark's ETC boson must be a
particle with technicolor strong interactions.  This means that the
model described above is not self-consistent.  Since this new
strongly-interacting particle generates mass for the $t$ but not the
$b$, it has the potential to give large contributions to other
relations that violate weak-interaction isospin.  In particular, it can
give an unwanted large correction to the relation $\mw = \mz
\cos\theta_w$ in \leqn{eq:s}.

The third problem relates directly to the technicolor sector itself. 
This issue arises from the precision electroweak measurments.  In
principle, the agreement of precision electroweak measurements with the
Standard Model is a strong constraint on any type of new physics.  The
constraint turns out to be especially powerful for technicolor.  To
explain this point, I would like to present some general formalism and
then specialize it to the case of technicolor.

At first sight, new physics can affect the observables of precision
electroweak physics through radiative corrections to the $SU(2)\times
U(1)$ boson propagators, to the gauge boson vertices, and to 4-fermion
box diagrams.  Typically, though, the largest effects are those from
vacuum polarization diagrams.  To see this, recall that almost all
precision electroweak observables involve 4-fermion reactions with
light fermions only.  (An exception is the $Z\to b\bar b$ vertex, whose
discussion I will postpone to Section 5.7.) In this case,  the vertex
and box diagrams involve only those new particles that couple directly
to the light generations. If the new particles are somehow connected to
the mechanism of $SU(2)\times U(1)$ breaking and fermion mass
generation, these couplings are necessarily small.  The vacuum
polarization diagrams, on the other hand, can involve all new particles
which couple to $SU(2)\times U(1)$, and can even be enhanced by color
or flavor sums over these particles.

The vacuum polarization corrections also can be accounted in a very
simple way.  It is useful, first, to write the $W$ and $Z$ vacuum
polarization amplitudes in terms of  current-current expectation values
for the $SU(2)$ and electromagnetic currents.  Use  the relation
\beq
       J_Z  = J_3 - s^2 J_Q \ ,
\eeq{eq:p6}
where $J_Q$ is the electromagnetic current, and $s^2 = \sin^2\theta_w$,
$c^2 = \cos^2\theta_w$.  Write
the weak coupling
constants explicitly in terms of $e$, $s^2$ and $c^2$.
Then the 
vacuum polarization amplitudes of $\gamma$, $W$, and $Z$ and the $\gamma Z$
mixing amplitude take the form
\beqa
      \Pi_{\gamma\gamma} & = & e^2 \Pi_{QQ}\CR
      \Pi_{WW} & = & {e^2\over s^2} \Pi_{QQ}\CR
    \Pi_{ZZ} &=& {e^2\over c^2s^2} (\Pi_{33} - 2s^2 \Pi_{3Q}+ s^4\Pi_{QQ}) \CR
      \Pi_{Z\gamma} &=& {e^2\over cs} (\Pi_{3Q} - s^2 \Pi_{QQ}) \ .
\eeqa{eq:q6}
The current-current amplitudes $\Pi_{ij}$ are functions of $(q^2/M^2)$,
where $M$ is the mass of the new particles whose loops contribute to
the vacuum polarizations.

If these new particles are too heavy to be found at the $Z^0$ or in the
early stages of LEP 2, the ratio $q^2/M^2$ is bounded for $q^2 = \mz^2$.
Then
it is reasonable to expand the current-current
expectation values a power series.  In making this expansion, it is
important to take into account that any amplitude involving an
electromagnetic current will vanish at $q^2 = 0$ by the standard QED
Ward identity.  Thus, to order $q^2$, we have six coefficients, 
\beqa
     \Pi_{QQ} & = & \phantom{\Pi_{11}(0) + } q^2 \Pi'_{QQ}(0) + \cdots \CR
     \Pi_{11} & = & \Pi_{11}(0) +  q^2 \Pi'_{11}(0) + \cdots \CR
      \Pi_{3Q} & = & \phantom{\Pi_{11}(0) + } q^2 \Pi'_{3Q}(0) + \cdots \CR
      \Pi_{33} & = & \Pi_{33}(0) +  q^2 \Pi'_{33}(0) + \cdots 
\eeqa{eq:r6}
To specify the coupling constants $g$, $g'$ and the scale $v$ of the
electroweak theory, we must measure three parameters.  The most
accurate reference values come from $\alpha$, $G_F$, and $m_Z$.  Three
of the coefficients in \leqn{eq:r6} are absorbed into these parameters.
This leaves three independent coefficients which can in principle be
extracted from experimental measurements. These are conventionally
defined \cite{PandT} as 
\beqa
      S &=&  16\pi \bigl[ \Pi'_{33}(0) - \Pi'_{3Q}(0) \bigr] \CR
      T &=&  {4\pi\over s^2 c^2 \mz^2}\bigl[ \Pi_{11}(0) - \Pi_{33}(0) \bigr]
                       \CR
      U &=&  16\pi \bigl[ \Pi'_{33}(0) - \Pi'_{11}(0) \bigr] 
\eeqa{eq:s6}
I include in these parameters only the contributions from new physics.
From the definitions, you can see that $S$ measures the overall
magnitude of $q^2/M^2$ effects, and $T$ measures the magnitude of
effects that violate the custodial  $SU(2)$ symmetry of the new particles. 
The third parameter $U$ requires both $q^2$-dependence and 
$SU(2)$ violation  and typically is small in
explicit models.

%%%%%%%%%%%%%%%%%%%%%%%%%%%%%%%%%%%%%%%%%%%%%%%%%%%%%%%%%%%%%%%%%%%%%%
\begin{figure}[t]
\begin{center}
\leavevmode
{\epsfxsize=2.5in\epsfbox{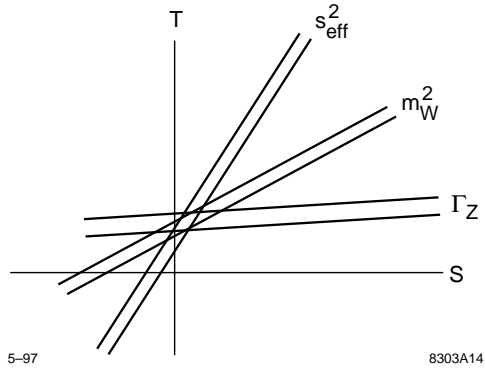}}
\end{center}
 \caption{Schematic determination of $S$ and $T$ from precision electroweak
             measurements. For each observable, the width of the 
band corresponds to the
          experimental error in its determination.}
\label{fig:twentyseven}
\end{figure}
%%%%%%%%%%%%%%%%%%%%%%%%%%%%%%%%%%%%%%%%%%%%%%%%%%%%%%%%%%%%%%%%%%%%%%

%%%%%%%%%%%%%%%%%%%%%%%%%%%%%%%%%%%%%%%%%%%%%%%%%%%%%%%%%%%%%%%%%%%%%%
\begin{figure}[tb]
\begin{center}
\leavevmode
{\epsfxsize=5in\epsfbox{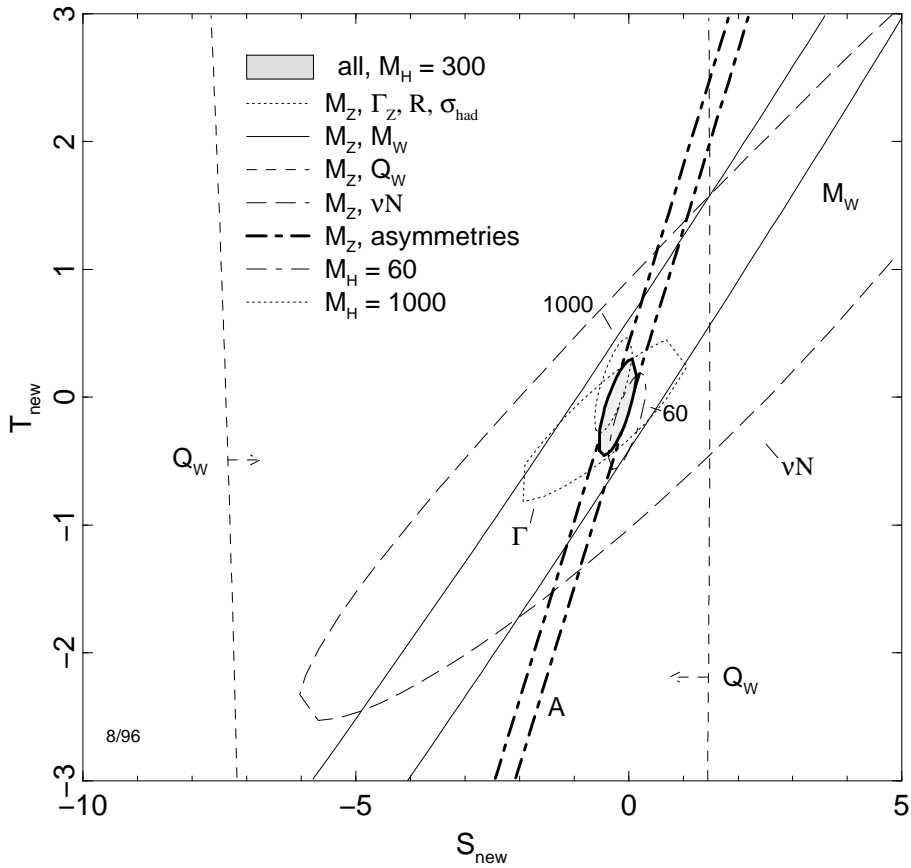}}
\end{center}
 \caption{Current determination of $S$ and $T$ by  
        a fit to the corpus of precision electroweak data, from 
 \protect\cite{LEPDG}.  The various ellipses show fits to a 
subset of the data, including the values of $\alpha$, $G_F$, and $\mz$
 plus those of one or
 several additional observables.} 
\label{fig:twentyeight}
\end{figure}
%%%%%%%%%%%%%%%%%%%%%%%%%%%%%%%%%%%%%%%%%%%%%%%%%%%%%%%%%%%%%%%%%%%%%%

By inserting the new physics contributions to the intermediate boson
propagators in weak interaction diagrams, we generate shifts from the
Standard Model predictions which are linear in $S$, $T$, and $U$.  For
example, the effective value of $\sstw$ governing the forward-backward
and polarization asymmetries at the $Z^0$ is shifted from its value
$(s^2)_\SM$, in the Minimal Standard Model, 
by 
\beq
    (s^2)_\eff - (s^2)_\SM = {\alpha\over c^2-s^2}
             \bigl[ {1\over 4} S - s^2 c^2 T \bigr] \ .
\eeq{eq:t6}
All of the standard observables except for $\mw$ and $\Gamma_W$ are
independent of $U$, and since $U$ is in any event expected to be small,
I will ignore it from here on.  In that case, any precision weak
interaction measurement restricts us to the vicinity of the line in the
$S$-$T$ plane.  The constraints that come from the measurements of
$(s^2)_{\mbox{eff}}$, $\mw$, and $\Gamma_Z$ are sketched in
Figure~\ref{fig:twentyseven}.  If these lines meet, they indicate
particular values of $S$ and $T$ which fit the deviations from the
Standard Model in the whole corpus of weak interaction data.
Figure~\ref{fig:twentyeight} shows such an $S$-$T$ fit to the data
available in the summer of 1996 \cite{LEPDG}.  The various curves show
fits to $\alpha$, $G_F$, $\mz$ plus a specific subset of the other
observables; the varying slopes of these constraints illustrate the
behavior shown in  Figure~\ref{fig:twentyseven}.

There is one important subtlety in the interpretation of the final
values of $S$ and $T$.  In determining the Minimal Standard Model reference
values for the fit, it is necessary to specify the  value of the top
quark mass and also a value for the mass of the Minimal Standard Model
Higgs boson. Raising $m_t$ gives the same physical effect as increasing
$T$; raising $m_H$ increases $S$ while slightly decreasing $T$.  Though
$m_t$ is known from direct measurements, $m_H$ is not.  The analysis of
Figure~\ref{fig:twentyeight} assumed $m_t = 175$ GeV, $m_H = 300$ GeV.
In comparing $S$ and $T$ to the predictions of technicolor models, it
is most straightforward to compute the difference between the
technicolor contribution to the vacuum polarization and that of a 1 TeV
Higgs boson. Shifting to this reference value, we have the experimental
constraint 
\beq
         S = -0.26 \pm 0.16 \ .
\eeq{eq:u6}
The negative sign indicates that there should be a smaller contribution
to the $W$ and $Z$ vacuum polarizations than that predicted by a 1 TeV
Standard Model Higgs boson.  This is in accord with the fact that a lower
value of the Higgs boson mass gives the best fit to the Minimal Standard
Model, as I have indicated in \leqn{eq:g2}.

In many models of new physics, the contributions to $S$ become small as
the mass scale $M$ increases, with the behavior $S \sim \mz^2/M^2$. 
This is the case, for example, in supersymmetry.  For example,
charginos of mass about 60 GeV can contribute to $S$ at the level of of
a few tenths of a unit, but heavier charginos have a negligible effect
on this parameter.  In  technicolor models, however, there is a new
strong interaction sector with resonances that can appear directly in
the $W$ and $Z$ vacuum polarizations. There is a concrete formula which
describes these effects.  Consider a technicolor theory with $SU(2)$
isospin global symmetry.  In such a theory, we can think about
producing hadronic resonances through $\ee$ annihilation. In the
standard parametrization, the cross section for $\ee$ annihilation to
hadrons through a virtual photon is given by the point cross section
for $\ee\to \mu^+\mu^-$ times a factor $R(s)$, equal asymptotically to
the sum of the squares of the quark charges.  Let $R_V(s)$ be the
analogous factor for a photon which couples to the isospin current
$J^{\mu 3}$ and so creates $I=1$ vector resonances only, and let
$R_A(s)$ be the factor for a photon which couples to the axial isospin
current $J^{\mu 5 3}$.  Then 
\beq
    S = {1\over 3\pi} \int^\infty_0 {ds\over s} \left[ R_V(s) - R_A(s) - 
           H(s)\right] \ ,
\eeq{eq:v6}
where $H(s) \approx \frac{1}{4}\theta(s-m_h^2)$ is the 
contribution of the Standard Model Higgs boson used to compute the 
reference value in \leqn{eq:t6}.
  In practice, this $H(s)$ gives  a small correction.  If one
evaluates $R_V$ and $R_A$ using the spectrum of QCD, scaled up
appropriately by the factor \leqn{eq:q}, one finds \cite{PandT}
\beq
      S = + 0.3 N_D {N_{TC}\over 3} \ ,
\eeq{eq:w6}
where $N_D$ is the number of weak doublets and $N_{TC}$ is the number
of technicolors.  Even for $N_D =1$ and $N_{TC} =3$, this is a
substantial positive value, one inconsistent with \leqn{eq:u6} at the 3
$\sigma$ level. Models with several technicolor weak doublets are in
much more serious conflict with the data.

These phenomenological problems of technicolor are challenging for the
theory, but they do not necessarily rule it out.  Holdom \cite{Holdom} 
has suggested a specific dynamical scheme which solves the first of these
three problems.  In estimating the scale of ETC interactions, we assumed that
the techniquark condensate falls off rapidly at high momentum, as the
quark condensate does in QCD.  If the techniquark mass term fell only
slowly at high momentum, ETC would have a larger influence at larger
values of $m_E$.  Then the flavor-changing direct effect of ETC on
light quark physics would be reduced. It is possible that such a
difference between technicolor and QCD  would also ameliorate the other
two problems I have discussed \cite{Appelquist}. In particular, if the
$J=1$ spectrum of technicolor models is not dominated
 by the low-lying $\rho$ and $a_1$ mesons, as is the case in QCD,
 there is a chance that
the vector and axial vector contributions to \leqn{eq:v6} would cancel
to a greater extent.

It is disappointing that theorists are unclear about the precise
predictions of technicolor models, but it is not surprising.
Technicolor relies on the presence of a new strongly-coupled gauge
theory. Though the properties of QCD at strong coupling now seem to be
well understood through numerical lattice gauge theory computations,
our understanding of strongly coupled field theories is quite
incomplete.  There is room for quantiatively and even qualitatively
different behavior, especially in theories with a large number of
fermion flavors. What the arguments in this section show is that
technicolor cannot be simply a scaled-up version of QCD.  It is a
challenge to theorists, though, to find the strong-interaction
 theory whose different
dynamical  behavior fixes the problems that extrapolation from 
 QCD would lead us to
expect.

\subsection{Direct probes of new strong interactions}

If the model-dependent constraints on technicolor have led us into a murky
theoretical situation, we should look for experiments that have a
directly, model-independent interpretation.  The guiding principle
of technicolor is that $SU(2)\times U(1)$ symmetry breaking is caused
by new strong interactions. We should be able to test this idea by
directly observing elementary particle reactions involving these new 
interactions.  In the next few sections, I will explain how these
experiments can be done.

In order to design experiments on new strong interactions, there are
two problems that we must discuss.  First,  the natural energy scale
for technicolor, and also for alternative theories with new strong
interactions, is of the order of 1 TeV.  Thus, to feel these
interactions, we will need to set up parton colllisions with energies
of order 1 TeV in the center of mass.  This energy range is well beyond
the capabilities of LEP 2 and the Tevatron, but it should be available
at the LHC and the $\ee$ linear collider.   Even for these facilities,
the experiments are challenging.  For the LHC, we will see that it
requires the full design luminosity.  For the linear collider, it
requires a center-of-mass energy of 1.5 TeV, at the top of the energy
range now under consideration.

%%%%%%%%%%%%%%%%%%%%%%%%%%%%%%%%%%%%%%%%%%%%%%%%%%%%%%%%%%%%%%%%%%%%%%
\begin{figure}[t]
\begin{center}
\leavevmode
\epsfbox{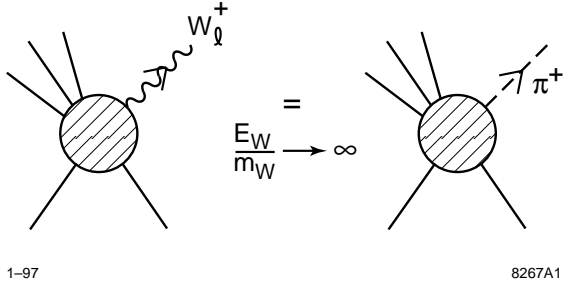}
\end{center}
 \caption{The Goldstone Boson Equivalence Theorem.}
\label{fig:twentynine}
\end{figure}
%%%%%%%%%%%%%%%%%%%%%%%%%%%%%%%%%%%%%%%%%%%%%%%%%%%%%%%%%%%%%%%%%%%%%%

Second, we need to understand which parton collisions we should study.
Among the particles that interact in high-energy collisions, do any
carry the new strong interactions?  At first it seems that all of the
elementary particles of collider physics are weakly coupled. But
remember that, in the models we are discussing,  the $W$ and $Z$ bosons
acquire their mass through their coupling to the new strong
interactions.  As  a part of the Higgs mechanism, these bosons, which
are massless and transversely polarized before symmetry breaking, pick
up longitudinal polarization states by combining with the Goldstone
bosons of the symmetry-breaking sector.  It is suggestive, then, that
at very high energy, the longitudinal polarization states of the $W$
and $Z$ bosons should show their origin and interact like the pions of
the strong interaction theory.  In fact, this correspondence can be
proved; it is called the Goldstone Boson Equivalence Theorem
\cite{GBET,GBETx,LQT,ChandG}.  The statement of the theorem is shown in
Figure~\ref{fig:twentynine}.

%%%%%%%%%%%%%%%%%%%%%%%%%%%%%%%%%%%%%%%%%%%%%%%%%%%%%%%%%%%%%%%%%%%%%%
\begin{figure}[t]
\begin{center}
\leavevmode
{\epsfxsize=3.75in\epsfbox{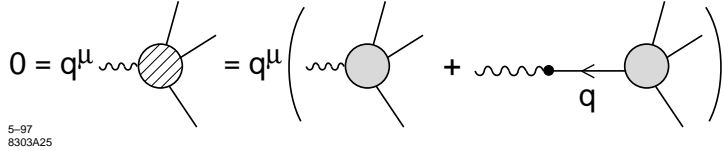}}
\end{center}
 \caption{Ward identity used in the proof of the Goldstone Boson 
Equivalence Theorem.}
\label{fig:thirty}
\end{figure}
%%%%%%%%%%%%%%%%%%%%%%%%%%%%%%%%%%%%%%%%%%%%%%%%%%%%%%%%%%%%%%%%%%%%%%

It is complicated to give a completely general proof of this theorem,
but it is not difficult to demonstate the simplest case. Consider a
process in which one $W$ boson is emitted. Since the $W$
couples to a conserved gauge current, the emission amplitude obeys a
Ward identity, shown in Figure~\ref{fig:thirty}.  We can analyze this
Ward identity as we did the analogous diagrammatic identity in
Figure~\ref{fig:twentyfive}.  The current which creates the $W$
destroys a state of the strong interaction theory; this is either a
massive state or a massless state consisting of one pion.  Call the
vertex from which the $W$ is created directly $\Gamma_W$, and call the
vertex for the creation of a pion $i\Gamma_\pi$. Then the Ward identity
shown in Figure~\ref{fig:thirty} reads 
\beq
    q_\mu \Gamma_W^\mu(q) + q_\mu \left( {g f_\pi q^\mu\over 2}\right)
     {i\over q^2} i \Gamma_\pi(q) = 0 \ .
\eeq{eq:x6}
Using \leqn{eq:h6}, this simplifies to 
\beq
      q_\mu \Gamma_W^\mu = \mw \Gamma_\pi \ .
\eeq{eq:y6}

To apply this equation, look at the explicit polarization vector
representing a vector boson of longitudinal polarization.  For a $W$
boson moving in the $\hat 3$ direction, $q^\mu = (E,0,0,q)$ with
$E^2-q^2 = \mw^2$, the longitudinal polarization vector is 
\beq
          \epsilon^\mu = \left( {q\over \mw}, 0 , 0 ,  {E\over \mw}
                    \right) \ .
\eeq{eq:z6}
This vector satisfies $\epsilon \cdot q = 0$.  At the same time, it
becomes increasingly close to $q^\mu/\mw$ as $E \to \infty$.  Because
of this, the contraction of $\epsilon^\mu$ with the first term in the
vertex shown in Figure~\ref{fig:thirty} is well approximated by
$(q_\mu/\mw)\Gamma_W^\mu$ in this limit, while at the same time the
contraction of $\epsilon^\mu$ with the pion diagram gives zero.  Thus,
$\Gamma_W$ is the complete amplitude for emission of a physical $W$
boson.  According to \leqn{eq:y6}, it satisfies 
\beq
          \epsilon_\mu \Gamma_W^\mu =  \Gamma_\pi
\eeq{eq:a7}
for $E \gg \mw$.  This is the precise statement of Goldstone boson
equivalence.

The Goldstone boson equivalence theorem tells us that the longitudinal
polarization states of $W^+$, $W^-$, and $Z^0$, studied in very high
energy reactions, are precisely the pions of the new strong
interactions. In the simplest technicolor models, these particles would
have the scattering amplitudes of QCD pions.  However, we can also
broaden our description to include more general models.  To do this, we
simply write the most general theory of pion interactions at energies
low compared to the new strong-interaction scale, and then reinterpret
the initial and final particles are longitudinally polarized weak
bosons.

This analysis is dramatically simplified by the observation we made below
\leqn{eq:t} that the new strong interactions should contain a global 
$SU(2)$ symmetry which remains exact when the weak interaction $SU(2)$ 
is spontaneously broken.  I explained
there that this symmetry is required to obtain the relation $\mw =
\mz\cos\theta_w$, which is a regularity of the weak boson mass
spectrum.   This unbroken symmetry shows up in technicolor models as
the manifest $SU(2)$ isospin symmetry of the techniquarks.

From here on, I will treat the pions of the new strong interactions as
 massless particles with an exact isospin $SU(2)$ symmetry.  The
pions form a triplet with $I=1$. Then a two-pion state has isospin 0,
1, or 2.  Using Bose statistics, we see that the three scattering
channels of lowest angular momentum are 
\beqa
         I = 0 &\quad& J= 0 \CR  I = 1 &\quad & J=1 \CR  
          I = 2 &\quad& J= 0 
\eeqa{eq:b7}
From here on, I will refer to these channels by their isospin value.
Using the analogy to the conventional strong interactions, it is
conventional to call a resonance in the $I=0$ channel a $\sigma$ and a
resonance in the $I=1$ channel a $\rho$ or techni-$\rho$.

Now we can describe the pion interactions by old-fashioned pion
scattering phenomenology \cite{Kallen}. As long we are at energies
sufficiently low that the process $\pi\pi \to 4 \pi$ is not yet
important, unitarity requires the scattering amplitude in the channel
$I$ to have the form 
\beq
    \M_I = 32\pi e^{i\delta_I} \sin \delta_I \cdot \cases{ 1 & $J=0$\cr
            3 \cos\theta & $J=1$\cr}\ ,
\eeq{eq:c7}
where $\delta_I$ is the phase shift in the channel $I$.  Since the
pions are massless, these can be expanded at low energy as 
\beq
    \delta_I = {s\over A_I} \left(1  + {s\over M_I^2} + \cdots \right)\ ,
\eeq{eq:d7}
where $A_I$ is the relativistic generalization of the scattering length
and $M_I$ similarly represents the effective range. The parameter $M_I$
is given this name because it estimates the position of the leading
resonance in  the channel $I$.  The limit $M_I\to \infty$ is called the
{\em Low Energy Theorem} (LET) model.

Because the pions are Goldstone bosons, it turns our that their
scattering lengths can be predicted in terms of the amplitude
\leqn{eq:b6} \cite{scleng}.  Thus, 
\beq
    A_I = \cases{ 16 \pi f_\pi^2 = (1.7\ \mbox{TeV})^2 & $I=0$\cr
 96 \pi f_\pi^2 = (4.3\ \mbox{TeV})^2 & $I=1$\cr
 -32 \pi f_\pi^2 & $I=2$\cr}\ .
\eeq{eq:e7}
Experiments which involve $WW$ scattering at
very high energy should give us the chance to observe these values of
$A_I$ and to measure the corresponding values of $M_I$.

The values of $A_I$ given in \leqn{eq:e7}
represent the basic assumptions about manifest and
spontaneously broken symmetry which are built into our analysis.  The
values of $M_I$, on the other hand, depend on the details of the
particular set of new strong interactions that Nature has provided. 
For example, in a technicolor model, the quark model of techicolor
interactions predicts that the strongest low-lying resonance should be
a $\rho$ ($I=1)$, as we see in QCD. In a model with strongly coupled
spin-0 particles, the strongest resonance would probably be a $\sigma$,
 an $I=0$
scalar bound state.  More generally, if we can learn
which channels have low-lying resonances and  what the masses of these
resonances are, we will have a direct experimental window into the
nature of the new interactions which break $SU(2)\times U(1)$.

\subsection{New strong interactions in $WW$ scattering}

How, then, can we create collisions of longitudinal $W$ bosons at TeV
center-of-mass energies?  The most straightforward method to create
high-energy $W$ bosons is to radiate them from  incident colliding
particles, either quarks at the LHC or electrons and positrons at the
linear collider.

%%%%%%%%%%%%%%%%%%%%%%%%%%%%%%%%%%%%%%%%%%%%%%%%%%%%%%%%%%%%%%%%%%%%%%
\begin{figure}[tb]
\begin{center}
\leavevmode
{\epsfysize=1.5in\epsfbox{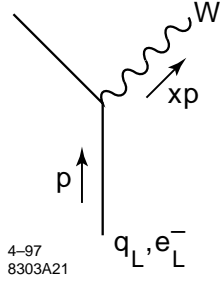}}
\end{center}
 \caption{Kinematics for the radiation of a longitudinal $W$ parton.}
\label{fig:thirtyone}
\end{figure}
%%%%%%%%%%%%%%%%%%%%%%%%%%%%%%%%%%%%%%%%%%%%%%%%%%%%%%%%%%%%%%%%%%%%%%

The flux of $W$ bosons associated with a proton or electron beam can be
computed by methods similar to those used to discuss parton evolution
in QCD \cite{PS,Dawson}.  We imagine that the $W$ bosons are emitted from the
incident fermion lines and come together in a collision process with
momentum transfer $Q$.  The kinematics of the emission process is shown
in Figure~\ref{fig:thirtyone}. The emitted bosons are produced with a
spectrum in  longitudinal momentum, parametrized by the quantity $x$,
the {\em longitudinal fraction}.  They also have a spectrum in
transverse momentum $p_\perp$. The emitted $W$ boson is off-shell, but
this can be ignored to a first approximation if $Q$ is much larger than
$(\mw^2 + p_\perp^2)^{1/2}$.  In this limit, the distribution of the
emitted $W$ bosons is described by relatively simple formulae.  Note
that an incident $d_L$ or $e^-_L$ can radiate a $W^-$, while an
incident $u_L$ or $e^+_R$ can radiate a $W^+$.

The distribution of transversely polarized $W^-$ bosons emitted from an
incident $d_L$ or $e^-_L$ is given by 
\beq
 \int dx\, 
 f_{e^-_L \to W^-_{tr} }(x,Q) = \int^{Q^2}_{0} {d p_\perp^2\over 
 p_\perp^2 + \mw^2}
         \int {dx\over x} \ {\alpha\over 4\pi s^2} {1 + (1-x)^2\over x} \ ,
\eeq{eq:f7}
where, as before, $s^2 = \sstw$. The integral over transverse momenta
gives an enhancement factor of $\log Q^2/\mw^2$, analogous to the
factor $\log s/m_e$ which appears in the formula for radiation of
photons in electron scattering processes. The distribution of
longitudinally polarized $W^-$ bosons has a somewhat different
structure, 
\beqa
 \int dx\, 
 f_{e^-_L \to W^-_{long} }(x,Q) &=& \int^{Q^2}_{0} {d p_\perp^2 \mw^2\over
 (p_\perp^2 + \mw^2)^2 }
         \int {dx\over x} \ {\alpha\over 2\pi s^2} {1-x\over x} \CR
       & = &   \int {dx\over x} \ {\alpha\over 2\pi s^2} {1-x\over x}     \ .
\eeqa{eq:g7}
This formula does not show the logarithmic distribution in $p_\perp$
seen in \leqn{eq:f7}; instead, it produces longitudinally polarized $W$
bosons at a characteristic $p_\perp$ value of order $\mw$.

\begin{figure}[t]
\begin{center}
\leavevmode
{\epsfxsize=4in\epsfbox{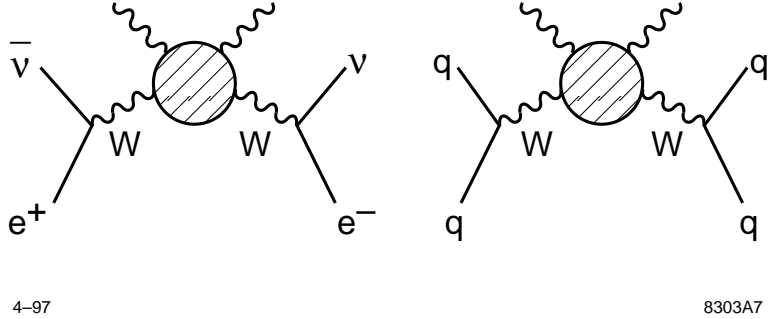}}
\end{center}
 \caption{Collider processes which involve $WW$ scattering.}
\label{fig:thirtytwo}
\end{figure}

When both beams radiate longitudinally polarized $W$ bosons, we can
study boson-boson scattering through the reactions shown in
Figure~\ref{fig:thirtytwo}.  In $pp$ reactions one can in principle
study all modes of $WW$ scattering, though the most complete simulations
have been done for the especially clean $I=2$ channel, $W^+W^+\to
W^+W^+$.  In $\ee$ collisions, one is restricted to the channels
$W^+W^- \to W^+W^-$ and $W^+W^- \to  Z^0 Z^0$.
 The diagrams in which a longitudinal
$Z^0$ appears in the initial state are suppressed by the small $Z^0$
coupling to the electron 
\beq
         { g^2(e^-_L \to e^-_L Z^0)\over g^2(e^-_L \to \nu W^-)} = 
      \left((\half - s^2)/cs\over 1/\sqrt{2}s \right)^2 = 0.2 \ . 
\eeq{eq:h7}
The $I=2$ process $W^-W^- \to W^-W^-$ could be studied in a dedicated
$e^-e^-$ collision experiment.

%%%%%%%%%%%%%%%%%%%%%%%%%%%%%%%%%%%%%%%%%%%%%%%%%%%%%%%%%%%%%%%%%%%%%%
%%%%%%%%%%%%%%%%%%%%%%%%%%%%%%%%%%%%%%%%%%%%%%%%%%%%%%%%%%%%%%%%%%%%%%

I will now briefly discuss the experimental strategies for observing
these reactions in the LHC and linear collider environments and present
some simulation results.  In the $pp$ reactions, the most important
background processes come from the important high transverse momentum
QCD processes which, with some probability, give final states that
mimic $W$ boson pairs.  For example, in the process $gg\to gg$ with a
momentum transfer of 1~TeV, each final gluon typically radiates gluons
and quarks before final hadronization, to produce a system of hadrons
with of order 100 GeV.  When the mass of this system happens to be
close to the mass of the $W$, the process has the characteristics of
$WW$ scattering. Because of the overwhelming rate for $gg\to gg$, all
studies of $WW$ scattering at hadron colliders have restricted
themselves to detection of one or both weak bosons in leptonic decay
modes.   Even with this restriction, the process $gg \to t\bar t$
provides a background of isolated lepton pairs at high transverse
momentum.  This background and a similar one from $q\bar q \to W +$
jets, with jets faking leptons, are controlled by requiring some
further evidence that the initial $W$ bosons are color-singlet systems
radiated from quark lines. To achieve this, one could require a forward
jet associated with the quark from which the $W$ was radiated, or a low
hadronic activity in the central rapidity region, characteristic of the
collision of color-singlet species.

%%%%%%%%%%%%%%%%%%%%%%%%%%%%%%%%%%%%%%%%%%%%%%%%%%%%%%%%%%%%%%%%%%%%%%
\begin{figure}[t]
\begin{center}
\leavevmode
{\epsfxsize=4.0in
\epsfbox{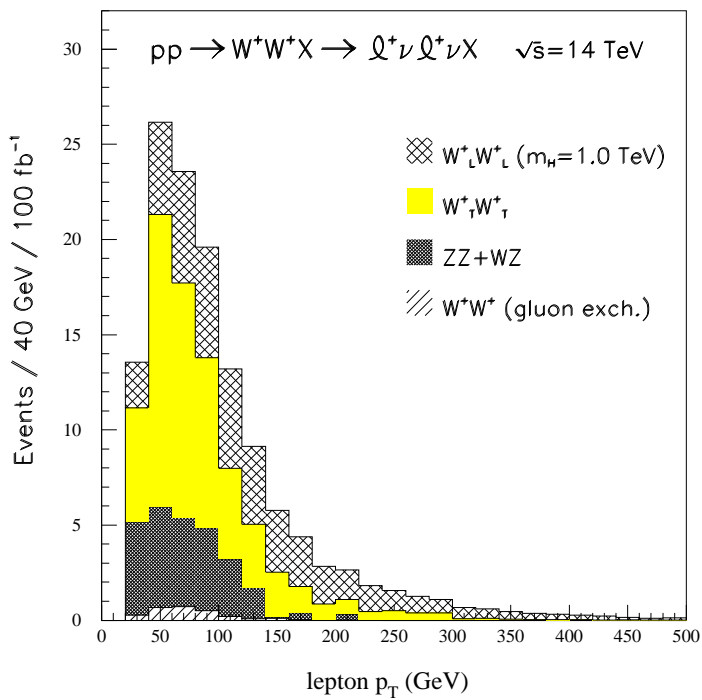}}
\end{center}
 \caption{Expected numbers of $W^+W^+ \to (\ell \nu)(\ell  \nu)$ events due
to signal and background processes, after all cuts, for a 100 fb$^{-1}$
 event sample at the LHC, from \protect\cite{Atlas}. The signal corresponds
to a Higgs boson of mass 1~TeV.}
\label{fig:thirtythree}
\end{figure}
%%%%%%%%%%%%%%%%%%%%%%%%%%%%%%%%%%%%%%%%%%%%%%%%%%%%%%%%%%%%%%%%%%%%%%

Figure~\ref{fig:thirtythree} shows a simulation by the ATLAS
collaboration of a search for new strong interactions in $W^+W^+$
scattering \cite{Atlas}. In this study, both $W$ bosons were assumed to
be observed in their leptonic decays to $e$ or $\mu$, and a forward jet
tag was required. The signal corresponds to a model with a 1~TeV Higgs
boson, or, in our more general terminology, a 1~TeV $I=0$ resonance. 
The size of the signal is a few tens of events in a year of running at
the LHC at high luminosity. Note that the experiment admits a
substantial background from various sources of tranversely polarized
weak bosons.  Though there is a significant excess above the Standard
Model expectation, the signal is not distinguished by a resonance peak,
and so it will be important to find experimental checks that the
backgrounds are correctly estimated. An illuminating study of the other
important reaction $pp \to Z^0 Z^0 + X$ is given in \cite{Baggerscrew}.

%%%%%%%%%%%%%%%%%%%%%%%%%%%%%%%%%%%%%%%%%%%%%%%%%%%%%%%%%%%%%%%%%%%%%%
\begin{figure}[p]
\begin{center}
\leavevmode
{\epsfxsize=4.0in\epsfbox{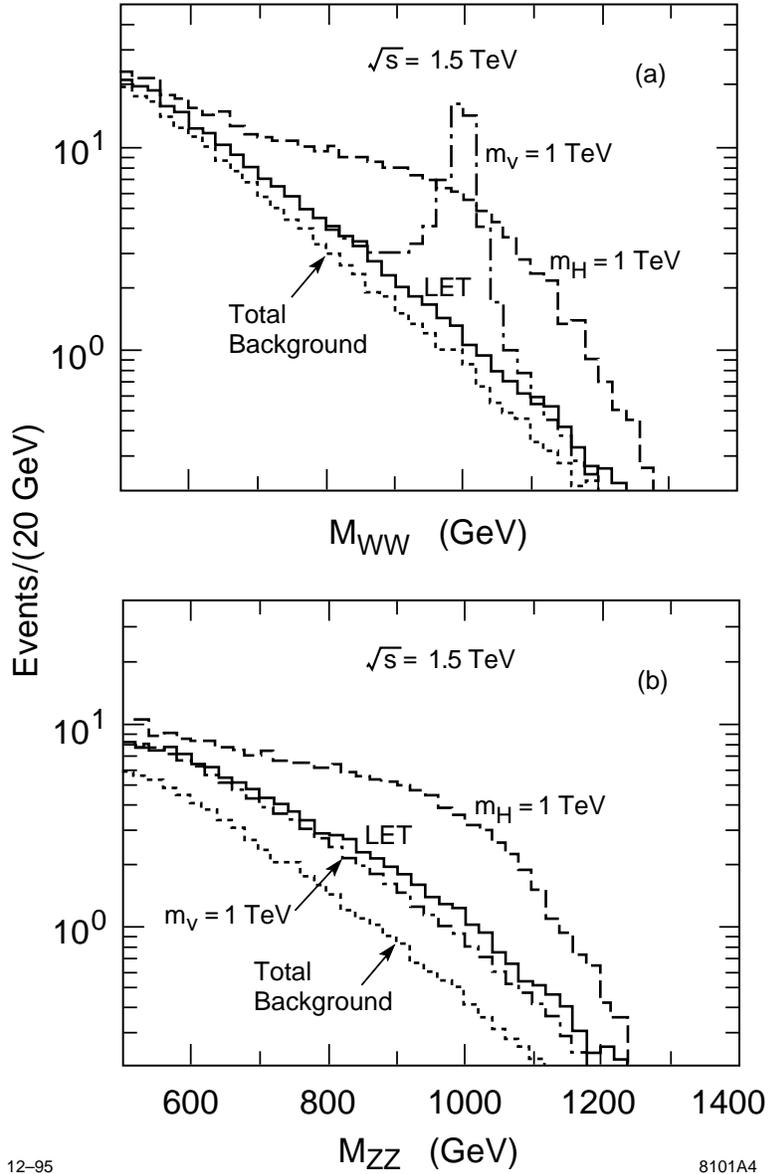}}
\end{center}
 \caption{Expected numbers of $W^+W^-$ and $ZZ \to 4$ jet events due
to signal and background processes, after all cuts, for a 200 fb$^{-1}$
 event sample at an $\ee$ linear collider at 1.5 TeV in the 
center of mass, from \protect\cite{Bargerscrew}.  Three different models
for the signal are compared to the Standard Model background.}
\label{fig:thirtyfour}
\end{figure}
%%%%%%%%%%%%%%%%%%%%%%%%%%%%%%%%%%%%%%%%%%%%%%%%%%%%%%%%%%%%%%%%%%%%%%

The $WW$ scattering experiment is also  difficult at an $\ee$ linear
collider.  A center of mass energy well above 1 TeV must be used, and
again the event rate is a few tens per year at high luminosity. The
systematic problems of the measurement are different, however, so that
the $\ee$ results might provide important new evidence even if a small
effect is first seen at the LHC.  In the $\ee$ environment, it is
possible to identify the weak bosons in their hadronic decay modes, and
in fact this is necessary to provide sufficient rate.  Since the
hadronic decay captures the full energy-momentum of the weak boson, the
total momentum vector of the boson pair can be measured.  This, again,
is fortunate, because the dominant backgrounds to $WW$ scattering
through new strong interactions come from the photon-induced processes
$\gamma\gamma\to W^+ W^-$ and $\gamma e \to Z W \nu$.  The first of
these backgrounds can be dramatically reduced by insisting that the
final two-boson system has  a transverse momentum between 50 and 300
GeV, corresponding to the phenomenon we noted in \leqn{eq:g7} that
longitudinally polarized weak bosons are typically emitted with a
transverse momentum of order $\mw$.  This cut should be accompanied by
a forward electron and positron veto to remove processes with an
initial photon which has been radiated from one of the fermion lines.

The expected signal and background after cuts, in $\ee \to \nu\bar\nu
W^+ W^-$ and $\ee \to \nu\bar\nu Z^0 Z^0$, at a center-of-mass energy
of 1.5 TeV, are shown in Figure~\ref{fig:thirtyfour}
\cite{Bargerscrew}.  The signal is  shown for a number of different
models and is compared to the Standard Model expectation for
transversely polarized boson pair production.  In the most favorable
cases of 1~TeV resonances in the $I=0$ or $I=1$ channel, resonance
structure is apparent in the signal, but in models with higher
resonance masses one must again rely on observing an enhancement over
the predicted Standard Model backgrounds.  At an $\ee$ collider, one
has the small advantage that these backgrounds come from electroweak
processes and can therefore be precisely estimated.

Recently, it has been shown that the process $WW \to t\bar t$ can be
observed at an $\ee$ linear collider at 1.5 TeV \cite{barklowtt}. This
reaction probes the involvement of the top quark in the new strong
interactions.  If the $W$ and top quark masses have a common origin,
the same resonances which appear in $WW$ scattering should also appear
in this reaction.  However, some models, for example, Hill's topcolor
\cite{topcolor}, attribute the top quark mass to interactions specific
to the third generation which lead to top pair condensation.  The study
of $WW\to t\bar t$ can directly address this issue experimentally.

\subsection{New strong interactions in $W$ pair-production}

\begin{figure}[tb]
\begin{center}
\leavevmode
\epsfbox{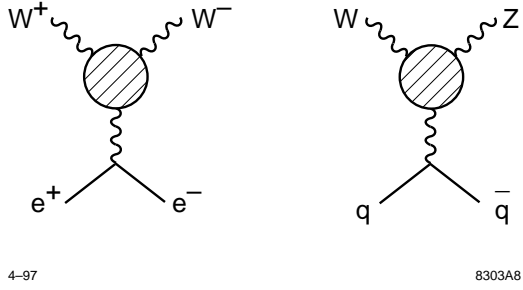}
\end{center}
 \caption{Collider processes which involve  vector boson pair-production.}
\label{fig:thirtyfive}
\end{figure}

In addition to providing direct $WW$ scattering processes, new strong
interactions can affect collider processes by creating a resonant
enhancement of fermion pair annihilation in to weak bosons. The most
important reactions for studying this effect are shown in
Figure~\ref{fig:thirtyfive}.  As with the processes studied in Section
5.4, these occur both in the $pp$ and $\ee$ collider environment.

%%%%%%%%%%%%%%%%%%%%%%%%%%%%%%%%%%%%%%%%%%%%%%%%%%%%%%%%%%%%%%%%%%%%%%
%%%%%%%%%%%%%%%%%%%%%%%%%%%%%%%%%%%%%%%%%%%%%%%%%%%%%%%%%%%%%%%%%%%%%%

The effect is easy to understand by a comparison to the familiar strong
interactions.   In the same way that the boson-boson scattering
processes described in the previous section were analogous to pion-pion
scattering, the strong interaction enhancement of $W$ pair production
is analogous to the behavior of the pion form factor. We might
 parametrize the
enhancement of the amplitude for fermion pair annihilation into
longitudinally polarized $W$ bosons by a form factor $F_\pi(q^2)$.  In
QCD, the pion form factor receives a large enhancement from the $\rho$
resonance. Similarly, if the new strong interactions contain a strong
$I=1$ resonance, the amplitude for longitudinally polarized $W$ pair
production should be multiplied by the factor 
\beq
      F_\pi(q^2) = { - M_1^2 + i M_1 \Gamma_1 \over 
 q^2- M_1^2 + i M_1 \Gamma_1 } \ ,
\eeq{eq:i7}
where $M_1$ and $\Gamma_1$ are the mass and width of the resonance.  If
there is no strong resonance, the new strong interactions still have an
effect on this channel, but it may be subtle and difficult to detect. 
A benchmark is that the phase of the new pion form factor is related to
the pion-pion scattering phase shift in the $I=1$ channel, 
\beq
      \arg F_\pi(s) = \delta_1(s) \ ;
\eeq{eq:j7}
this result is true for any strong-interaction model as long as $\pi\pi
\to 4 \pi$ processes are not important at the given value of $s$
\cite{BjD}.

\begin{figure}[tb]
\begin{center}
\leavevmode
{\epsfxsize=5in\epsfbox{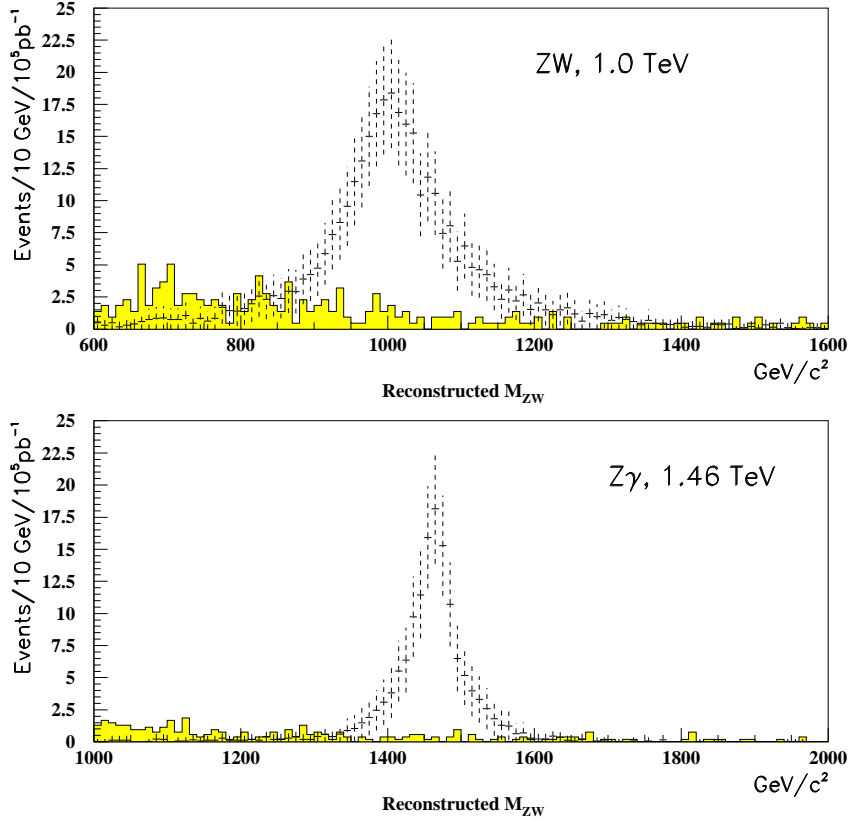}}
\end{center}
\caption{Reconstructed masses at the LHC
for new strong interaction resonances 
decaying into gauge boson pairs, from \protect\cite{Atlas}:
(a) a 1~TeV techni-$\rho$ resonance decaying into $WZ$ and observed in the 
3-lepton final state; (b) a 1.46~TeV techni-$\omega$ decaying into $\gamma Z$
and observed in the $\gamma \ell^+\ell^-$ final state.}
\label{fig:thirtynine}
\end{figure}

At the LHC, an $I=1$ resonance in the new strong interactions can be
observed as an enhancement in $pp\to W Z + X$, with both $W$ and $Z$
decaying to leptons, as long as the resonance is sufficiently low in
mass that its peak occurs before the $q\bar q$ luminosity spectrum cuts
off.  The ATLAS collaboration has demonstrated a sensitivity up to
masses of 1.6 TeV~\cite{wulz}. The signal for a 1 TeV resonance is quite
 dramatic,
as demonstrated in Figure~\ref{fig:thirtynine}.

%%%%%%%%%%%%%%%%%%%%%%%%%%%%%%%%%%%%%%%%%%%%%%%%%%%%%%%%%%%%%%%%%%%%%%
%%%%%%%%%%%%%%%%%%%%%%%%%%%%%%%%%%%%%%%%%%%%%%%%%%%%%%%%%%%%%%%%%%%%%%

Also shown in this figure is an estimate of a related effect that
appears in some but not all models, the production of an $I=0$, $J=1$
resonance analogous to the $\omega$ in QCD, which then decays to 3 new
pions or to $\pi \gamma$.   Though the first of these modes is not
easily detected at the LHC, the latter corresponds to the final state
$Z^0 \gamma$, which can be completely reconstructed if the $Z^0$ decays
to $\ell^+\ell^-$.

At an $\ee$ collider, the study of the new pion form factor can be
carried a bit farther.  The process $\ee\to W^+ W^-$ is the most
important single process in high-energy $\ee$ annihilation, with a
cross section greater than that for all annihilation processes to quark
pairs.  If one observes this reaction in the topology in which one $W$
decays hadronically and the other leptonically, the complete event can
be reconstructed including the signs of the $W$ bosons.  The $W$ decay
angles contain information on the boson polarizations.  So it is
possible to measure the pair production cross section to an accuracy of
a few percent, and also to extract the contribution from $W$ bosons
with longitudinal polarization.  The experimental techniques for this
analysis have been reviewed in \cite{barklowM}.

%%%%%%%%%%%%%%%%%%%%%%%%%%%%%%%%%%%%%%%%%%%%%%%%%%%%%%%%%%%%%%%%%%%%%%
\begin{figure}[t]
\begin{center}
\leavevmode
{\epsfxsize=3.5in\epsfbox{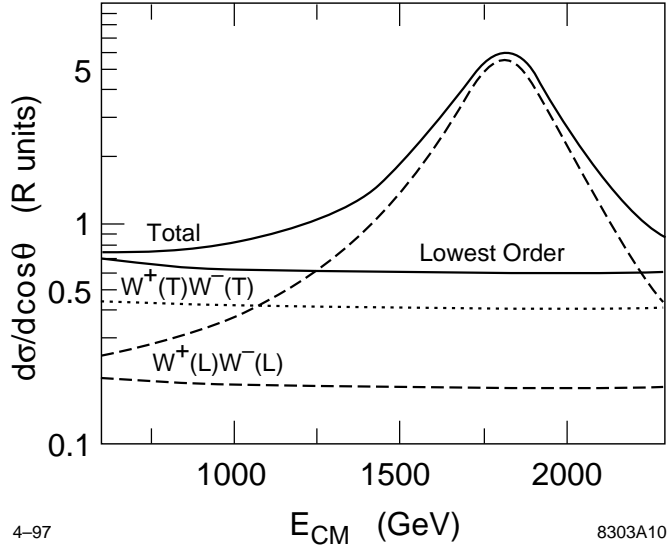}}
\end{center}
 \caption{Technirho resonance effect on the differential cross section
      for $\ee\to W^+W^-$   at $\cos\theta = -0.5$.  The figure shows
        the effect on the various $W$ polarization states.} 
\label{fig:thirtyseven}
\end{figure}
%%%%%%%%%%%%%%%%%%%%%%%%%%%%%%%%%%%%%%%%%%%%%%%%%%%%%%%%%%%%%%%%%%%%%%

\begin{figure}[tb]
\begin{center}
\leavevmode
\epsfbox{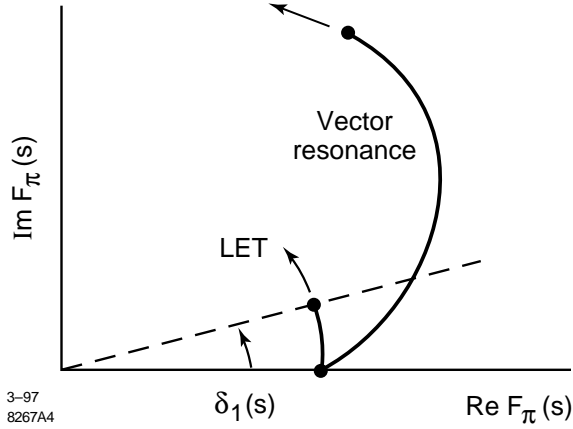}
\end{center}
 \caption{Dependence of $F_\pi(s)$ on energy, in models without and 
       with a new strong interaction resonance in the $I=J=1$ channel.}
\label{fig:thirtysix}
\end{figure}

Because an $I=1$ resonance appears specifically in the pair-production
of longitudinally polarized $W$ bosons, the resonance peak in the cross
section has associated with it an effect in the $W$ polarizations which
is significant even well below the peak.  This effect is seen in
Figure~\ref{fig:thirtyseven}, which shows the differential cross
section for $W$ pair production at a fixed angle as a function of
center-of-mass energy, in a minimal technicolor model with the $I=1$
technirho resonance at 1.8 TeV.  By measuring the amplitude for
longitudinal $W$ pair production accurately, then, it is possible to
look for $I=1$ resonances which are well above threshold.  In addition,
measurement of 
the interference between the transverse and longitudinal $W$ pair
production amplitudes 
allows one to determine the phase of the new pion form
 factor~\cite{barklowM}.  This effect is present even in models with no
resonant behavior, simply by virtue of the relation \leqn{eq:j7} and
the model-independent leading term in \leqn{eq:d7}.
Figure~\ref{fig:thirtysix} shows the behavior of the new pion form
factor as an amplitude in the complex plane as a function of the
center-of-mass energy in the nonresonant and resonant cases.

%%%%%%%%%%%%%%%%%%%%%%%%%%%%%%%%%%%%%%%%%%%%%%%%%%%%%%%%%%%%%%%%%%%%%%
%%%%%%%%%%%%%%%%%%%%%%%%%%%%%%%%%%%%%%%%%%%%%%%%%%%%%%%%%%%%%%%%%%%%%%

%%%%%%%%%%%%%%%%%%%%%%%%%%%%%%%%%%%%%%%%%%%%%%%%%%%%%%%%%%%%%%%%%%%%%%
\begin{figure}[tb]
\begin{center}
\leavevmode
\epsfbox{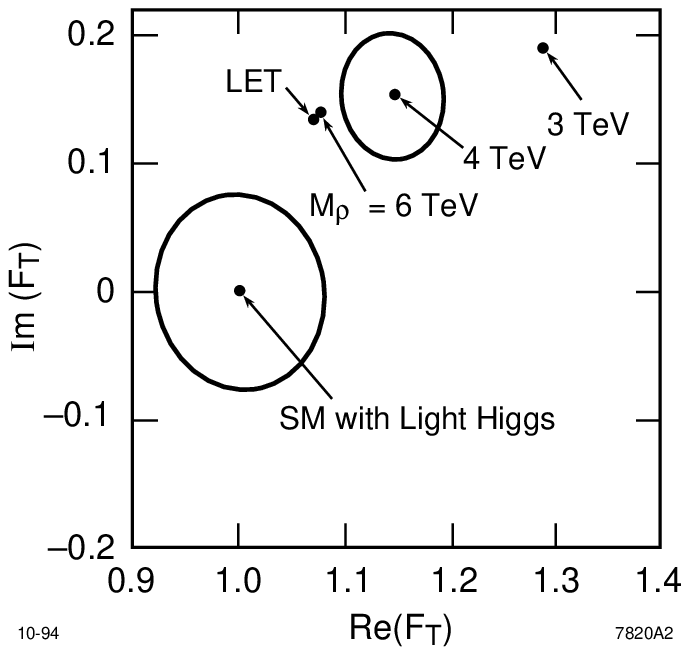}
\end{center}
 \caption{Determination of the new pion form factor an an $\ee$ linear
collider at 1.5 TeV with an unpolarized data sample of 200 fb$^{-1}$, 
 from \protect\cite{barklowM}.  The  simulation results are compared to 
 model with a high-mass $I=1$ resonance and the model-independent
 contribution to pion-pion scattering.  The contour about the 
  light Higgs point (with no new strong interactions) is a 95\% confidence
 contour; that about the point $M = 4$ TeV is a 68\% confidence contour.}
\label{fig:thirtyeight}
\end{figure}
%%%%%%%%%%%%%%%%%%%%%%%%%%%%%%%%%%%%%%%%%%%%%%%%%%%%%%%%%%%%%%%%%%%%%%

The expectations for the measurement of the new pion form factor at a
1.5 TeV linear collider, from simulation results of Barklow
\cite{barklowM}, are show in Figure~\ref{fig:thirtyeight}.  The
estimated sensitivity of the measurement is compared to the
expectations from a model incorporating the physics I have just
described~\cite{mycoll}. A nonresonant model with scattering in the
$I=1 $ channel given only by the scattering length term in \leqn{eq:d7}
is already distinguished from a model with no new strong interactions
at the 4.6 $\sigma$ level, mainly by the measurement of the imaginary
part of $F_\pi$. In addition, the measurement of the resonance effect
\leqn{eq:i7} in the real part of $F_\pi$ can distinguish the positions
of $I=1$ resonances more than a factor two above the collider
center-of-mass energy.

\subsection{Overview of $WW$ scattering experiments}

It is interesting to collect together and summarize the various probes
for resonances in the new strong interactions that I have described in
the previous two sections. I have described both direct studies of $WW$
scattering processes and indirect searches for resonances through their
effect on fermion annihilation to boson pairs.  With the LHC and the
$\ee$ linear collider, these reactions would be studied in a number of
channels spanning all of the cases listed in \leqn{eq:b7}.  Of course,
with fixed energy and luminosity, we can only probe so far into each
channel.  It is useful  to express this reach quantitatively and to ask
whether it should give a sufficient picture of the resonance structure
that might be found.

There is a well-defined way to estimate how far one must reach to 
have interesting sensitivity to new resonances. The model-independent
lowest order expressions for the $\pi\pi$ scattering amplitudes
\beq
    \M_I = 32\pi e^{i\delta_I} {s\over A_I} \cdot \cases{ 1 & $J=0$\cr
            3 \cos\theta & $J=1$\cr}\ ,
\eeq{eq:k7}
violate unitarity when $s$ becomes sufficiently large, and this gives
a criterion for the value of $s$ by which new resonances must appear
\cite{LQT}.  The unitarity violation begins for $s = A_I/2$; with the 
values of the $A_I$ given in \leqn{eq:e7}, we find the bounds
\beq
I=0\ : \quad \sqrt{s} < 1.3 \ \mbox{TeV} \ , \qquad
I=1\ : \quad \sqrt{s} < 3.0 \ \mbox{TeV} \ .
\eeq{eq:l7}
For comparison, if we scale up the QCD resonance masses by the factor
\leqn{eq:q}, we find a techni-$\rho$ mass of 2.0~TeV, well below the 
the $I=1$ unitarity bound given in \leqn{eq:l7}.  It is 
interesting to compare these goals to the reach expected for the 
experiments we have described.

One of the working groups at the recent Snowmass summer study addressed
the question of estimating the sensitivity to new strong interaction
resonances in each of the boson-boson scattering  channels that will be
probed by  the high-energy colliders \cite{Persis}. Their results are
reproduced in Table 1.  Results are given for experiments at the LHC
and at a 1.5 TeV $\ee$ linear collider, with luminosity samples of 100
fb$^{-1}$ per experiment.  The method of the study was to use 
simulation data from the literature
to estimate the sensitivity to the parameters $M_I$ in
\leqn{eq:d7}, allowing just this one degree of freedom per channel. 
Situations with multiple resonances with coherent or cancelling effects
were not considered. Nevertheless, the determination of these basic 
parameters should give a
general qualitative picture of the new strong
interactions. The estimates of the sensitivity to these 
parameters go well beyond the goals set in
\leqn{eq:l7}.

%%%%%%%%%%%%%%%%%%%%%%%%%%%%%%%%%%%%%%%%%%%%%%%%%%%%%%%%%%%%%%%%%%%%%%%%%%%
\begin{table*}[ht]
\begin{center}
\caption[*]{LHC and linear collider (`NLC') sensitivity to resonances
in the new strong interactions, from \protect\cite{Persis}. `Reach'
gives the value of the resonance mass corresponding to an enhancement
of the cross section for boson-boson scattering at the 95\%\ confidence
level obtained in Section VIB2. `Sample' gives a representative set of
errors for the determination of a resonance mass from this enhancement.
`Eff. $\L$ Reach' gives the estimate of the resonance mass for a 95\%\
confidence level enhancement. All of these estimates are based on
simple parametrizations in which a single resonance dominates the
scattering cross section.\bigskip}
\label{tab:Peskin}
\begin{tabular}{cccccc}
\hline
\hline
Machine & Parton Level Process & I & Reach & Sample & Eff. $\L$ Reach \\ 
\hline \\
LHC & $qq' \to qq'ZZ$ & 0 & 1600 & $1500^{+100}_{-70}$& 1500 \\ \\
LHC & $q \bar q \to WZ$ & 1 & 1600 & $1550^{+50}_{-50}$ & \\ \\
LHC & $qq' \to qq'W^+W^+$ & 2 & 1950 & $2000^{+250}_{-200}$&  \\ \\
NLC & $e^+e^- \to \nu \bar \nu ZZ$ & 0 & 1800 & $1600^{+180}_{-120}$&
 2000 \\ \\
NLC & $e^+e^- \to \nu \bar \nu t \bar t$ & 0 & 1600 & $1500^{+450}_{-160}$&
\\ \\
NLC & $e^+e^- \to W^+W^-$ & 1 & 4000 & $3000^{+180}_{-150}$ \\ \\
\hline
\hline
\end{tabular}
\end{center}
\end{table*}
%%%%%%%%%%%%%%%%%%%%%%%%%%%%%%%%%%%%%%%%%%%%%%%%%%%%%%%%%%%%%%%%%%%%%%%%%
 
If new strong interactions are found, further experiments at higher
energy will be necessary to characterize them precisely.  Eventually,
we will need to work out the detailed hadron  spectroscopy of these new
interactions, as was done a generation ago for QCD.  Some techniques
for measuring this spectrum seem straightforward if the high energy
accelerators will be available.  For example, one could measure the
spectrum of $J=1$ resonances from the cross section for $\ee$ or
$\mu^+\mu^-$ annihilation to multiple longitudinal $W$ and $Z$ bosons.  I
presume that there are also elegant spectroscopy experiments that can
be done in high-energy $pp$ collisions, though these have not yet been
worked out.  It may be interesting to 
think about this question.
 If the colliders of the next generation do
discover these new strong interactions, the new spectroscopy  will be
 a central issue of particle physics twenty years from now.

\subsection{Observable effects of extended technicolor}

Beyond these general methods for observing new strong interactions,
which apply to any model in which electroweak symmetry breaking has a
strong-coupling origin, each specific model leads to its own
model-dependent predictions. Typically, these predictions can be tested
at energies below the TeV scale, so they provide phenomena that can
be explored before the colliders of the next generation reach their
ultimate energy and luminosity. On the other hand, these predictions
are specific to their context. Excluding one such  phenomenon rules out
a particular model but not necessarily the whole class of
strong-coupled theories.  We have seen an example of this already in
Section 5.2, where the strong constraints on technicolor models from
precision electroweak physics force viable models to have particular
dynamical behavior but do not exclude these models completely.

In this section, I would like to highlight three such predictions
specifically associated with technicolor theories.  These three
phenomena illustrate the range of possible effects that might be found.
A systematic survey of the model-dependent predictions of models of
strongly-coupled electroweak symmetry breaking is given in
\cite{Persis}.

All three of these predictions are associated with the extended
technicolor mechanism of quark and lepton mass generation described at
the end of Section 5.1 and in Figure~\ref{fig:twentysix}. To see the first
prediction, 
 note from the figure that the Standard Model quantum numbers of
the external fermion must be carried either by the techniquark or by
the ETC gauge boson.  The simplest possibility is to assign the
techniquarks the quantum numbers of a generation of quarks and leptons
\cite{FS}. Call these fermions $(U,D,N,E)$.  The pions of the
technicolor theory, the Goldstone bosons of spontaneously broken chiral
$SU(2)$, have the quantum numbers 
\beq
       \pi^+ \sim  \bar U \gamma^5 D + \bar N \gamma^5 E \ , 
    \pi^0 \sim   \bar U \gamma^5 U - \bar D \gamma^5 D + \bar N \gamma^5 N 
 - \bar E \gamma^5 E \ . 
\eeq{eq:m7}
But the theory contains many more pseudoscalar mesons.  In fact, in the
absence of the coupling to $SU(3)\times SU(2)\times U(1)$, the model
has the global symmetry $SU(8)\times SU(8)$ (counting each techniquark
as three species), which would be spontaneously broken to a vector
$SU(8)$ symmetry by dynamical techniquark mass generation.  This would
produce an $SU(8)$ representation of Goldstone bosons, 63 in all.  Of
these, three are the Goldstone bosons eaten by the $W^\pm$ and $Z^0$ in
the Higgs mechanism.  The others comprise four color singlet bosons,
for example, 
\beq
      P^+ \sim  \frac{1}{3} \bar U \gamma^5 D - \bar N \gamma^5 E \ , 
\eeq{eq:n7}
four color triplets, for example,
\beq
      P_3 \sim \bar U \gamma^5 E \ ,
\eeq{eq:o7}
and four color octets, for example, 
\beq
      P_8^+ \sim \bar U \gamma^5 t^a D \ ,
\eeq{eq:oo7}
where $t^a$ is a $3\times 3$ $SU(3)$ generator.  These additional
particles are known as {\em pseudo-Goldstone bosons} or, more simply,
{\em technipions}.

Phenomenologically, the technipions resemble Higgs bosons with the same
Standard Model  quantum numbers.  They are produced in $\ee$
annihilation at the same rate as for pointlike charged bosons. The idea
of Higgs bosons with nontrivial color is usually dismissed in studies
of the Higgs sector because this structure is not `minimal'; however,
we see that these objects appear naturally from the idea of
technicolor.  The colored objects are readily pair-produced at proton
colliders, and the neutral isosinglet color-octet state can also be
singly produced through gluon-gluon fusion \cite{LaneColl}.

The masses of the technipions arise from Standard Model radiative
corrections and from ETC interactions; these are expected to be of the
order of a few hundred GeV.  Technipions decay by a process in which
the techniquarks exchange an ETC boson and convert to ordinary quarks
and leptons.  This decay process favors decays to heavy flavors, for
example, $P^+_8 \to \bar t b$.  In this respect, too, the technipions
resemble Higgs bosons of a highly nonminimal Higgs sector resulting
from an underlying composite structure.

%%%%%%%%%%%%%%%%%%%%%%%%%%%%%%%%%%%%%%%%%%%%%%%%%%%%%%%%%%%%%%%%%%%%%%
\begin{figure}[t]
\begin{center}
\leavevmode
{\epsfxsize=4.5in\epsfbox{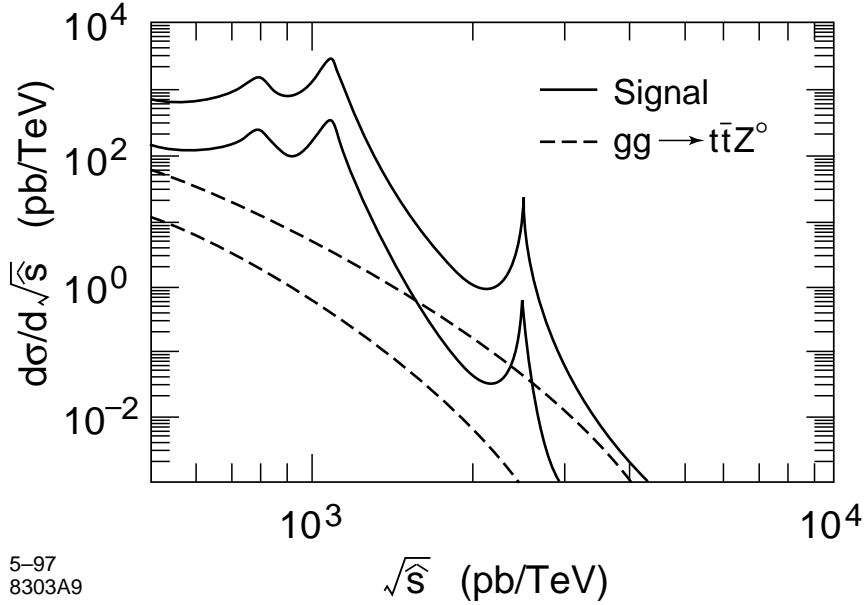}}
\end{center}
 \caption{Cross section for the production of ETC boson pair states in 
       $pp$ collisions, from \protect\cite{AandW}.  The $\bar E E$ states
       are observed as $t \bar t Z^0$ systems of definite 
        invariant mass.  The two
       sets of curves correspond to signal and Standard Model background
       (with the requirement $|p_\perp(t) | > 50$ GeV) for $pp$ 
        center-of-mass energies of 10 and 20 TeV.}
\label{fig:forty}
\end{figure}
%%%%%%%%%%%%%%%%%%%%%%%%%%%%%%%%%%%%%%%%%%%%%%%%%%%%%%%%%%%%%%%%%%%%%%

If ETC bosons are needed to generate mass in technicolor  models, it is
interesting to ask whether these bosons can be observed directly. In
\leqn{eq:n6}, I showed that the ETC boson associated with the top quark
should have a mass of about 1 TeV, putting it within the mass range
accessible to the LHC.  Arnold and Wendt considered a particular
signature of ETC boson pair production at hadron colliders
\cite{AandW}. They assumed (in contrast to the assumptions of the
previous few paragraphs) that the ETC bosons carry color; this allows
these bosons to be pair-produced in gluon-gluon collisions.  Because
ETC bosons carry technicolor, they will not be produced as free
particles; rather, the ETC boson pair will form a technihadron $\bar E
E$.  These hadrons will decay when the ETC boson emits a top  quark and
turns into a techniquark, $E \to  T \bar t$. When both ETC bosons have
decayed, we are left with  a technicolor pion, which is observed as a
longitudinally polarized $Z^0$.  The full reaction is 
\beq
 gg \to \bar E E \to \bar E T + \bar t \to Z^0 +  t  + \bar t  \ ,
\eeq{eq:p7}
in which the $t Z^0$ system and the $Z^0 t \bar t$ systems both form
definite mass combinations corresponding to technihadrons.  The cross
section for this reaction is shown in Figure~\ref{fig:forty}.  Note
that the multiple peaks in the signal show
contributions from both the $J=0$ and the $J=2$ bound
states of ETC bosons.

A second manifestation of ETC dynamics is less direct, but it is
visible at lower eneriges.  To understand this effect, go back to the
elementary ETC gauge boson coupling that produces the top quark mass,
\beq
   \Delta \L = g_E E_\mu \bar Q_L \gamma^\mu T_L \ ,
\eeq{eq:q7}
where $Q_L = (t,b)_L$ and $T_L = (U,D)_L$.  If we put this interaction
together with a corresponding coupling to the right-handed quarks, we
obtain the term \leqn{eq:k6} which leads to the fermion masses.  On the
other hand, we could contract the vertex \leqn{eq:q7} with its own
Hermitian conjugate.  This gives the vertex 
\beq
     i \Delta\L =  (i g_E \bar Q_L\gamma^\mu T_L) {-i\over -m_E^2}
 (i g_E \bar T_L \gamma_\mu Q_L)  \ .
\eeq{eq:r7}
By a Fierz transformation \cite{PS}, this expression can be rearranged
into
\beq
     i \Delta\L = {-ig_E^2\over m_E^2} ( \bar Q_L\gamma^\mu \tau^a Q_L) 
            ( \bar T_L\gamma_\mu \tau^a T_L)  \ , 
\eeq{eq:rr7}
where $\tau^a$ are the weak isospin matrices.  The last factor gives
just the technicolor currents which couple to the weak interaction
vector bosons. Thus, we can replace this factor by 
\beq
  \bar T_L\gamma_\mu \tau^3 T_L \to  {1\over 4}{e\over cs} f_\pi^2 Z_\mu
\eeq{eq:s7}
Then this term has the interpretation of a technicolor modification of
the $Z\to b\bar b$ and $Z\to t\bar t$ vertices \cite{ChivSS}.

It is not difficult to estimate the size of this effect.  Writing the
new contribution to the $Z^0$ vertex together with the Standard Model
contributions, we have
 \beq
   \Delta\L = {e\over cs} Z_\mu \bar Q_L\gamma^\mu \left\{ \tau^3 - s^2 Q 
           - {g_E^2\over 2 m_E^2} f_\pi^2 \tau^3 \right\}  Q_L \ .
\eeq{eq:t7}
For the left-handed $b$, $\tau^3 = -\half$, and so the quantity in brackets is
\beqa
  g_L^b &=&  -\half + \frac{1}{3}s^2 + \frac{1}{4} {g_E^2\over m_E^2} f_\pi^2 
    \CR
        &=& \bigl( g_L^b \bigr)_\SM \left( 1 - \frac{1}{2}
                   {m_t \over 4\pi f_\pi} \right) \ ,
\eeqa{eq:u7}
where in the last line I have used \leqn{eq:n6} to estimate
$g_E/m_E$. The value of the correction, when squared, is about 6\% and
would tend to decrease the branching ratio for $Z^0 \to b\bar b$.

%%%%%%%%%%%%%%%%%%%%%%%%%%%%%%%%%%%%%%%%%%%%%%%%%%%%%%%%%%%%%%%%%%%%%%
\begin{figure}[tb]
\begin{center}
\leavevmode
{\epsfxsize=2.25in\epsfbox{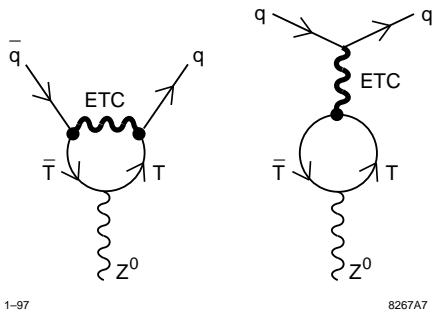}}
\end{center}
 \caption{Modification of the $Z^0 b \bar b$ and $Z^0 t \bar t$
 vertices by ETC interactions.}
\label{fig:fortyone}
\end{figure}
%%%%%%%%%%%%%%%%%%%%%%%%%%%%%%%%%%%%%%%%%%%%%%%%%%%%%%%%%%%%%%%%%%%%%%

The effect that we have estimated is that of the first diagram in
Figure~\ref{fig:fortyone}.  In more complicated models of ETC
\cite{CST,KH,Guo}, effects corresponding to both of the diagrams shown in
the figure contribute, and can also have either sign.  Typically, the
two types of diagrams cancel in the $Z^0 b \bar b$ coupling and add in
the $Z^0 t \bar t$ coupling \cite{Mur}. Thus, it is interesting to
study this effect experimentally in $\ee$ experiments both at the $Z^0$
resonance and at the $t \bar t$ threshold.
 
\subsection{Recapitulation}

In this section, I have discussed the future experimental program of
particle physics for the case in which electroweak symmetry breaking
has its origin in new strong interactions.  We have discussed
model-independent probes of the new strong interaction sector and
experiments which probe specific aspects of technicolor models. In this
case, as opposed to the case of supersymmetry, some of the most
important experiments can only be done at very high energies and
luminosities, corresponding to the highest values being considered for
the next generation of colliders.  Nevertheless, I have argued that, if
plans now proposed can be realized, these experiments form a rich
program which provides a broad experimental view of the new
interactions.

Two sets of contrasting viewpoints appeared in our analysis.  The first
was the contrast between experiments that test model-dependent as
opposed to model-independent conclusions.  The search for technipions,
for corrections to the $Z t \bar t$ vertex, and for other specific
manifestations of technicolor theories can be carried out at energies
well below the 1 TeV scale.  In fact, the precision electroweak
experiments and the current precision determination of the $Z^0 \to
b\bar b $ branching ratio already strongly constrain technicolor
theories.  However, such constraints can be evaded by clever
model-building.  If an anomaly predicted by technicolor is found, it
will be important and remarkable.  But in either case, we will need to
carry out the TeV-energy experiments to see the new interactions
directly and to clearly establish their properties.

The second set of contrasts, which we saw also in our study of
supersymmetry, comes from the different viewpoints offered by $pp$ and
$\ee$ colliders. In the search for anomalies, the use of both types of
experiments clearly offers a broader field for discovery.  But these
two types of facilities also bring different information to the more
systematic program of study of the new strong interactions summarized
in Table 1.  The table makes quantitative the powerful capabilities of
the LHC to explore the new strong interaction sector. But it also shows
that an $\ee$ linear collider  adds to the LHC an exceptional
sensitivity in the $I=1$ channel, reaching well past the unitarity
bound, and sensitivity to the process $W^+W^- \to t\bar t$, which tests
the connection between the new strong interactions and the top quark
mass generation. Again in this example, we see how the LHC and the
linear collider, taken together, provide the information for a broad
and coherent picture of physics beyond the standard model.

\section{Conclusions}

This concludes our grand tour of theoretical ideas about what physics
waits for us at this and the next generation of high-energy colliders. 
I have structured my presentation around two specific concrete models
of new physics---supersymmetry and technicolor.  These models contrast
greatly in their details and call for completely different experimental
programs.  Nevertheless, they have some common features that I would
like to emphasize.

First of all, these models give examples of solutions to the problem I
have argued is the highest-priority problem in elementary particle
physics, the mechanism of electroweak symmetry breaking.  Much work has
been devoted to `minimal' solutions to this problem, in which the
future experimental program should be devoted to finding a few, or even
just one, Higgs scalar bosons.  It is possible that Nature works in
this way. But, for myself, I do not believe it.  Through these
examples, I have tried to explain a very different view of electroweak
symmetry breaking, that this phenomenon arises from a new principle of
physics, and that its essential simplicity is found not by counting the
number of particles in the model but by understanding that the model is
built around a coherent physical mechanism.  New principles have deep
implications, and we have seen in our two examples that these can lead
to a broad and fascinating experimental program.

If my viewpoint is right, these new phenomena are waiting for us,
perhaps already at the LEP 2 and Tevatron experiments of the next few
years, and at the latest at the LHC and the $\ee$ linear collider. If
the new physical principle that we are seeking explains the origin of
$Z$ and $W$ masses, it cannot be too far away.  In each of the models
that I have discussed, I have given a quantitative estimate of the
energy reach required.  At the next generation of colliders, we will be
there.

For those of you who are now students of elementary particle physics,
this conclusion comes with both discouraging and encouraging messages. 
The discouragement comes from the long time scale required to construct
new accelerator facilities and to carry out the large-scale experiments
that are now required on the frontier.  Some of your teachers can
remember a time when a high-energy physics experiment could be done in
one year. Today, the time scale is of order ten years, or longer if the
whole process of designing and constructing the accelerator is
considered.

The experiments that I have described put a premium not only on high
energy but also high luminosity.  This means that not only the
experiments but also the accelerator designs required for these studies
will require careful thinking and brilliant new ideas.  During the
school, Alain Blondel was fond of repeating, `Inverse picobarns must be
earned!'  The price of inverse femtobarns is even higher.  Thus, I
strongly encourage you to become involved in the problems of
accelerator design and the interaction of accelerators with
experiments, to search for solutions to the challenging problems that
must be solve to carry out experiments at 1 TeV and above.

The other side of the message is filled with promise.  If we can have
the patience to wait over the long time intervals that our experiments
require, and to solve the technical problems that they pose, we will
eventually arrive at the physics responsible for electroweak symmetry
breaking. If the conception that I have argued for in these lectures is
correct, this will be genuinely a new fundamental scale in physics,
with new interactions and a rich phenomenological structure. 
Though the experimental discovery and clarification
of this structure will be complex, the accelerators planned for the
next generation---the LHC and the $\ee$ linear collider---will provide
the powerful tools and analysis methods that we will require.  This is
the next frontier in elementary particle physics, and it is waiting for
you.  Enjoy it when it arrives!

\section*{Acknowledgments}

I am grateful to Belen Gavela, Matthias Neubert, and Nick Ellis for
inviting me to speak at the European School, to Alain Blondel,
Susannah Tracy, and Egil Lillestol for providing the very pleasant
arrangements for the school, to the students at the school for their
enthusiasm and for their criticisms,
and to Erez Etzion, Morris Swartz,
 and Ian Hinchliffe for comments on the manuscript.
 This work was supported by the Department
of Energy under contract DE--AC03--76SF00515.

\end{document}